\documentstyle[11pt,aaspp4,flushrt,psfig]{article} 
\font\fiverm=cmr5             \font\sevenrm=cmr7     
     \font\eightrm=cmr8           
              
          \font\sixrm=cmr6

\def\app{Astropart. Phys.}
\def\annrev{Ann. Rev. Astron. Astrophys.}
\def\jetp{Sov. Phys. JETP}
\def\pr{Phys. Rev.}
\def\rmp{Rev. Mod. Phys.}
\def\jpG{J. Phys. G: Nucl. Part. Phys.}
\def\teq#1{$\, #1\,$}                           % text equation
\def\dover#1#2{\hbox{${{\displaystyle{#1} \vphantom{(} }\over{
   \displaystyle{#2} \vphantom{(} }}$}}
\def\erg{\varepsilon}

\def\today{\ifcase\month\or
  January\or February\or March\or April\or May\or June\or
  July\or August\or September\or October\or November\or
  December\fi
  \space\number\day, \number\year}
{\catcode`\@=11
\gdef\SchlangeUnter#1#2{\lower2pt\vbox{\baselineskip 0pt\lineskip0pt
   \ialign{$\m@th#1\hfil##\hfil$\crcr#2\crcr\sim\crcr}}}}
\def\gtrsim{\mathrel{\mathpalette\SchlangeUnter>}}
\def\lesssim{\mathrel{\mathpalette\SchlangeUnter<}}

\begin{document}
%
%      WARNING: THE FOLLOWING MACROS ARE ESSENTIAL FOR AASTEX
%            INTERPRETATION OF THIS DOCUMENT:    DO NOT DELETE!!
%
\newcommand{\vol}[2]{$\,$\rm #1\rm , #2.}                 
\newcommand{\figureout}[2]{\centerline{}
   \centerline{\psfig{figure=#1,width=5.5in}}
    \figcaption{#2}\clearpage } 
\newcommand{\figureoutsmall}[2]{\centerline{\psfig{figure=#1,width=5.0in}}
    \figcaption{#2}\clearpage } 
\newcommand{\figureoutvsmall}[2]{\centerline{\psfig{figure=#1,width=4.0in}}
    \figcaption{#2}\clearpage } 
%
%      END OF ESSENTIAL MACROS.
%
%\vphantom{p}
%\vskip -45pt
%\centerline{\hskip 5.0truein\hfill\today}
%\vskip 15pt 

\title{RADIO TO GAMMA-RAY EMISSION FROM SHELL-TYPE\\
       SUPERNOVA REMNANTS: PREDICTIONS FROM\\
       NON-LINEAR SHOCK ACCELERATION MODELS}
   \author{Matthew G. Baring\altaffilmark{1}}
   \affil{Laboratory for High Energy Astrophysics, Code 661, \\
          NASA Goddard Space Flight Center, Greenbelt, MD 20771, U.S.A.\\
          \it baring@lheavx.gsfc.nasa.gov\rm}
   \altaffiltext{1}{Universities Space Research Association}
   \authoraddr{Laboratory for High Energy Astrophysics, Code 661,
      NASA Goddard Space Flight Center, Greenbelt, MD 20771, U.S.A.}
\vskip 6pt
   \author{Donald C. Ellison \& Stephen P. Reynolds}
   \affil{Department of Physics, North Carolina State University,\\
          Box 8202, Raleigh NC 27695, U.S.A.\\
          \it don\_ellison@ncsu.edu, steve\_reynolds@ncsu.edu\rm}

\and

   \author{Isabelle A. Grenier \& Philippe Goret}
   \affil{Service d'Astrophysique, CEA, DSM, DAPNIA,\\
          Centre d'Etudes de Saclay, 91191 Gif-sur-Yvette, France\\
          \it isabelle.grenier@cea.fr, goret@sapvxg.saclay.cea.fr\rm}

%\slugcomment{To appear in \it The Astrophysical Journal\rm , *** issue.}
%\slugcomment{Accepted for publication in \it The Astrophysical Journal\rm}
%
%\clearpage

\begin{abstract}  
Supernova remnants (SNRs) are widely believed to be the principal
source of galactic cosmic rays, produced by diffusive shock
acceleration in the environs of the remnant's expanding blast wave.
Such energetic particles can produce gamma-rays and lower energy
photons via interactions with the ambient plasma.  The recently
reported observation of TeV gamma-rays from SN1006 by the CANGAROO
Collaboration, combined with the fact that several unidentified EGRET
sources have been associated with known radio/optical/X-ray-emitting
remnants, provides powerful motivation for studying gamma-ray emission
from SNRs.  In this paper, we present results from a Monte Carlo
simulation of non-linear shock structure and acceleration coupled with
photon emission in shell-like SNRs.  These non-linearities are a
by-product of the dynamical influence of the accelerated cosmic rays on
the shocked plasma and result in distributions of cosmic rays which
deviate from pure power-laws.  Such deviations are crucial to
acceleration efficiency considerations and impact photon intensities
and spectral shapes at all energies, producing GeV/TeV intensity ratios
that are quite different from test particle predictions.  The Sedov
scaling solution for SNR expansions is used to estimate important shock
parameters for input into the Monte Carlo simulation.  We calculate ion
(proton and helium) and electron distributions that spawn neutral pion
decay, bremsstrahlung, inverse Compton, and synchrotron emission,
yielding complete photon spectra from radio frequencies to gamma-ray
energies.  The cessation of acceleration caused by the spatial and
temporal limitations of the expanding SNR shell in moderately dense
interstellar regions can yield spectral cutoffs in the TeV energy range
that are consistent with Whipple's TeV upper limits on those EGRET
unidentified sources that have SNR associations.  Supernova remnants in
lower density environments generate higher energy cosmic rays that
produce predominantly inverse Compton emission observable at super-TeV
energies, consistent with the SN1006 detection.  In general, sources in
such low density regions will be gamma-ray dim at GeV energies.
\end{abstract}  
%
%\clearpage 
%  
\section{INTRODUCTION}
 \label{sec:intro}

It is widely believed that supernova remnants (SNRs) are the primary
sources of cosmic-ray ions and electrons up to energies of at least
$\sim 10^{15}$ eV, where the so-called {\it knee} in the spectrum marks
a deviation from almost pure power-law behavior.  Such cosmic rays are
presumed to be generated by diffusive (also called first-order Fermi)
acceleration at the remnants' forward shocks.  These cosmic rays can
generate gamma rays via interactions with the ambient interstellar
medium, including nuclear interactions between relativistic and cold
interstellar ions, by bremsstrahlung of energetic electrons colliding
with the ambient gas, and inverse Compton (IC) emission off background
radiation.  Rudimentary models of gamma-ray production in supernova
remnants involving nuclear interactions date back to the early work of
Higdon \& Lingenfelter (1975) and Chevalier (1977), and later Blandford
\& Cowie (1982).  These preceded the first tentative associations of
two COS-B gamma-ray sources (Pollock 1985) with the remnants $\gamma$
Cygni and W28.  The study of gamma-ray SNRs remained quietly in the
background until the ground-breaking observational program of the EGRET
experiment aboard the Compton Gamma Ray Observatory (CGRO).  This
provided a large number of unidentified sources in the super-50 MeV
band, seen both in and above the galactic plane, which led to the
suggestion (e.g. Sturner \& Dermer 1995) of a possible supernova origin
(see Mukherjee, Grenier, \& Thompson 1997 for a discussion of this
population).  A handful of these EGRET sources have significant
associations with relatively young SNRs (Esposito et al. 1996).

Following the EGRET advances, the modeling of gamma-ray emission from
supernova remnants began in earnest with the paper of Drury, Aharonian,
\& V\"olk (1994), who computed (as did Naito and Takahara 1994) the
photon spectra expected from the decay of neutral pions generated in
collisions of shock-accelerated ions with those of the interstellar
medium (ISM).  These works assumed that the ions have power-law spectra
extending to $\sim 10^{14}$ eV or beyond.  Since then, a number of
alternative models examining other (i.e. electromagnetic) radiation
processes have been presented.  These include the work of Gaisser,
Protheroe, \& Stanev (1998) and Sturner et al.~(1997), who did not
treat non-linear effects from efficient particle acceleration, and the
recent analysis of Berezhko \& V\"olk (1997), who solved the fully
momentum-dependent diffusion-convection equation and included
non-linear effects, but treated ion injection in a parametric fashion.
Here, we include the injection and acceleration of both ions and {\it
thermal electrons} in non-linear shocks and describe cases where
emission from energetic electrons is likely to dominate ion emission,
as expected in low density regions where flat inverse Compton gamma-ray
components become important.  This fact was exploited by Mastichiadis
\& de Jager (1996) and Pohl (1996) to propose that SN1006 should
exhibit such a component, a timely prediction given the subsequent
report of a spatially-resolved (to the NW rim of the shell) detection
of SN1006 by the CANGAROO experiment (Tanimori et al. 1997) at energies
above 1.7 TeV.  Non-linear spectral models of diffusive shock
acceleration that include the back-reaction of the accelerated
particles on the shock structure do not produce exactly power-law
particle distributions, and generate electron and ion spectra which
differ considerably from each other.  Central to predictions of photon
emission from SNR shock acceleration are the details of this spectral
curvature, the maximum energies of the ions and electrons, their
relative abundances, and the enhancements of density of heavy ions
relative to protons caused by non-linear shock modification.

In this paper, we calculate {\it both} the ion and electron spectra
resulting from {\it non-linear}, cosmic ray-modified SNR shocks and,
using these, compute photon emission over the entire range of the
electromagnetic spectrum from radio waves to gamma-rays.  The
non-linear shock model we use is described in detail in Ellison, Jones,
\& Reynolds (1990), Jones \& Ellison (1991) and Ellison, Baring, \&
Jones (1996), and consists of a Monte Carlo simulation of the transport
of particles through a steady-state, plane-parallel shock.  The chief
advantage this technique has over other non-linear shock models is that
particle injection can be treated in a largely self-consistent fashion,
and this feature has been tested against particle distributions
observed at the Earth's bow shock (Ellison, M\"obius, \& Paschmann
1990) and in linear applications to interplanetary shocks (e.g. Baring,
et al. 1997).  The non-linear aspects of shock acceleration have been
shown to be consistent with the spectral index and curvature inferred
for relativistic electrons emitting radio synchrotron radiation in the
Tycho and Kepler remnants (Reynolds \& Ellison, 1992), and have very
recently been shown to produce the observed cosmic ray chemical
composition if normal (i.e. cosmic abundance) ISM gas and dust are
accelerated in smoothed SNR shocks (Meyer, Drury, \& Ellison 1997;
Ellison, Drury, \& Meyer 1997).  By imposing mass, momentum, and energy
conservation, we obtain a steady-state solution which self-consistently
includes {\it ion} injection and acceleration and simultaneously yields
the average shock structure and complete particle distribution
functions.  As with previous applications of our Monte Carlo technique,
we assume that the scattering properties of the particles, thermal and
energetic, obey simple scattering laws, in accord with heliospheric
shock observations and results of plasma simulations.  Via this
prescription, our ion injection model embodies the essential plasma
properties at least as far as gross observables are concerned.

A new feature included here is a parametric model for {\it thermal
electron} injection and suprathermal scattering.  With two additional
parameters, we calculate the complete distributions and absolute
acceleration efficiencies of both ions and electrons in shocks capable
of accelerating particles to TeV and super-TeV energies.  This allows
us to simultaneously describe photon emission from ions and electrons,
over a wide range of photon energies, in a single shock with a single
set of environmental and shock parameters.    The electron-to-proton
ratio at energies above $\sim 1$ GeV depends strongly on these
parameters and, in conjunction with multi-wavelength observations of
SNRs, permits us to constrain these parameters.  Our model spectra
include synchrotron radiation, bremsstrahlung, inverse Compton
scattering off the cosmic microwave background, and the decay of
neutral pions produced in collisions of nucleons.  In this latter
nuclear process, we find that contributions from accelerated alpha
particles are comparable to those from protons despite a cosmic
abundance ratio of \teq{\sim 0.1}, mostly since heavier elements are
accelerated more efficiently by the Fermi mechanism in non-linear
shocks.

The steady-state Monte Carlo technique precludes an exact dynamical
description for a remnant as it expands into the ISM.  To model such
time dependence, we use standard Sedov solutions for SNR evolution in
homogeneous media to estimate the shock speed \teq{V_{\rm sk}} and
radius \teq{R_{\rm sk}} at any age.  We then calculate the maximum
energy \teq{E_{\rm max}} to which particles can be accelerated
according to the well-known Fermi acceleration formula resulting from
the diffusion approximation.  Our best-approximation steady-state model
is then obtained by including an upstream free escape boundary set by
the diffusion length of the particles with \teq{E_{\rm max}}.  In this
way we incorporate into our Monte Carlo simulation the most essential
aspects and consequences of time-dependence in SNR expansions:
\teq{V_{\rm sk}} and \teq{E_{\rm max}} are the two key quantities for
any non-linear model of acceleration in expanding SNR shocks.  This
simple picture using the Sedov evolution does not account for the
energy from cosmic rays escaping upstream from the expanding shock:
hence a real SNR shock will expand less rapidly than the Sedov solution
predicts.  In addition, our picture omits adiabatic gains/losses
interior to the shock, which can be influential in determining
\teq{E_{\rm max}} (Berezhko \& V\"olk 1997).  Despite these
shortcomings, we argue that this hybrid technique overcomes the major
deficiencies of our steady-state and plane shock approximations and
contains the essential non-linear effects that are expected in
efficient particle acceleration.  Work is currently in progress
(Berezhko \& Ellison 1998, in preparation) to detail the differences
between this  steady-state, planar Monte Carlo model and the
time-dependent, spherical SNR shock model of Berezhko, Yelshin, \&
Ksenofontov (1996).  Preliminary results suggest the differences are
small.

We report here the parameters required for the maximum particle
energies obtained by diffusive shock acceleration to be consistent with
current upper limits from the Whipple and HEGRA atmospheric
\v{C}erenkov telescopes on remnants with putative EGRET source
associations.  We find that maximum particle energies of a few TeV for
ISM densities near $\sim 1$ proton cm$^{-3}$ are not inconsistent with
Fermi acceleration at supernova remnant shocks.  At the same time, our
models predict relatively weak emission in the 100 MeV--10 TeV range
for low upstream densities, \teq{\lesssim 1}cm$^{-3}$, which poses no
problem if the unidentified EGRET sources catalogued by Esposito et
al.  (1996) are not connected with shell emission.  This scenario
permits acceleration to much higher energies, though we find it
difficult, in accord with many previous papers, to accelerate up to the
{\it knee} in the cosmic ray spectrum at \teq{\sim 10^{15}}eV without
resorting to unrealistically low ISM densities.  The lower density
models may be most appropriate to sources out of the galactic plane
like SN1006, where super-50 TeV electrons are inferred from both X-ray
(presumably synchrotron) and TeV gamma-ray (probably inverse Compton)
observations (see Koyama et al. 1995 and Tanimori et al. 1998,
respectively).  Taken together, these complementary results illustrate
a general property of our modeling, and also of non-linear shock
acceleration theories in general, namely an anti-correlation between
the EGRET band flux and the maximum energy of cosmic ray acceleration.
We also produce broad-band emission spectra for a range of parameters
and specifically compare our results to observations of IC 443.  In the
gamma-ray band, we have little trouble reproducing the spectral index
and flux of the EGRET source 2EG J0618+2234 provided that inverse
Compton emission is relatively unimportant, a conclusion reached by
Gaisser, Protheroe, \& Stanev (1998).  However, we find that the
unusually flat radio spectrum of IC 443 is {\it not} well modeled with
standard Fermi acceleration without resorting to extremely inefficient
electron scattering.  If this flat radio spectrum cannot be attributed
to thermal contamination or free-free absorption, then it is probable
that some source other than particle acceleration at the SNR blast wave
is responsible for the photon emission.

The results and discussions of this paper identify a number of
important issues that should form focuses of future theoretical
research, including the radial/angular extent of X-ray and gamma-ray
emission, the modeling of flat spectrum radio sources, spatial
variations in radio, X-ray and gamma-ray spectral indices, identifying
which physical processes are responsible for the non-thermal X-ray and
gamma-ray flux, the role of magnetic-field obliquity around the shell,
the \teq{e/p} ratio and a more complete description of electron
injection, and cosmic ray abundances and production up to the knee.
Efforts in this direction should anticipate the expected improvements
in the near future in sensitivity and angular resolution of
ground-based and satellite X-ray and gamma-ray telescopes.

\section{FERMI ACCELERATION IN SUPERNOVA REMNANTS}
 \label{sec:model}

\subsection{The Monte Carlo Calculation of Fermi Acceleration}
 \label{sec:MonteCarlo}

Apart from the electron injection model introduced here, the Monte
Carlo simulation we use to model the diffusive shock acceleration of
ions has been described in detail elsewhere (e.g. Jones \& Ellison
1991; Baring, Ellison, \& Jones  1993; Ellison, Baring, \& Jones
1996).  It is a kinematic model, closely following Bell's (1978)
approach to diffusive acceleration, where the simulation is used to
calculate, in effect, solutions to a Boltzmann equation for particle
transport involving a collision operator, without making any assumption
concerning the isotropy of particle distributions.  Particles are
injected at a position far upstream and allowed to convect into the
shock (i.e. mimicking the interstellar medium exterior to a remnant
that is overtaken by the blast wave), diffusing between postulated
scattering centers (presumably magnetic irregularities in the
background plasma and self-generated turbulence) along the way.  As
particles diffuse between the upstream and downstream regions, they
continually gain energy (for a simulation example, see Fig.~3 of
Baring, Ellison, \& Jones 1994) in accord with the Fermi mechanism.
Our models here are restricted to infinite plane, steady-state, {\it
parallel} shocks where the angle \teq{\Theta_{\hbox{\sevenrm B}n}}
between the upstream magnetic field and the shock normal is assumed to
be zero everywhere.  As detailed below, the maximum linear scale in the
shock, the diffusion length of the highest-energy particles, is always
less than 1/4 of the shock radius, $R_{\rm sk}$, and almost all of the
flow deceleration (in the shock frame) occurs within $\Delta R/R_{\rm
sk} < 0.1$.  In a Sedov spherical blast wave, post-shock expansion and
velocity gradients can also affect the particle distribution, causing
adiabatic losses (Berezhko 1996); we expect these effects to be
confined to the highest-energy particles, so that the maximum energies
we obtain at later times may be slightly inaccurate.  However, at
earlier times the diffusion length scale is even less than $0.1 R_{\rm
sk}$, and the plane-shock approximation should be accurate.  The
non-linear Monte Carlo  technique has been generalized to oblique shock
geometry (i.e. shocks with \teq{\Theta_{\hbox{\sevenrm B}n} >0^\circ};
e.g.  Ellison, Baring, \& Jones 1996), and shock obliquity undoubtedly
plays an important role in supernova remnant considerations, for
example in the work of Fulbright \& Reynolds (1990) and Reynolds
(1996); treatment of it is deferred to future work.

\subsubsection{Particle Scattering}
 \label{sec:scatter}

We assume that particles of speed \teq{v} (measured in the local plasma
frame) scatter isotropically in this plasma frame with an exponential
distribution about a collision time \teq{t_c=\lambda/v} for mean free
paths \teq{\lambda}.  The particles make large angle scatterings,
mimicking diffusion in strongly turbulent plasmas where $\vert\delta
{\bf B}\vert /\vert {\bf B}\vert \sim 1$.  Such strong turbulence is
commonly observed in heliospheric shock environments (e.g. Hoppe et
al.  1981; Tsurutani, Smith, \& Jones 1983; Balogh et al.  1993), and
has been inferred near young supernova remnants (Achterberg, Blandford,
\& Reynolds 1994).  Note that Ellison, Baring, \& Jones (1996) observed
that the acceleration process was only weakly dependent on the type of
the scattering as long as $\Theta_{\hbox{\sevenrm B}n}$ is not close to
$90^\circ$:  pitch-angle diffusion (small-angle scattering) and
large-angle scattering generated similar particle distributions for a
wide range of shock parameters.  In the upstream region, the scattering
centers move at a speed, $v_{\hbox{\sixrm A}}$  relative to the bulk
flow speed $u(x)$, where $v_{\hbox{\sixrm A}} = B/\sqrt{4\pi n_{\rm p}
m_{\rm p}} \simeq 2.2\, (B/\mu {\rm G}) (n_{\rm p}/{\rm
cm}^{-3})^{-1/2}$  \hbox{km~s$^{-1}$}, is the Alfv\'en speed (here, $B$
is the magnetic field, $n_{\rm p}$ is the proton number density, and
$m_{\rm p}$ is the proton mass.  Thus, the scattering is inelastic in
the plasma frame (unless particle speeds far exceed the Alfv\'en speed)
and energy can be transferred from the superthermal population to the
background thermal gas. This effect is discussed in more detail in
Section 2.1.3.

We adopt a phenomenological mean free path to describe the complicated
plasma microphysics in a very simple prescription.  Specifically, we
take the scattering mean free path parallel to the mean magnetic field,
\teq{\lambda_i}, of {\it all} ions, thermal and super-thermal,
including both protons and heavier species, to be
\begin{equation}
   \lambda_i\; =\;\eta \, r_{\rm g}\ ,
 \label{eq:mfp}
\end{equation}
where $r_{\rm g} = pc/(QeB)$ is the gyroradius and \teq{\eta} is a
constant, independent of ion species, energy, and position relative to
the shock.  Here, \teq{p} is the particle momentum measured in the
local plasma frame, \teq{Q} is the charge number, and \teq{-e} is the
electronic charge.  The {\it spatial diffusion coefficient} along the
field is then \teq{\kappa=\lambda_i v/3}.  Note that for shocks of
speed $V_{\rm sk}$, \teq{\kappa /V_{\rm sk}} approximates the upstream
diffusion length scale.  The minimum  value of $\eta$ is unity, the
so-called Bohm limit, where diffusion is comparable along and
perpendicular to the mean magnetic field: this corresponds to strong
turbulence, \teq{\vert\delta {\bf B}\vert /\vert {\bf B}\vert \sim 1}.
The aptness of a power-law prescription, \teq{\lambda_i\propto
p^\alpha}, to shocked plasma environments is supported by particle
observations at the Earth's bow shock (where \teq{1/2 < \alpha < 3/2},
Ellison, M\"obius, \& Paschmann 1990), deductions from ions accelerated
in solar particle events (Mason, Gloeckler, \& Hovestadt 1983, where
\teq{1/2 < \alpha < 4/5}), and also from turbulence in the
interplanetary magnetic field (Moussas et al. 1992).  On the
theoretical side, plasma simulations (Giacalone, Burgess, \& Schwartz
1992) suggest a mean free path obeying \teq{\lambda_i\propto p^\alpha}
with \teq{\alpha \sim 2/3}.  We believe that the assumption of
\teq{\lambda_i\propto p} is simple, physically realistic, and
representative of the Fermi acceleration process.

\subsubsection{Electron Scattering and Injection Model}
 \label{sec:electrons}

The injection of thermal electrons into the Fermi process is poorly
understood.  This is a prominent problem in astrophysics in general,
and for supernova remnants in particular, given few or no palpable
radiative signatures of energetic ions.  In contrast, evidence of
non-thermal electrons in remnants is common, including ubiquitous
observations of radio synchrotron emission, and now detections of
non-thermal X-rays (Koyama et al.~1995; Keohane et al.  1997; Allen et
al. 1997) from three SNRs and the report of gamma-rays from SN1006
(Tanimori et al.  1997, 1998).  While protons resonantly create and
scatter off Alfv\'en waves at all energies from thermal upwards, and so
have a ready supply of magnetic turbulence for providing spatial
diffusion (for example see Lee's 1982 model of the Earth's bow shock),
it is unclear whether there is a significant presence in shocked
plasmas of the much shorter wavelength waves, i.e. whistlers, that
resonate with thermal and suprathermal electrons.  Levinson (1992,
1996) included whistlers in his quasi-linear theory description of wave
generation and electron diffusion and acceleration at kinetic energies
between $\sim 5$ keV and 3 MeV.  Levinson's model therefore may not
describe electron injection from truly thermal energies in the case of
old SNRs, where the plasma temperatures are well below 5 keV, but may
apply to younger remnants with shock velocities of several thousand km
s$^{-1}$.  Galeev, Malkov \& V\"olk (1995) have suggested that oblique,
lower hybrid waves can be excited by ion beams and that these waves can
then accelerate thermal electrons. However, this mechanism remains
highly speculative and is restricted, in any case, to
quasi-perpendicular shocks.  Hence, below a few keV, the situation
remains inconclusive, being complicated by the fact that whistlers can
be strongly damped in warm or hot plasmas; Alfv\'en modes usually
escape this fate.  We note that the low energy electron injection issue
can be circumvented in alternative scenarios, such as that put forward
by Ellison, Jones, \& Ramaty (1990) and Chan \& Lingenfelter (1993),
where the decay of nucleosynthetic material provides an injection of
MeV leptons into the Fermi process.

Our electron model assumes that, at high electron momenta, the electron
mean free path \teq{\lambda_{\rm e}} is proportional to the electron
gyroradius, exactly as in equation~(\ref{eq:mfp}). At lower momenta,
however, we modify equation~(\ref{eq:mfp}) by introducing an arbitrary
momentum, \teq{p_{\rm crit}} [or, equivalently, kinetic energy
\teq{E_{\rm crit} = \sqrt{p_{\rm crit}^2 c^2 + (m_{\rm e} c^2)^2}
-m_{\rm e} c^2}], below which electrons have a constant mean free path,
i.e.
\begin{equation}
   \lambda_{\rm e} =\;\cases{
    \eta\, r_{\rm g,e}(p_{\rm crit}) \ = \ {\rm constant}
                     \; ,& $\quad p\leq p_{\rm crit}\vphantom{\bigl)}$ \cr
    \eta\, r_{\rm g,e}(p)\; ,& $\quad p>p_{\rm crit}\vphantom{\bigl)}$ \ , \cr}
 \label{eq:mfpe}
\end{equation}
where $r_{\rm g,e}(p) = pc/(eB)$ is the electron gyroradius for the
electron momentum \teq{p}.  Keeping $\lambda_{\rm e}$ constant below
$p_{\rm crit}$ is one way of describing inefficient scattering at low
energies.  In addition, we inject electrons, not with typical upstream
thermal energies (e.g.  $kT\sim$ few eV) or downstream Rankine-Hugoniot
temperatures, but at some fraction of the {\it downstream proton}
temperature $T_{\rm p,DS}$ which results from the thermalization over
the {\it subshock velocity jump}:  $kT_{\rm p,DS} \sim m_{\rm p}
(V_{\rm sub} - u_2)^2$, where $V_{\rm sub}$ is the flow speed at the
subshock, $u_2= V_{\rm sk}/r$ is the downstream flow speed, and $r$ is
the overall shock compression ratio.  In plasma shocks, electrons are
in fact heated at the subshock by plasma processes (e.g.  Cargill \&
Papadopoulos 1988) such as electric fields induced by particle motions
and electron-proton charge separations, Given that this heating takes
place mainly at the subshock, which can be much weaker than the overall
shock (e.g. $r_{\rm sub} \sim 2.5$, see Figure~\ref{fig:fourprofiles}
below), we parameterize the electron heating by setting the downstream
thermal electron temperature, $T_{\rm e,DS}$, to
\begin{equation}
   \dover{3}{2} \, k \, T_{\rm e,DS} \; =\;
   f_{\rm e} \,  {1\over 2} \,  m_{\rm p} \,
   \bigl(\Delta V_{\rm sub}\bigr)^2 \ ,
 \label{eq:electrontemp}
\end{equation}
where $\Delta V_{\rm sub} = V_{\rm sub} - u_2$.  In principle, the
parameter $f_{\rm e} \leq 1$ can be varied to match X-ray
observations.  In practice,  we inject electrons in our simulation far
{\it upstream} with a temperature $T_{\rm e,inj} \equiv T_{\rm e,DS}$,
since the energy the electrons gain in their first crossing of the
shock  from compression is generally much less than \teq{k \, T_{\rm
e,DS}}.  This prescription results in electron temperatures consistent
with those deduced from observations of thermal X-ray emission in SNRs
(for a recent collection of observational studies, see Zimmermann,
Tr\"umper, \& Yorke 1996).  Note that we treat electrons as test
particles and do not include any influence they have on the shock
dynamics.  The relaxation of this approximation is deferred to future
work, though electrons are dynamically unimportant for most of our
models and likely to be so for most astrophysical conditions.

We believe that equations~(\ref{eq:mfpe}) and (\ref{eq:electrontemp})
constitute a simple model for electron injection that addresses the
most salient features of Levinson's (1992) developments, without adding
unnecessary parameters whose determination is beyond current
observational capabilities.  In particular, equation~(\ref{eq:mfpe})
models the expected inefficiency of electron scattering compared to
ions at thermal and suprathermal energies.  Our prescription
guarantees, through the parameters $p_{\rm crit}$ and $f_{\rm e}$,
acceptable injection efficiencies for thermal electrons in smoothed
non-linear shocks, primarily because they need to sample long length
scales in order to experience the greatest possible compressive power
of the shock.
\footnote[1]{
We note that the results we present here for electron injection
efficiencies are in partial disagreement with our earlier results in
Ellison \& Reynolds (1991). In that paper (specifically Figure 7 in
that paper), we claimed that electron injection efficiencies were
extremely sensitive to the injection energy and that electrons injected
with energies less that several 10's of keV would fall many orders of
magnitude below protons in the super-GeV domain. Our current results
for the electron injection efficiency, are somewhat less  sensitive to
injection parameters and do not show the strong decrease in electron
normalization compared to protons we claimed earlier.  We believe our
current results are correct and that our previous claim was an error.
In any case, this error is restricted to the lowest energy electrons
and the {\it shape} of the electron spectra above $\sim 100$ keV is
unaffected.  In particular, our modeling of the radio synchrotron
emission (Reynolds \& Ellison 1992) is unaffected by this since only
the shape of the electron spectrum at relativistic energies was used.
}

\placefigure{fig:TestProfile}
Since this is the first presentation of our electron injection model,
we give a simple example to illustrate its basic properties. Using the
artificial shock flow profile shown in Figure~\ref{fig:TestProfile}, we
inject and accelerate electrons and protons keeping the shock profile
fixed, i.e. we are doing a test-particle example and do not attempt to
find a self-consistent solution. Both electrons and protons are
injected at $x=0$ in this test case with $\delta$-function
distributions at 1 keV; in our self-consistent models addressed later,
we always inject particles far upstream.  The protons are scattered
using equation~(\ref{eq:mfp}) and the electrons using
equation~(\ref{eq:mfpe}).  The densities in scalar momentum space,
$f(|{\bf p}|)$, are shown in Figure~\ref{fig:TestSpectra}, where we
plot $|{\bf p}|^{2.5} f(|{\bf p}|)$ to flatten the spectra.  The top
curve (solid line) is the proton distribution, while the lower dashed
curve is the  electron distribution with $p_{\rm crit}=0$ (i.e. for
this example, electrons and protons have identical functions for their
mean free paths).  From test-particle Fermi acceleration theory (e.g.
Blandford \& Ostriker 1978), we expect that
\begin{equation}
   f(|{\bf p}|) \ d|{\bf p}| \;\propto\;
   |{\bf p}|^{-\sigma} \ d|{\bf p}| \qquad {\rm with} \qquad
   \sigma \; =\; \dover{r_{\rm eff} + 2}{r_{\rm eff} - 1} \ ,
 \label{eq:FofP}
\end{equation}
where $f(|{\bf p}|) d|{\bf p}|$ is the number density between $|{\bf
p}|$ and $|{\bf p}| + d|{\bf p}|$ and $r_{\rm eff}$ is the effective
compression ratio ``felt'' by particles with a particular upstream
diffusion length.  For a spatial diffusion coefficient
\teq{\kappa=\lambda v/3}, the upstream diffusion length, $L_{\rm D}$,
is approximated by
\begin{equation}
   L_{\rm D} \;\simeq\; {\kappa \over {u(x)}} \;\simeq\;
   \dover{\lambda v}{3 u(x)} \, =\, \dover{\eta r_{\rm g} v}{3 u(x)}\ .
 \label{eq:DiffLength}
\end{equation}
For low energy particles with $-0.2\, \eta \, r_{\rm g1} < -L_{\rm D} <
0 $, $r_{\rm eff}=2$ ($\sigma = 4$) (note that we define the upstream
direction to be negative in Figure~\ref{fig:TestProfile}). As particles
increase in energy, the magnitude of $L_{\rm D}$ increases and for
$-10\, \eta\, r_{\rm g1} < -L_{\rm D} < -0.2\, \eta\, r_{\rm g1}$,
particles will feel $r_{\rm eff} \simeq 3$ ($\sigma = 2.5$).  For
$-L_{\rm D} < -10\, \eta\, r_{\rm g1}$, $r_{\rm eff} = 4$ ($\sigma =
2$).  The different $r_{\rm eff}$'s translate into spectral breaks
which are clearly visible in Figure~\ref{fig:TestSpectra}.  The heavy
vertical lines are calculated from equation~(\ref{eq:DiffLength}) and
indicate the momenta corresponding to $L_{\rm D} = -0.2 \, \eta\,
r_{\rm g1}$ and $-10 \, \eta\, r_{\rm g1}$ for electrons and protons.
The spectral breaks occur within a factor of two of $L_{\rm D}$
predicted by equation~(\ref{eq:DiffLength}).  Note that the
distributions plotted are calculated downstream from the shock and the
fact that the proton ``thermal'' peak is at a much higher momentum than
the electron peak reflects both the fact that if protons and electrons
have the same energy, the proton momentum will be $\sqrt{m_{\rm p} /
m_{\rm e}}$ greater, and that protons receive a much larger energy
boost in a single shock crossing than do electrons.  The dotted curve
in Figure~\ref{fig:TestSpectra} shows $f(|{\bf p}|)$ for electrons with
$p_{\rm crit} =1.5\times 10^{-3} \, m_{\rm p} c$ (i.e. $E_{\rm crit} =
3$ MeV). With this $p_{\rm crit}$, and keeping $E_{\rm inj}=1$ keV, the
upstream electron diffusion length, at injection, is
\begin{equation}
   L_{\rm D} \;\simeq\;
   \dover{\eta\, r_{\rm g}(p_{\rm crit}) v_{\rm inj}}{3 u(x)}
   \;\simeq\; -0.26 \ \eta\, r_{\rm g1}\ .
\end{equation}
The position $-0.26\, \eta\, r_{\rm g1}$ is indicated in
Figure~\ref{fig:TestProfile} by an arrow.  Thus, the lowest energy
electrons will have an upstream diffusion length such that $r_{\rm eff}
= 3$ and this is reflected in the fact that the spectrum at the lowest
energies has $\sigma \simeq 2.5$.  Once electrons obtain $-L_{\rm D} <
-10 \, \eta r_{\rm g1}$, their slope flattens to $\sigma = 2$, as
occurred with the dashed curve.

\placefigure{fig:TestSpectra}
One important consequence of our choice of \teq{\lambda\propto r_{\rm
g}} is that, except for the possibility that the values for \teq{\eta}
may differ, non-relativistic electrons and protons of a given energy
\teq{E} have {\it the same upstream diffusion length}, feel the same
effective compression ratio, and hence will attain the same slope
(provided \teq{E \ge E_{\rm crit}}, since then \teq{\eta_{\rm
e}=\eta_{\rm p}}); such equality is true also for fully relativistic
particles.  Deviations from this behavior arise in the
trans-relativistic regime, thereby generating an adjustment in the
relative normalizations of the distributions of electrons and protons.
Electrons at \teq{E<E_{\rm crit}} will possess steeper distributions
than those at \teq{E>E_{\rm crit}}.  This simple picture can be altered
by other injection conditions if, for example as indicated in
Figure~\ref{fig:TestSpectra}, the proton energy after a single shock
crossing is well above the electron energy.  Note that in a smooth
shock, the {\it less} efficiently electrons are scattered, the {\it
more} efficiently they will be accelerated.  This behavior is clearly
illustrated in Figure~\ref{fig:TestSpectra}, where the slope obtained
at a particular energy is determined by the effective compression ratio
that a particle feels as it scatters back and forth across the shock.
The further upstream a particle diffuses, the greater the effective
compression ratio and the flatter the subsequent spectrum at a given
energy, corresponding to greater acceleration efficiency.  At electron
energies well above an MeV, the combined effect of $E_{\rm crit}$ and
$f_{\rm e}$ is just to scale the intensity of the electron spectrum:
the larger $E_{\rm crit}$ (corresponding, say, to greater damping of
whistler waves) and/or $f_{\rm e}$, the more efficiently the electrons
are injected and accelerated, but the spectral shape stays constant at
high energies.

\subsubsection{Smoothed, Non-Linear Shocks}
 \label{sec:non-linear}

While most applications of shock acceleration theory to astrophysics
are {\it test-particle} ones (e.g. see Jones and Ellison 1991; Baring
1997, for discussions), non-linear effects become important in strong
shocks when the energy density in accelerated particles is comparable
to the thermal gas pressure.  If this is the case, the flow
hydrodynamics are modified by the backpressure of the accelerated
particles, forcing the upstream plasma to decelerate forming a
precursor to the discontinuous viscous subshock.  In our Monte Carlo
simulation, the spatial structure of the shock is determined by
iteration of both the average flow speed throughout the shock and the
overall compression ratio, until the mass, momentum, and energy fluxes
are constant everywhere; typical velocity profiles are depicted in
Figure~\ref{fig:fourprofiles} (discussed below).  The non-linearity of
this problem is manifested through the feedback of the particles on the
flow velocity, which in turn determines the shape of the particle
distribution.  The net effect that emerges is one where the overall
compression ratio, from far upstream to far downstream of the
discontinuity, {\it exceeds} that of the test-particle scenario.  This
phenomenon was identified by Eichler (1984), and Ellison \& Eichler
(1984), and arises because (i) high energy particles escape from the
shock which reduces the overall energy density and pressure allowing
the compression of the downstream gas to increase, and (ii) production
of relativistic particles softens the equation of state of the gas,
also allowing the net compression to increase.  Losses due to cooling
can also generate very large compression ratios during the radiative
phase of supernova remnant evolution.  As far as shock dynamics are
concerned, radiative cooling via the escape of photons is completely
analogous to the escape of particles.

We note that the results presented here include non-adiabatic heating
of the upstream thermal gas through the generation and dissipation of
Alfv\'en waves in a manner similar to that assumed by McKenzie \&
V\"olk (1984) and Markiewicz, Drury, \& V\"olk (1990).  The overall
acceleration efficiency depends critically on the strength of the
subshock which, in turn, depends on the amount of heating in the
precursor. If heating is minimal, as with adiabatic compression, the
subshock will be strong and the injection and acceleration of particles
at the subshock will be efficient. This will result in a large escaping
energy flux at the highest energies and a large overall compression
ratio. On the other hand, if heating beyond that from adiabatic
compression takes place, the subshock will be weaker, injection and
acceleration will be less, and the escaping energy flux and overall
compression ratio will be lower. This effect is discussed in detail in
Berezhko, Yelshin, \& Ksenofontov (1996) and we use their technique for
approximating the heating due to Alfv\'en wave dissipation.  Briefly,
it is assumed that cosmic rays generate Alfv\'en waves which rapidly
saturate. At this point, the background gas is heated {\it at the same
rate} as energy from the cosmic rays is transfered to the Alfv\'en
waves, independent of the details of the damping mechanism. In
addition, we follow Berezhko, Yelshin, \& Ksenofontov and assume that
the upstream Alfv\'en waves propagate primarily toward the shock so
that the speed of the upstream scattering centers responsible for
particle acceleration is reduced by the Alfv\'en speed.  Downstream, we
assume the waves are frozen in the fluid.  An important difference
between our treatment of Alfv\'en wave dissipation and that of
Berezhko, Yelshin, \& Ksenofontov is that they assume that the Alfv\'en
waves saturate at $\delta B \sim B$, i.e. the Bohm limit, while we keep
$\eta$ in equation~(\ref{eq:mfp}) a free parameter.  As Berezhko,
Yelshin, \& Ksenofontov describe, the effects of this heating are most
important at high Alfv\'en Mach numbers and can dramatically reduce the
overall compression ratio from values $\sim 100$  in weak magnetic
field conditions to values not much above the test-particle value in
strong magnetic fields.  A paper detailing the implementation of this
effect in the Monte Carlo simulation is in preparation (Berezhko \&
Ellison 1998).

As long as the diffusion coefficient is an increasing function of
energy, pure power laws are not produced in non-linear shocks.  Since
higher energy particles have longer diffusion lengths, they sample a
broader portion of the flow velocity profile, and feel larger
compression ratios.  Consequently, these particles have a flatter
power-law index than those at lower energies, thereby driving the
pressure in a non-linear fashion.  The severity of the non-linearity is
determined by the overall scale of the shock precursor which couples to
the diffusion length $d_{\rm max} \sim \kappa(E_{\rm max})/u_{\rm sk}$
of the highest energy particles in the system.  This defines the scale
of the turbulent foreshock region, beyond which waves generated by the
shocked plasma do not penetrate into the ISM.  We discuss how the
maximum energy \teq{E_{\rm max}} is determined in Section~2.3 below;
for now, we remark that acceleration can be limited by particles
escaping if they diffuse sufficiently far ahead of the shock, of the
order of a few tenths of the shock radius.  More commonly for young
SNRs such as Cas A, the finite age of a SNR shock limits the time
available for particle acceleration, giving a lower maximum energy.

In the Monte Carlo simulation, we limit the acceleration by introducing
an upstream free escape boundary (FEB) at the distance,
$d_{\hbox{\sevenrm FEB}} = d_{\rm max}$, ahead of the shock.  Shocked
particles reaching the FEB stream freely across it and are lost from
the system (i.e. to the interstellar medium outside) effectively
truncating the acceleration process.  This boundary could correspond to
the finite curvature of a real SNR shock, in which case it would scale
as some fraction of the shock radius, or it could correspond to the
finite-age limit as we discuss below.  Note that the relevant size of
the acceleration region could be further constrained by the presence of
dense neutral or incompletely ionized material (Drury, Duffy, \& Kirk
1996), since such regions strongly suppress wave generation.  We model
the downstream region as a uniform flow, ignoring the radial dependence
of the flow speed that emerges from scaling solutions (e.g. Sedov
1959).  This approximation is acceptable as long as $d_{\rm max} \ll
R_{\rm sk}$, and if this applies, adiabatic losses/gains are small, so
we neglect them in this paper.  Note that Berezhko (1996) finds that
downstream adiabatic heating and geometrical effects in a spherical
blast wave increase the maximum energy somewhat above the finite-age
limit in a plane shock.  Berezhko contends that such increases arise
implicitly because \teq{d_{\rm max}} approaches \teq{\sim 0.1R_{\rm
sk}}.  Such modifications to our approach may prove necessary at late
times, as discussed in Section~2.3.2.

\subsection{SNR Blast Waves and the Sedov Solution}
 \label{sec:Sedov}

Initial ejection velocities in supernovae are of order \teq{5 \times
10^3 - 2 \times 10^4}  \hbox{km~s$^{-1}$}\ (e.g. Chevalier 1981), and
the ejecta push a blast wave into the ISM. Initially, the forward
moving blast wave expands relatively freely, but the actual evolution
depends on the spatial density profile of the ejecta and of the
surrounding, pre-supernova material; for power-law variations of ejecta
density with radius, Chevalier (1982) found a self-similar driven wave
solution in which the outer blast wave radius, \teq{R_{\rm sk}}, varies
as a power, $m$,  of time between 0.57 and~1 (i.e. \teq{R_{\rm
sk}\propto t^m} and \teq{V_{\rm sk}\propto t^{m-1}}) depending on the
power-law exponents in ejecta and circumstellar medium.  As the
swept-up mass, $M_{\rm swept}$, increases, eventually this self-similar
evolution is broken, and a gradual transition takes place toward a full
Sedov self-similar solution, i.e. \teq{m = 0.4} (e.g. Cioffi, McKee, \&
Bertschinger 1988).  Such a transition, at times $t \sim t_{\rm
trans}$, marks the epoch where the mass, \teq{4\pi R_{\rm sk}^3 \,
\rho_1 /3}, of the interstellar medium (of mass density \teq{\rho_1})
that has been swept up by the blast wave has come to dominate the mass,
$M_{\rm ej}$, of the supernova ejecta.

While the evolution of any real SNR may be extremely complex,
particularly due to variations in the pre-supernova environment, the
gross features of the evolution after \teq{t_{\rm trans}} can be
modeled simply with the standard Sedov (1959) relations for shock speed
and radius.  For times $t_{\hbox{\fiverm SNR}} > t_{\rm trans}$, but
before the shock becomes radiative, we assume the outer shock radius,
$R_{\rm sk}$, and speed, $V_{\rm sk}$, obey these relations, i.e.
\begin{equation}
   R_{\rm sk} \;\simeq\;
     \xi \left( \dover{{\cal E}_{\hbox{\sixrm SN}}}{\rho_1} \right)^{1/5}
     t_{\hbox{\fiverm SNR}}^{2/5} \ , \qquad
   V_{\rm sk} \;\simeq\;
     \dover{2}{5} \xi
     \left ( {{\cal E}_{\hbox{\sixrm SN}} \over \rho_1} \right )^{1/5}
     t_{\hbox{\fiverm SNR}}^{-3/5} \ ;
   \qquad t_{\hbox{\fiverm SNR}} > t_{\rm trans}\ .
 \label{eq:Sedovsol}
\end{equation}
In these expressions, ${\cal E}_{\hbox{\sixrm SN}}$ is the energy of
the supernova explosion $\xi=1.15$ (e.g. Shu 1992), and energy losses
from cosmic rays escaping from the shock are neglected.  We then make
the following definitions for the transition between the quasi-free
expansion and Sedov phases, using the criterion that the swept-up mass
equal the ejected mass:  \teq{M_{\rm swept} \simeq (4\pi/3)\, R_{\rm
trans}^3 \, \rho_1 = M_{\rm ej}}, defining $R_{\rm trans}$ as the
radius of the outer forward shock at the transition, and where
$\rho_1=1.4 \, n_{\rm p,1} m_{\rm p}$ is the unshocked ISM density
($n_{\rm p,1}$ is the unshocked proton number density and hereafter we
assume the ISM has cosmic abundances).  It follows that
\begin{equation}
   R_{\rm trans} \;\equiv\;
     \left( \dover{3}{4\pi} \dover{M_{\rm ej}}{\rho_1} \right)^{1/3}
   \; \simeq\; 1.9
     \left( \dover{n_{\rm p,1}}{{\rm cm}^{-3}} \right)^{-1/3}
     \left( \dover{M_{\rm ej}}{M_{\odot}} \right)^{1/3} \ {\rm pc} \ .
 \label{eq:Rtrans}
\end{equation}
Using $R_{\rm trans}$, we define the time of the transition from the
standard Sedov solution, i.e.
\begin{equation}
   t_{\rm trans} \;\equiv\;
     \left( \dover{R_{\rm trans}}{\xi} \right)^{5/2}
     \left( \dover{{\cal E}_{\hbox{\sixrm SN}}}{\rho_1} \right)^{-1/2}
   \; \simeq\; 90
     \left( \dover{n_{\rm p,1}}{{\rm cm}^{-3}} \right)^{-1/3}
     \left( \dover{{\cal E}_{\hbox{\sixrm SN}}}{10^{51}{\rm erg}}\right)^{-1/2}
     \left( \dover{M_{\rm ej}}{M_{\odot}} \right)^{5/6} \ {\rm yr} \ ,
 \label{eq:Ttrans}
\end{equation}
and the shock speed at the transition as,
\begin{equation}
   V_{\rm trans} \;\equiv\; \dover{2}{5} \xi
     \left( \dover{{\cal E}_{\hbox{\sixrm SN}}}{\rho_1} \right)^{1/5}
     t_{\rm trans}^{-3/5} \; \simeq\; 8200
     \left(\dover{{\cal E}_{\hbox{\sixrm SN}}}{10^{51}{\rm erg}} \right)^{-1/2}
     \left( \dover{M_{\rm ej}}{M_{\odot}} \right)^{-1/2}
     \ {\rm  \hbox{km~s$^{-1}$}} \ .
 \label{eq:Vtrans}
\end{equation}
If we assume that ${\cal E}_{\hbox{\sixrm SN}}=10^{51}$ erg and $n_{\rm
p,1} = 1$ \hbox{cm$^{-3}$}, then SN Ia with ejected mass of \teq{M_{\rm
ej}\sim M_{\odot}} have  $V_{\rm trans} \sim 8000$ \hbox{km~s$^{-1}$},
while SN II with \teq{M_{\rm ej}\sim 4M_{\odot}} have $V_{\rm trans}
\sim 4000$ \hbox{km~s$^{-1}$}.

While the above definitions are clearly approximations and alternative
ones could be made, these are simple, they model the most prominent
features of SNRs in homogeneous media, and they are appropriate to the
accuracy of current observations and model approximations.

\subsection{Acceleration Times and Maximum Particle Energies}
 \label{sec:acctime-Emax}

The maximum energy that can be attained by ions in diffusive shock
acceleration is determined by one of two approaches: (i) by equating
the acceleration time as a function of energy to the age of the remnant
(for the free expansion or early Sedov phase), or (ii) if the diffusion
length of the highest energy particles is comparable to the shock
radius (which occurs later in the Sedov phase), by capping that length
at some fraction of the shock radius, namely 25\%.

\subsubsection{Maximum Energy as a Function of Time}
 \label{sec:Emax}

Consider particles (possibly thermal) injected into the acceleration
process at a momentum \teq{p_i} in a shock with \teq{u_1} (\teq{u_2})
representing the upstream (downstream) component of flow speed normal
to the shock in its rest frame (which is uniquely defined since we
consider plane-parallel shocks), and \teq{\kappa_1} and \teq{\kappa_2}
being the upstream and downstream spatial diffusion coefficients in the
direction normal to the shock.  Here and elsewhere, the subscript 1 (2)
always implies quantities determined far upstream (downstream) from the
shock, and the negative $x$-direction will denote the shock normal.
Using the diffusion equation, the standard form for the acceleration
time, \teq{\tau_a}, to a given momentum \teq{p_{\rm max}}, is found to
be (e.g. Forman \& Morfill 1979; Drury 1983)
\begin{equation}
   \tau_a (p)\; =\; \dover{3}{u_1-u_2} \,\int_{p_i}^{p_{\rm max}}
     \biggl( \dover{\kappa_1}{u_1} + \dover{\kappa_2}{u_2} \biggr)
     \,\dover{dp'}{p'} \ .
 \label{eq:acctime}
\end{equation}
Since the interstellar medium is effectively stationary in the
observer's frame relative to the expanding shock front, \teq{u_1\simeq
V_{\rm sk}}.  Equation~(\ref{eq:acctime}) is strictly valid only in the
diffusion approximation (i.e. for $v\gg u_1$) and hence is appropriate
for our applications to relativistic energies here.

If we use equation~(\ref{eq:mfp}), relate the upstream and downstream
diffusion coefficients via \teq{\kappa_2=g \kappa_1} (defining $g$),
and assume that $V_{\rm sk}$ is constant in time (corresponding, for
example, to the free expansion phase of a SNR), the inversion of
equation~(\ref{eq:acctime}) yields a rate of energy gain
\begin{equation}
   \dover{dE}{dt} \;\simeq\; 300 \ \dover{r -1}{r (1 + gr)} \,\dover{Q}{\eta}
     \left( \dover{B_1}{3 \mu {\rm G}} \right)
     \left( \dover{V_{\rm sk}}{10^3 \,\hbox{km~s$^{-1}$}} \right)^2
     \ {\rm eV} \ {\rm s}^{-1} \ .
 \label{eq:dEdTshock}
\end{equation}
Here \teq{B_1} is the far upstream (interstellar) magnetic field and we
have assumed $\eta =$ constant across the shock.  For a maximum energy
$E_{\rm max}$ (much larger than the injection energy) corresponding to
\teq{p_{\rm max}}, this integrates to give an acceleration time
\begin{equation}
   \tau_a \;\simeq\; 106 \ \dover{r (1 + gr)}{r-1} \,\dover{\eta}{Q}
     \left( \dover{B_1}{3 \mu {\rm G}} \right)^{-1}
     \left( \dover{V_{\rm sk}}{10^3 \, \hbox{km~s$^{-1}$}} \right)^{-2}
     \left( \dover{E_{\rm max}}{{\rm 1 TeV}} \right) \ {\rm yr} \ ,
 \label{eq:taua1}
\end{equation}
For \teq{g=0}, particles spend virtually no time in the downstream
region; for \teq{g=1}  the mean free path is independent of the
upstream or downstream region; and for \teq{g=1/r} (\teq{r=u_1/u_2} is
the shock compression ratio), the mean free path is inversely
proportional to the background density, so that particles spend similar
times on either side of the shock.  Generally we favor \teq{g=1/r},
which assumes that the field turbulence that is responsible for
particle diffusion traces the plasma density. Such an assumption
(adopted for example by Draine \& McKee 1993) is suggested by the
expected field compression (\teq{B_2/B_1=r}) at quasi-perpendicular
shocks, and appears to be supported by Ulysses magnetometer data at
highly oblique interplanetary shocks (Baring et al.  1997).  Note,
however, that particles of the highest energies would be expected to
spend somewhat more time diffusing in the upstream region outside the
expanding shell due to its convex shape, thereby favoring \teq{g>1/r}
scenarios.  Note also that the proportionality \teq{\tau_a\propto
E_{\rm max}} is a consequence of the \teq{\lambda\propto r_{\rm g}}
assumption.  It follows that if the shock speed is constant, the
maximum energy obtainable for a SNR shock of an age
\teq{t_{\hbox{\fiverm SNR}}} is
\begin{equation}
   E_{\rm max}(t_{\hbox{\fiverm SNR}}) \;\simeq\;
     E_{\rm trans} \dover{t_{\hbox{\fiverm SNR}}}{t_{\rm trans}} \ ; \qquad 
     t_{\hbox{\fiverm SNR}} < t_{\rm trans} \ ,
 \label{eq:Enmax1}
\end{equation}
where
\begin{equation}
   E_{\rm trans} \;\equiv\; 60 \ \dover{r - 1}{r(1 + gr)}\,
     \dover{Q}{\eta}\, \biggl(\dover{B_1}{3\mu {\rm G}}\biggr)\,
     \Bigl( \dover{n_{\rm p,1}}{1\, {\rm cm}^{-3}} \Bigr)^{-1/3}\;
     \left( \dover{{\cal E}_{\hbox{\sixrm SN}}}{10^{51}{\rm erg}}\right)^{1/2}
     \,\biggl( \dover{M_{\rm ej}}{M_{\odot}} \biggr)^{-1/6} \ {\rm TeV}
 \label{eq:Etrans}
\end{equation}
is the maximum energy ions achieve at $t_{\rm trans}$ and is obtained
from equation~(\ref{eq:taua1}) using our definitions of $V_{\rm
trans}$, $t_{\rm trans}$, and $R_{\rm trans}$.  As an example, for
\teq{r=9}, \teq{g=1/r}, \teq{\eta=10}, ${\cal E}_{\hbox{\sixrm
SN}}=10^{51}$ erg, \teq{M_{\rm ej} =M_{\odot}}, \teq{B_1 = 3\times
10^{-6}}G, and \teq{n_{\rm p,1} = 1}cm$^{-3}$, we find that $V_{\rm
trans} \simeq 8200$ \hbox{km~s$^{-1}$}, $t_{\rm trans} \simeq 90$ yr,
and  $E_{\rm trans} \simeq 2.6$ TeV for protons {\it and} electrons.
Berezhko (1996) obtains a similar form to equation~(\ref{eq:Etrans}),
with a slightly larger coefficient because of his treatment of
expansion and the associated adiabatic effects.

From equation~(\ref{eq:Etrans}) it is clear
that particularly energetic supernova explosions or large ISM field
strengths are required to generate cosmic rays above \teq{10^{14}}eV
and subsequently populate the ``knee'' in the cosmic ray distribution.
Note also, that $E_{\rm trans}$ has a fairly weak dependence on the ISM
density, but one which becomes important at very low densities.
Furthermore, equation~(\ref{eq:Etrans}) depends on the charge of the
species but not the mass and hence is identical for protons and
electrons, provided that \teq{p_{\rm max}} far exceeds the critical
electron momentum \teq{p_{\rm crit}} discussed above.  The maximum
energy in equation~(\ref{eq:Enmax1}) does not result in abrupt cutoffs
to the distributions of the accelerated populations, but rather marks
the energy about which quasi-exponential turnovers appear: spatial
diffusion near the FEB smears out the energy of the cutoffs.  If we use
Equations~(\ref{eq:Ttrans}), (\ref{eq:Enmax1}), and (\ref{eq:Etrans}) to
estimate $E_{\rm max}$ for very young remnants like SN1987A, it quickly
becomes clear that these remnants will take several decades to
accelerate particles to energies beyond a few TeV.  For \teq{{\cal
E}_{\rm SN}\sim 10^{51}}erg, \teq{M_{\rm ej}\sim M_{\odot}}, and
$t_{\hbox{\fiverm SNR}}=10$ yr, \teq{E_{\rm max}\sim 7} TeV.  Protons
of this energy produce pion-decay photons of substantially lower energy
(by a factor of a few); similarly, electron bremsstrahlung photons are
on average about one-third the energy of the primary electrons.  IC
photons produced by 7 TeV electrons have energies of less than 1 TeV.
Thus we expect that SN1987A will not be a bright TeV gamma-ray source
anytime in the next decade, contrary to the suggestion of Kirk, Duffy,
\& Ball (1995).

In determining \teq{E_{\rm max}} at any time, it must be noted that it
is some weighted function of the injection history, as well as the
acceleration history of the highest energy particles.  Hence, besides
the unknowns in the shock processes, the SNR remnant environment and
its evolution, the fact that the rate per unit area at which particles
are injected into the shock is dependent on a remnant's evolutionary
phase complicates the picture.  This rate is proportional to
\teq{R_{\rm sk}^2 V_{\rm sk}}, an intrinsic variation that provides a
rapid increase in injection during the free expansion phase.  However,
in the Sedov phase, the number of protons per unit time that are
crossed by the shock is
\begin{equation}
   \dover{dN_{\rm p}}{dt} \;\simeq\; 6\times 10^{47}
     \left( \dover{n_{\rm p,1}}{{\rm cm}^{-3}} \right)^{2/5}
     \left( \dover{{\cal E}_{\hbox{\sixrm SN}}}{10^{51}{\rm erg}} \right)^{3/5}
     \left( \dover{t_{\hbox{\fiverm SNR}}}{10^3 \hbox{\rm yr}} \right)^{-1/5}
     \qquad {\rm sec}^{-1} \ ,
 \label{eq:crossrate}
\end{equation}
a slowly decaying function of time.  For purposes of our $E_{\rm max}$
estimate, we will {\it neglect} particles accelerated during the free
expansion phase and assume that the rate at which the shock overtakes
ambient particles in the Sedov phase is independent of time.  This
assumption marks an important distinction between our calculation and
Berezhko's (1996).  He obtains considerably higher maximum energies
because of strong acceleration in the free-expansion phase, which we
neglect here because of the relatively small number of particles
injected then.  Hence, in the Sedov phase, if we continue to assume
that $\lambda = \eta\, r_{\rm g}$, equation~(\ref{eq:dEdTshock})
yields:
\begin{equation}
   E_{\rm max}(t_{\hbox{\fiverm SNR}}) \;\simeq\; 5 \,
     E_{\rm trans} \left[ 1 - \left(
     \dover{t_{\hbox{\fiverm SNR}}}{t_{\rm trans}} \right)^{-1/5} \right]\ ;
     \qquad t_{\hbox{\fiverm SNR}} > t_{\rm trans} \ ,
 \label{eq:Emax}
\end{equation}
where we have assumed that $(r-1)/[r(1+gr)]$ is a weakly varying
function of time.  In fact, the compression ratio, $r$, will vary with
time as the shock Mach number and $E_{\rm max}$ change, but this
variation will become smaller at later times.  This solution for
$E_{\rm max}$ includes the acceleration history of particles at all
times during the Sedov phase, and asymptotically approaches $5 \,
E_{\rm trans}$ at late times.  The further inclusion of particles
accelerated during the free expansion phase would only modify this
formula to \teq{E_{\rm trans} [6-5 (t_{\hbox{\fiverm SNR}} /t_{\rm
trans})^{-1/5}]}, a small change, but one that encompasses the highest
energy cosmic rays.  The reader is referred to Berezhko \& V\"olk
(1997) for an estimate of how these highest energy particles influence
the $\gamma$-ray emission.

\subsubsection{Maximum Energy as a Function of Shock Precursor Scale}
 \label{sec:Emax2}

As the remnant evolves, the maximum extent of the precursor outside the
outer shock will be determined by the diffusion length ahead of the
shock, $d_{\hbox{\sevenrm FEB}}$, of the highest energy particles in
the system at its current age, whose energy is $E_{\rm
max}(t_{\hbox{\fiverm SNR}})$ (as determined from
equation~[\ref{eq:Emax}]).  Since the diffusion scale is $\sim
\kappa_1/V_{\rm sk}$, one obtains, for fully relativistic particles,
\begin{equation}
   d_{\hbox{\sevenrm FEB}} \;\sim\;
     \dover{\eta r_{\rm g,max} c}{3 V_{\rm sk}}
   \; =\; \dover{\eta}{3 Q e B_1} \dover{E_{\rm max}^{\rm age}}{V_{\rm sk}}
   \quad , \quad
   E_{\rm max}^{\rm age} \; =\;
     E_{\rm max}(t_{\hbox{\fiverm SNR}}) \ ,
 \label{eq:dFEB}
\end{equation}
where we have used the superscript `age' to indicate that the maximum
energy is determined by the age of the remnant.  The distance to the
FEB defines the full width of the shock precursor, and must be a small
fraction of $R_{\rm sk}$ in order for the plane-parallel shock
simulation to be applicable to quasi-spherical shells.  In the Sedov
phase, the combination of equations~(\ref{eq:Sedovsol}),
(\ref{eq:Etrans}), and~(\ref{eq:Emax}) yields
\begin{equation}
   {d_{\hbox{\sevenrm FEB}} \over R_{\rm sk}} \;\simeq\;
     \dover{2}{3} \, \dover{r-1}{r (1 + gr)}
     \left( \dover{t_{\hbox{\fiverm SNR}}}{t_{\rm trans}} \right)^{1/5}
     \left[ 1-\left( \dover{t_{\hbox{\fiverm SNR}}}{t_{\rm trans}} 
          \right)^{-1/5} \right] \ ,
 \label{eq:FEBratio}
\end{equation}
and it is clear that if this phase lasts long enough,
$d_{\hbox{\sevenrm FEB}}/ R_{\rm sk}$ will become greater than unity,
rendering our scheme for the termination of acceleration inconsistent.
A similar $t_{\hbox{\fiverm SNR}}^{1/5}$ dependence was noted by Kang
\& Jones (1991).  The precise age at which $d_{\hbox{\sevenrm FEB}}/
R_{\rm sk} = 1$ is a strong function of the assumed value of $g$, and
to a lesser extent of \teq{r}. In practice, we  place our upstream free
escape boundary at a distance $d_{\hbox{\sevenrm FEB}}$
(equation~[\ref{eq:dFEB}]) ahead of the shock if $d_{\hbox{\sevenrm
FEB}} / R_{\rm sk} < f < 1$. If $d_{\hbox{\sevenrm FEB}} / R_{\rm sk} >
f$, we set $d_{\hbox{\sevenrm FEB}} = f R_{\rm sk}$.  In this case, the
maximum energy is given by
\begin{eqnarray}
   E_{\rm max}^{\rm size} & \simeq &
     f \dover{3 Q e}{\eta} B_1 V_{\rm sk} R_{\rm sk}\nonumber\\[-5.5pt]
 &&  \label{eq:EmaxSize}\\[-5.5pt]
   & \sim & 270 \, f \dover{Q}{\eta}
     \left( \dover{B_1}{3 \mu {\rm G}} \right)
     \left( \dover{n_{\rm p,1}}{{\rm cm}^{-3}} \right)^{-2/5}
     \left( \dover{{\cal E}_{\hbox{\sixrm SN}}}{10^{51}{\rm erg}} \right)^{2/5}
     \left( \dover{t_{\hbox{\fiverm SNR}}}{10^3 {\rm yr}} \right )^{-1/5}
     {\rm TeV}\ ,\nonumber
\end{eqnarray}
where we use the superscript `size' to indicate that this energy is
constrained by the remnant size.

The arbitrary factor $f$, which we set equal to $1/4$ in all of the
work here, is included to ensure that the self-generated magnetic
turbulence from the highest energy particles is localized  (as
discussed just below) to the precursor of the spherical shock.  The
transition at $d_{\hbox{\sevenrm FEB}}/ R_{\rm sk} = f = 1/4$ is, in
effect, a transition from age-limited acceleration at early times to
size-limited acceleration at late times.  By our definitions, this
transition occurs when \teq{d_{\hbox{\sevenrm FEB}} /(f R_{\rm sk})=1},
that is, when \teq{t_{\hbox{\fiverm SNR}} /t_{\rm trans} =(3 f r^2 g +
3 f r + 2 r -2)^5/(2 r - 2)^5}.  For $r\sim 9$, $f=1/4$, and $g=1/r$,
the transition occurs at $t_{\hbox{\fiverm SNR}}/t_{\rm trans} \gtrsim
20$, i.e. acceleration of the highest-energy particles ceases when
their diffusion lengths approach $f R_{\rm sk}$ for much of the Sedov
phase, the scenario that Berezhko, Yelshin, \& Ksenofontov (1994)
prefer.  However, if $g=1$, the acceleration is age-limited for a much
larger range of times (up to $t_{\hbox{\fiverm SNR}}/t_{\rm trans} \sim
3\times 10^{3}$), as in the considerations of Lagage \& Cesarsky
(1983).  We do not attempt to model acceleration into the radiative
phase which begins at approximately \teq{t_{\rm rad} \simeq 2.9\times
10^4 \, (n_{\rm p,1}/{\rm cm}^{-3})^{-9/17}\, ({\cal E}_{\rm
SN}/10^{51}{\rm erg})^{4/17}} yr (Blondin et al. 1997), and therefore
is well beyond the ages of the young remnants considered here.

\placefigure{fig:figEmax}
Hence, the combination of size-limited and space-limited acceleration
is implemented in our steady-state Monte Carlo model by placing a free
escape boundary at a distance, $d_{\hbox{\sevenrm FEB}}$, upstream from
the shock, where
\begin{equation}
   d_{\hbox{\sevenrm FEB}} \; =\;
     \min \left\{ \dover{\eta}{3 Q e B_1}
     \dover{E_{\rm max}^{\rm age}}{V_{\rm sk}} \ , \ f R_{\rm sk} \right\}
   \ , \quad f \; =\; \dover{1}{4} \ .
 \label{eq:dFEBtwo}
\end{equation}
The maximum energies produced by this procedure are shown in
Figure~\ref{fig:figEmax}.  The lower three thick curves all have $\eta
= 10$, $B_1 = 3\times 10^{-6}$ G, $n_{\rm p,1} = 1$ \hbox{cm$^{-3}$},
${\cal E}_{\hbox{\sixrm SN}} = 1\times 10^{51}$ erg, and $M_{\rm ej} =
M_{\odot}$ with choices for \teq{g} as indicated.  The thick solid line
labeled (b) has the same $\eta$, $B_1$, ${\cal E}_{\hbox{\sixrm SN}}$,
and $M_{\rm ej}$ parameters as above with $g = 1/r$ and $n_{\rm p,1} =
0.01$\hbox{cm$^{-3}$}.  For the uppermost solid line labeled (c), which
shows maximum energies well above $10^{15}$ eV, we have selected
parameters that are especially tuned to yield a high maximum energy,
i.e. are appropriate for particle acceleration up to the ``knee'' in
the cosmic ray spectrum.  In this case, we have used $\eta=1$ (i.e.
Bohm diffusion with \teq{\lambda\sim r_{\rm g}}), $g=1/r$, $B_1 =
10\times 10^{-6}$ G, $n_{\rm p,1} = 10^{-3}$\hbox{cm$^{-3}$}, ${\cal
E}_{\hbox{\sixrm SN}} = 10\times 10^{51}$ erg, and $M_{\rm ej}=
10M_{\odot}$.  For convenience, we have taken $r=8.5$ in all plots
although the actual $r$ will depend on the particular parameters used.

While clearly an approximation, we believe this scheme for setting the
maximum energy by converting the time-dependent effects of Sedov
dynamics into size-limited acceleration is accurate enough to allow us
to describe the essential non-linear effects in a convenient and
realistic fashion.  It differs qualitatively from the time-dependent
analyses of Berezhko (1996) and other researchers (e.g. Drury,
Markiewicz, \& V\"olk 1989; Kang \& Jones 1991; Berezhko, Yelshin, \&
Ksenofontov 1994), principally because of omissions such as adiabatic
energy gains in the decelerated remnant interior downstream of the
outer shock, the complex interplay between geometry and acceleration
history, and the contributions of free expansion phase cosmic ray
acceleration to that in the Sedov epoch.  Berezhko (1996) observes that
these additional features can increase the maximum energy of
acceleration by factors of 2--3, which will in turn affect the
non-linear feedback between the Fermi process and the hydrodynamics.
However, we note that a significant contribution to this increase in
\teq{E_{\rm max}} above our estimates may be due to Berezhko's
assumption that cosmic rays find the interior of the remnant
impenetrable due to the establishment of large scale hydrodynamic
turbulence via Raleigh-Taylor instabilities, and Berezhko's consequent
imposition of a downstream reflecting boundary.  In fact, it is
probable that the presence of such turbulence and associated field
amplification (e.g. see Jun \& Norman 1996) will reduce the
scale-length for diffusion, thereby increasing downstream escape of
cosmic rays and lowering the maximum energy.  Therefore, clearly there
is some degree of subjectivity in the choice of \teq{E_{\rm max}},
being not tightly-constrained by observations; our own choice is
motivated by its convenience.

\subsubsection{Loss Processes for Electrons}
 \label{sec:loss}

Of the various loss processes that influence the particles accelerated
at young SNRs, only synchrotron and inverse Compton losses for
electrons are important for a wide range of parameters (Coulomb losses
only become important for old remnants, e.g. Sturner et al.  1997).
The rate of synchrotron energy loss in a field $B$ is given by Lang
(1980).  For inverse Compton (IC) losses under the conditions we
envision, the most important source of seed photons is the primordial
cosmic microwave background radiation; other radiation fields
contribute less than 20\% (e.g. Gaisser, Protheroe, \& Stanev 1998) to
IC cooling.  The IC loss rate is given by a similar expression to the
synchrotron loss rate, obtained simply by substituting the radiation
energy density for magnetic field energy density.  The field strength
with the same energy density as the 2.73 K background radiation is
$B_{\rm cbr} \simeq 3.32\times 10^{-6}$ G, leading to the compact
formula describing both synchrotron and inverse Compton losses:
\begin{equation}
   \left( \dover{dE}{dt} \right)_{\rm tot} \;\simeq\;  -0.034 \
     \left( {\sqrt{B^2 + B_{\rm cbr}^2} \over {3\mu {\rm G}} } \right)^2
     \left( \dover{E}{{1 \rm TeV}} \right)^2 \ {\rm eV} \ {\rm s}^{-1} \ .
 \label{eq:TotLossRate}
\end{equation}

In oblique shocks, a particle undergoing acceleration spends time both
upstream and downstream in magnetic fields of varying strength. This
makes it impossible to get a precise measure of the loss rate without
detailed knowledge of the shock geometry.  In the general oblique case,
we can set $B=\Gamma B_1$ with $1 \leq \Gamma \le r$, and $\Gamma = r$
gives an upper limit to the loss rate.  Here, however, we model only
parallel shocks where the mean magnetic field doesn't vary through the
shock and equation~(\ref{eq:TotLossRate}) can be used directly with
$B=B_1$.  By comparing equations~(\ref{eq:dEdTshock}) and
(\ref{eq:TotLossRate}) we obtain an upper limit to the electron cutoff
energy in the Sedov phase, i.e.
\begin{eqnarray}
   E_{\rm cutoff} &\simeq &\ 180 \
     \left[ \dover{r -1}{r (1 + gr)} \, \dover{Q}{\eta} \right]^{1/2}
     \left( \dover{B_1}{3 \mu {\rm G}} \right)^{1/2}
     \left( {\sqrt{\, \Gamma^2 B_1^2 + B_{\rm cbr}^2}
       \over {3 \mu {\rm G}}}\right)^{-1}
     \left( \dover{n_{\rm p,1}}{{\rm cm}^{-3}} \right)^{-1/5}\nonumber\\[-5.5pt]
  && \label{eq:Cutoff}\\[-5.5pt]
   &&\qquad \times \ \
     \left( \dover{{\cal E}_{\hbox{\sixrm SN}}}{10^{51}{\rm erg}} \right)^{1/5}
     \left( \dover{t_{\hbox{\fiverm SNR}}}{10^3 {\rm yr}} \right)^{-3/5}
     \ {\rm TeV} \ .\nonumber
\end{eqnarray}
For our Monte Carlo calculations, we implement
equation~(\ref{eq:Cutoff}) with $\Gamma = 1$.  A similar cutoff energy
was obtained by Sturner et al.~(1997).  In Figure~\ref{fig:figEmax}
\ we show equation~(\ref{eq:Cutoff}) as light dotted lines. The
leftmost dotted line gives the electron cutoff energy for the lower
solid line example (a), the middle dotted line gives $E_{\rm cutoff}$
for the $n_{\rm p,1} = 0.01$ \hbox{cm$^{-3}$}, $\eta=10$ example (b),
and the rightmost dotted line gives $E_{\rm cutoff}$ for the $n_{\rm
p,1} = 10^{-3}$ \hbox{cm$^{-3}$}, $\eta=1$ example (c).  Clearly,
electron acceleration will be essentially unaffected in the high
density ISM throughout the Sedov phase, but can be severely truncated
in lower density regions at all times.  In the highest $E_{\rm max}$
example in Figure~\ref{fig:figEmax}, where parameters were chosen to
optimize cosmic ray production, protons can obtain energies two orders
of magnitude higher than electrons during most of the SNR evolution.

\section{PHOTON PRODUCTION MECHANISMS}
 \label{sec:mechanisms}

Having outlined the relevant processes involved in energetic particle
production, we now describe how photons are produced once particle
distributions are obtained. Unless otherwise stated, all photon
energies are expressed as $\erg_{\gamma}$, the gamma ray energy in
units of $m_{\rm e} c^2$, and we assume cosmic abundance for helium
(i.e. $n_{\rm He,1} = 0.1 n_{\rm p,1}$) and that the helium is fully
ionized so that the electron number density is $n_{\rm e,1} = 1.2\,
n_{\rm p,1}$.  For the purposes of this paper, we neglect contributions
from species other than protons, fully stripped helium ions, and
electrons.  For a shock-accelerated distribution of electrons or ions,
\teq{(dJ/dE)_{\rm e,i}}, the number density per unit {\it kinetic}
energy is \teq{(4\pi /v_{\rm e,i}) (dJ/dE)_{\rm e,i}}, and the number
of photons emitted per unit volume per second in the range
\teq{\erg_{\gamma}} to \teq{\erg_{\gamma} + d\erg_{\gamma}} takes the
form
\begin{equation}
   \dover{dn_{\gamma}(\erg_{\gamma} )}{dt} \; =\; \int_{0}^{\infty} 
     \dover{dn_{\gamma}(E_{\rm e,i} ,\,\erg_{\gamma} )}{dt} \;
     \left[ \left( \dover{4\pi}{v_{\rm e,i}} \right)
            \left( \dover{dJ}{dE} \right)_{\rm e,i} \right]\, dE_{\rm e,i} \ ,
 \label{eq:emissiontot}
\end{equation}
where \teq{dn_{\gamma}(E_{\rm e,i} ,\,\erg_{\gamma} )/ dt} is the
emissivity of a single particle, either an electron or an ion, of
kinetic energy \teq{E_{\rm e,i}}.  We consider four processes: pion
decay emission, bremsstrahlung, inverse Compton scattering, and
synchrotron radiation, and note that, as discussed in Section~3.1
below, emission due to secondary electrons (i.e. pairs produced via the
decay of \teq{\pi^{\pm}} created in ion-ion collisions) is negligible.
If a source is at a distance $d_{\hbox{\fiverm SNR}}$, with an emission
volume $V_{\hbox{\fiverm SNR}}$, the number of photons per unit area
per second per unit photon energy arriving at Earth is
$[dn_{\gamma}(\erg_{\gamma} ) / dt]  V_{\hbox{\fiverm SNR}} /(4 \pi
d_{\hbox{\fiverm SNR}}^2 )$.

\subsection{Pion Decay Radiation}
 \label{sec:pion}

We first consider \teq{\pi^0} decay emission resulting from pions
generated in ion--ion collisions: \teq{p+p \to \pi^0 \to \gamma +
\gamma}, etc.  This process has been popular in discussions of
gamma-ray emission from supernova remnants dating from the early work
of Higdon \& Lingenfelter (1975) to the more extensive analyses of
Drury, Aharonian, \& V\"olk (1994) and Naito \& Takahara (1994), and
plays a prominent role in the modeling of the diffuse galactic
gamma-ray background (Bertsch et al.  1993, Hunter et al. 1997).  Here,
we adopt a modified scaling model for pion production, the details of
which can be found in Baring \& Stecker (1998).

The scaling concept, originally devised by Feynman (1969), is usually
invoked at cosmic ray hadron energies above 10 GeV.  It assumes that
the Lorentz invariant differential cross-section \teq{E\,
d^3\sigma/dp^3} approaches a function that is independent of the fast
hadron's energy as it tends to infinity.  A variety of scaling
descriptions exist, and all of them, including ours, determine this
function via empirical fits to experimental data.  Scaling models work
well for kinetic energies up to hundreds of GeV, beyond which quantum
chromodynamics (QCD) becomes important and scaling violations ensue.
Such deviations from scaling behavior in this regime are accounted for
in a fairly simple manner in the model of Baring \& Stecker (1998) by
adopting non-scaling corrections to the total cross-section which
extend the usefulness into the super-TeV range.  At low hadron kinetic
energies, corresponding to photons produced between $\sim 20$ MeV and
200 MeV, an isobaric model (Stecker 1971) that is discussed at length
in Baring \& Stecker (1998) is appropriate.  We adopt non-scaling
corrections in this regime also, so that our computations only
underestimate photon spectra by $\sim 20$--30\%.  Minor improvements to
this can be achieved using a hybrid scaling-isobaric approach (see
Dermer 1986a,b; Baring \& Stecker 1998).

Consider first proton--proton collisions.  Baring \& Stecker (1998) use
a {\it radial scaling} (RS) model, where the Lorentz invariant
differential cross-section \teq{E\, d^3\sigma/dp^3} is approximated by
a function that depends on two variables: (i) the component of pion
momentum \teq{p_t^{\ast}} transverse to the cosmic ray proton (or ion)
beam direction, and (ii) the radial scaling variable
\teq{x_{\hbox{\fiverm R}}}, which is the ratio of the center-of-mass
(CM) frame pion Lorentz factor \teq{\gamma^{\ast}_{\pi}} to
\teq{\gamma^{\ast}_{\pi ,{\rm max}}}, the maximum possible value of
\teq{\gamma^{\ast}_{\pi}}.  Note that hereafter, asterisks denote
quantities evaluated in the CM frame of the colliding protons.  Pion
production kinematics dictate that \teq{\gamma^{\ast}_{\pi ,{\rm max}}
= [2(\gamma_{\rm p} -1)+\mu_{\pi}^2]/[2\mu_{\pi}\sqrt{2(\gamma_{\rm p}
+1)}]} for \teq{\mu_{\pi} = m_{\pi}/m_{\rm p} =0.1438}.  Baring \&
Stecker (1998) adopt the form for \teq{E\, d^3\sigma/dp^3} obtained by
Tan \& Ng (1983), which was applied separately to \teq{\pi^+} and
\teq{\pi^-} production data: the average of these is taken as the
cross-section for \teq{\pi^0} creation by invoking isospin conservation
in strong interactions. For given transverse momentum \teq{p_t^{\ast}}
and \teq{\gamma_{\pi}^{\ast}}, there are two solutions for the pion
Lorentz factor \teq{\gamma_{\pi}}, namely \teq{\gamma_{\pi}^{\pm}
=\gamma_{\rm cm}\, ( \gamma_{\pi}^{\ast} \pm \beta_{\rm cm}\,
p_l^{\ast})}, where \teq{p_l^{\ast}=\sqrt{(\gamma_{\pi}^{\ast})^2
-1-(p_t^{\ast})^2}}.  Here \teq{p_l^{\ast}} is the {\it longitudinal}
momentum of the pion, i.e.  along the direction defined by one of the
incoming protons.  The Lorentz factor \teq{\gamma_{\rm
cm}=\sqrt{(\gamma_{\rm p} +1)/2}} is that corresponding to the boost
between the CM and laboratory (i.e.  ISM) frames.  The expression for
the differential photon production rate is then
\begin{eqnarray}
   \dover{dn_{\gamma}(\erg_{\gamma})}{d\erg_{\gamma}} \; & = &\;
     4\pi\, n_{\rm p}\, \gamma^{\ast}_{\pi ,{\rm max}}\, c\;
   \dover{m_{\rm e}}{m_{\pi}}
     \int^{\infty}_{\gamma_{\hbox{\fiverm TH}}} d\gamma_{\rm p} \;
     \beta_{\rm p}\; \psi (\gamma_{\rm p} )\; N_{\rm p}(\gamma_{\rm p} )    
   \nonumber\\[-5.5pt]
 && \label{eq:ngammadot}\\[-5.5pt]
   && \int dp_l^{\ast}\, dx_{\hbox{\fiverm R}}\,
     \Biggl\{ \dover{\theta (\gamma_{\pi}^+ -\gamma_-)}{\vphantom{\Bigl(}
       \sqrt{ {(\gamma_{\pi}^+)}^2-1}} +
            \dover{\theta (\gamma_{\pi}^- -\gamma_-)}{\vphantom{\Bigl(}
       \sqrt{ {(\gamma_{\pi}^-)}^2-1}} \Biggr\} \;
     \biggl( E\,\dover{d^3\sigma}{dp^3}\biggr)_{\rm RS}
       (x_{\hbox{\fiverm R}} ,\, p_t^{\ast}) \; ,\nonumber
\end{eqnarray}
where \teq{n_{\rm p}} is the ambient proton density, and \teq{\gamma_-
= (m_{\rm e}/m_{\pi}) \, [\erg_{\gamma} +m_{\pi}^2/(4m_{\rm
e}^2\erg_{\gamma})]} is the minimum pion Lorentz factor permitted by
kinematics for pions decaying to produce photons of energy
\teq{\erg_{\gamma}}.  The \teq{\theta} functions in
equation~(\ref{eq:ngammadot}) are Heaviside step functions that are
zero for $x<0$ and unity otherwise.  The proton distribution in
equation~(\ref{eq:ngammadot}) can be obtained from omni-directional
fluxes produced in the Monte Carlo shock simulations via \teq{N_{\rm
p}(\gamma_{\rm p} )=[4\pi m_{\rm p} c^2/v_{\rm p}] (dJ/dE)_{\rm p}}.
Also, \teq{\gamma_{\hbox{\fiverm TH}}\simeq 1.298} is the proton
Lorentz factor corresponding to the threshold of pion production.  The
limits on the \teq{x_{\hbox{\fiverm R}}} integration are defined by
\teq{1/\gamma^{\ast}_{\pi ,{\rm max}}\leq x_{\hbox{\fiverm R}}\leq 1},
while those for \teq{p_l^{\ast}} are given by the constraints that
\teq{\gamma_{\pi}^{\pm} \geq \gamma_-}, according to the appropriate
term in equation~(\ref{eq:ngammadot}).  For monoenergetic protons, this
differential spectrum closely resembles results produced by the PYTHIA
code (described in St\"ostrand \& van Zijl 1987) in the range
\teq{10<\gamma_{\rm p} <1000}, and for power-law protons it compares
well with predictions of Gaisser, Protheroe, \& Stanev (1998).

In the strong interaction, neutrons interact with virtually the same
properties as protons.  Hence heavier species such as alpha particles
basically provide additional supplies of {\it nucleons}, and the
collisions of individual nucleons can be described in the above
manner.  However, care must be taken to account for the nuclear binding
of heavier elements.  Orth \& Buffington (1976) posited that the
inclusive cross-section for cosmic rays of mass number \teq{A_{\rm cr}}
colliding with ISM target nuclei of mass number \teq{A_{\hbox{\fiverm
ISM}}} is
\begin{equation}
   \sigma\;\sim\;
      \Bigl(A_{\rm cr}^{3/8}+A_{\hbox{\fiverm ISM}}^{3/8}-1\Bigr)^2
     \;\sigma_{pp\to \pi^0X}\quad .
 \label{eq:alphacorr}
\end{equation}
While experimental data on inelastic collisions involving nuclei
heavier than hydrogen are sparse, this prescription is appropriate for
proton-helium interactions, but its accuracy is unclear for heavier
nuclei such as Fe, which are therefore omitted from consideration
here.

\subsubsection{Secondary electron production}
 \label{sec:elecprod}

Primary electrons, i.e. those directly accelerated by the Fermi
process, dominate the contributions of electron emission.  Secondary
electrons and positrons are produced via the decay of charged pions
that are created in \teq{pp} and \teq{p\alpha} collisions, and the
cross-section for these modes is comparable to that of the neutral pion
modes.  Hence, the rate of pair production in \teq{pp} collisions is
roughly \teq{dn_{\pm}(E_{\rm e} )/dt\, \sim\, n_{\rm p}^2 c\,
\sigma_{pp\to\pi^0 X}(f E_{\rm e} )}, where \teq{f} is a factor of the
order of a few that accounts for the pion production inelasticity.  If
the pairs are permitted to build up without escape for the entire
remnant lifetime, then one obtains the maximal estimate for the
secondary pair density:  \teq{n_{\pm}(E_{\rm e} )\sim t_{\hbox{\fiverm
SNR}}\, n_{\rm p}^2 c\,\sigma_{pp\to\pi^0 X}(fE_{\rm e} )}.
Remembering that the timescale for \teq{pp} collisions is \teq{t_{\rm
pp}\sim (n_{\rm p} c \sigma_{pp\to\pi^0 X})^{-1}}, and that the primary
electron density is \teq{n_{\rm e}\sim n_{\rm p}}, then the accumulated
pair density can be written as \teq{n_{\pm}\sim n_{\rm e}
t_{\hbox{\fiverm SNR}} /t_{\rm pp}}.  The timescale \teq{t_{\rm pp}}
for collisions involving 1 GeV--1 TeV protons is typically of the order
of \teq{10^7} years, immediately leading to the conclusion that the
primary electron density far exceeds that of the secondaries in young
SNRs.  Hence the contribution of secondaries to the bremsstrahlung,
inverse Compton and synchrotron emission can be neglected.  This fact
was pointed out by Mastichiadis (1996) for the specific case of
synchrotron radiation.

\subsection{Bremsstrahlung}
 \label{sec:brems}

Electrons will produce bremsstrahlung radiation as they scatter off the
ambient gas.  Hence, the rate of photon production, $dn_\gamma (E_{\rm
e} ,\,\erg_{\gamma} ) /dt$,  in the energy interval between
$\erg_{\gamma}$ and $\erg_{\gamma} +d\erg_{\gamma}$ by an electron of
{\it kinetic energy} $E_{\rm e}$ takes the standard form
\begin{equation}
   \dover{dn_{\gamma}(E_{\rm e} ,\,\erg_{\gamma})}{dt} \; =\;
      v_{\rm e}\Bigl[\, (n_{\rm p}+4 n_{\rm He})\,
      \sigma_{\rm e-p} (E_{\rm e} ,\,\erg_{\gamma} ) +
      n_{\rm e} \sigma_{\rm e-e} (E_{\rm e} ,\,\erg_{\gamma} )\, \Bigr] \ ,
 \label{eq:dndtbrems}
\end{equation}
where $\erg_{\gamma}$ is the gamma-ray energy in units of $m_{\rm e}
c^2$.  Here the electron-ion cross-section, \teq{\sigma_{\rm e-p}
(E_{\rm e} ,\,\erg_{\gamma} )}, differential in photon energy (i.e.
integrated over photon and final electron angles), is the famous
Bethe-Heitler cross-section (Bethe \& Heitler 1934; see also Jauch \&
Rohrlich 1980), evaluated in the Born approximation; it is used for any
electron energy, relativistic or non-relativistic.  We note that the
ultrarelativistic form for the e-p cross-section [e.g. see
equation~(15-101) of Jauch \& Rohrlich 1980] is the same as
\teq{\sigma_1} given in equation~(A2) in the Appendix.  The electron
speed \teq{v_{\rm e}} is the relative velocity in bremsstrahlung
collisions.  The Bethe-Heitler formula applies to both protons and
alpha particles, with a charge-dependence \teq{\sigma_{\rm e-p}\propto
Z^2}.  This leads to the \teq{n_{\rm p}+4 n_{\rm He}} factor
multiplying \teq{\sigma_{\rm e-p}} in equation~(\ref{eq:dndtbrems}).
Coulomb corrections to the Bethe-Heitler cross-section, such as through
the Sommerfeld-Elwert factor (Elwert 1939), become important only for
projectile electron speeds considerably less than $c/10$; we omit them
from our considerations since they are only marginally important given
our electron temperatures of a few keV.

The situation for the electron-electron cross-section, $\sigma_{\rm
e-e} (E_{\rm e} ,\,\erg_{\gamma} )$, is more complicated.  The full
quantum electrodynamical expression for the angle-integrated
cross-section, differential in photon energy, was first derived by Haug
(1975).  The result is extremely long (over a page of algebra), and is
numerically cumbersome given that it contains terms that are
individually divergent (to several orders) in photon energy as
\teq{\erg_{\gamma}\to 0}.  This unwieldiness motivates us to use other
expressions that are derived from asymptotic limits for
non-relativistic (Fedyushin 1952; Garibyan 1953) and ultrarelativistic
(Baier, Fadin, \& Khoze 1967) electrons.  These are presented in detail
in the Appendix, and we choose to switch between the two asymptotic
regimes at an energy of \teq{E_{\rm e} =2}MeV.

The relative importance of electron-electron and electron-ion
bremsstrahlung in the gamma-ray range can be quickly deduced from the
cross-sections listed in the Appendix.  For \teq{\gamma_{\rm e}\gg 1}
and \teq{\erg_{\gamma}\gg 1}, the e-e cross-section in equation~(A1) is
dominated by the \teq{\sigma_1} term, which coincides with the
ultrarelativistic limit of the e-p cross-section [see equation~(15-101)
of Jauch \& Rohrlich 1980].  Hence the emissivities depend only on the
target's charge, not its mass.  The ratio of electron-electron to
electron-ion contributions to bremsstrahlung is then simply
\teq{(n_{\rm p,1} + 2n_{\rm He,1})/(n_{\rm p,1} + 4n_{\rm He,1}) \simeq
0.86} for photon energies above $\sim 10$ MeV.  When the bremsstrahlung
collisions involve non-relativistic species (i.e. for X-ray
production), the situation changes and the electron-ion contributions
dominate the emission, since the dipole contribution to
non-relativistic electron-electron bremsstrahlung vanishes.  Note also
that {\it inverse bremsstrahlung}, i.e. radiation produced in
collisions between high energy protons and ISM electrons, contributes
insignificantly to the emissivity.  This can be seen from the formulae
of Jones (1971):  for photon energies \teq{\erg_{\gamma}\ll 1}, the
differential cross-section for inverse bremsstrahlung is
\teq{\sigma_{\rm p-e} \simeq (16 \, \alpha \, r_0^2/3)\,\log \, (0.68
\, \gamma_{\rm p}/\erg_{\gamma} )}.  This can be compared with the
Bethe-Heitler cross-section (e.g. Jauch \& Rohrlich 1980) for normal
bremsstrahlung, which in the \teq{\erg_{\gamma}\ll 1} limit for
\teq{\gamma_{\rm e}\gg 1} becomes \teq{\sigma_{\rm e-p}\simeq (16 \,
\alpha \, r_0^2/3)\,\log (2 \, \gamma_{\rm e}^2/\erg_{\gamma} )}; it is
clear that for \teq{m_{\rm e}\gamma_{\rm p}/m_{\rm p}\sim\gamma_{\rm
e}\gg 1}, inverse bremsstrahlung provides only minor contributions,
unless the \teq{e/p} cosmic ray ratio is quite small.

\subsection{Inverse-Compton Production of Gamma-Rays}
 \label{sec:inverseComp}

Inverse-Compton (IC) radiation is dominated by cosmic microwave
background (CMB) photons, with the less well-determined IR/optical
backgrounds that are local to typical remnants contributing generally
about 10\%--15\% of the IC flux (e.g. see Gaisser, Protheroe, \& Stanev
1998).  For electrons below $\sim 10$ TeV, the scatterings always occur
well in the Thomson limit and the photon energy in the electron rest
frame is much less than the electron rest energy:  \teq{4\gamma_{\rm e}
\erg_s\ll 1}, where the seed photon energy (in the lab frame) is
$\erg_s m_{\rm e} c^2$.  However, we must allow for higher energies
than this.  For the CMB, \teq{\erg_s m_{\rm e} c^2 \sim 3k T = 7.1
\times 10^{-4}}eV at the mean energy of the 2.73 K blackbody
distribution, so the Thomson limit is strictly valid only for electron
Lorentz factors obeying \teq{\gamma_{\rm e} \ll 3.5\times 10^8} or
$E_{\rm e} \ll 2\times 10^{14}$eV.  While this is often satisfied for
our calculations, Klein-Nishina corrections do become important for
$E_{\rm e} \gtrsim 3\times 10^{13}$eV.  This introduces both electron
recoil effects, that limit the maximum energy of the upscattered
photons to less than \teq{\gamma_{\rm e}}, and a drop in the
cross-section with increasing electron energy.  In our calculations, we
use the angle-integrated Klein-Nishina cross-section, differential in
the final energy of the photons, as derived by Jones (1968, see also
Blumenthal \& Gould 1970), the standard result adopted by other authors
(e.g. Sturner et al. 1997):
\begin{equation}
   \sigma_{\hbox{\fiverm K-N}}(\erg_s,\,\gamma_{\rm e} ;\erg_{\gamma} )\; =\;
     \dover{2\pi r_0^2}{\erg_s\gamma_{\rm e}^2}\,
     \biggl[ 2q\,\log_{\rm e}q +1+q-2q^2
          +\dover{\Gamma^2 q^2(1-q)}{2(1+\Gamma q)}\biggr]\ ,
 \label{eq:icjones}
\end{equation}
with \teq{\Gamma =4\erg_s\gamma_{\rm e}} being the parameter that
governs the importance (when \teq{\Gamma\gtrsim 1}) or otherwise of
photon recoil and Klein-Nishina effects, and with
\begin{equation}
   q\; =\;\dover{\erg_{\gamma}}{4\erg_s\gamma_{\rm e}
      (\gamma_{\rm e} -\erg_{\gamma} )}\ , \quad 0\leq q\leq 1\ ,
\end{equation}
where \teq{\erg_s m_{\rm e} c^2} is the initial photon energy,
\teq{\erg_{\gamma} m_{\rm e} c^2} is the upscattered (final) photon
energy, and \teq{\gamma_{\rm e} =(E_{\rm e} + m_{\rm e} c^2)/m_{\rm e}
c^2}, as for bremsstrahlung.  The constant \teq{r_0=e^2/(m_{\rm e}
c^2)} is the classical electron radius (\teq{\sigma_{\hbox{\fiverm
T}}=8\pi r_0^2/3} is the Thomson cross-section).  This result assumes
isotropic soft photon fields, the case for the CMB radiation.  Clearly
the Klein-Nishina decline is manifested through the last term in
equation~(\ref{eq:icjones}), while the recoil is embedded in the
\teq{q} parameter.  For any seed-photon (dimensionless) energy
\teq{\erg_s}, the maximum scattered photon energy is determined by
setting \teq{q=1}, giving \teq{\erg_{\gamma}\leq \Gamma\gamma_{\rm e}
=4 \gamma_{\rm e}^2\erg_s} in the Thomson limit and
\teq{\erg_{\gamma}\leq\gamma_{\rm e}} in the extreme Klein-Nishina
limit.

The inverse Compton emissivity for isotropic photon fields can then be
written down quickly (e.g. Blumenthal \& Gould 1970, or see Reynolds
1982, for representations in terms of frequency-dependent photon
intensities):
\begin{equation}
   \dover{dn_{\gamma}(\erg_{\gamma})}{dt} \; =\;
     c \int N_{\rm e}(\gamma_{\rm e})\, d\gamma_{\rm e}
     \int d\erg_s n_{\gamma}(\erg_s)\;
     \sigma_{\hbox{\fiverm K-N}}(\erg_s,\,\gamma_{\rm e} ;\erg_{\gamma} ) \ ,
 \label{eq:icspec}
\end{equation}
where \teq{n_{\gamma}(\erg_s)} is the distribution of seed photons, and
$N_{\rm e}(\gamma_{\rm e})$ is the electron energy distribution, that
can be expressed in terms of the $(dJ/dE)_{\rm e}$ that is computed
from our shock acceleration code:  $N_{\rm e}(\gamma_{\rm e}) = [4 \pi
m_{\rm e} c^2/v_{\rm e}] (dJ/dE)_{\rm e}$.  We use this expression with
a blackbody photon distribution
\begin{equation}
   n_{\gamma}(\erg_s)\; =\;
     n_{\hbox{\fiverm BB}}(\erg_s) \equiv
     \dover{\erg_s^2}{\pi^2\lambda_c^3}\,
     \dover{1}{e^{\erg_s/\Theta} - 1}\ ,
     \quad \Theta = \dover{kT}{m_{\rm e} c^2}\ ,
 \label{eq:blackbody}
\end{equation}
with $T = 2.73$ K so that \teq{\Theta =4.6\times 10^{-10}}.  Here
\teq{\lambda_c=\hbar /(m_{\rm e} c)} is the Compton wavelength.  This
form of the blackbody distribution is most appropriate for gamma-ray
applications, and is simply obtained (e.g. see Rybicki \& Lightman
1979) from the more familiar textbook form that uses photon
frequencies.  By using the multi-component fit to background soft
photon distributions in Figure~1 of Gaisser, Protheroe, \& Stanev
(1998), we determined that the CMB population contributes \teq{\sim
90\%} of the inverse Compton flux; for simplicity, hereafter, we use
just this background in all our IC flux calculations.

\subsection{Synchrotron Radiation}
 \label{sec:synch}

Using standard references such as Pacholczyk (1970) or Rybicki \&
Lightman (1979), the synchrotron emissivity from an electron Lorentz
factor distribution \teq{N_{\rm e}(\gamma_{\rm e})=[4 \pi m_{\rm e}
c^2/v_{\rm e}] (dJ/dE)_{\rm e}} can quickly be written down:
\begin{equation}
   \erg_{\gamma}\,\dover{dn_{\gamma}(\erg_{\gamma} )}{dt}\; =\;
     \dover{\sqrt{3}}{2\pi}\,\alpha\,\dover{e\, B_\perp}{m_{\rm e} c}
     \int_0^\infty N_{\rm e}(\gamma_{\rm e})\, F(x)\, d\gamma_{\rm e}
     \ \quad {\rm cm}^{-3} \, {\rm s}^{-1} \ ,
 \label{eq:synch}
\end{equation}
where
\begin{equation}
   F(x)\;\equiv\; x \int_x^\infty K_{5/3}(z) dz \ ; \quad
   x\; =\; \dover{\nu}{\nu_c}\; =\; \dover{\erg_{\gamma}}{\erg_c} \ ,
\end{equation}
is the well-known synchrotron spectral function for monoenergetic
electrons, with $K_{5/3}$ being the modified Bessel function.  Here
\teq{\erg_c = 3.398 \times 10^{-14} B_\perp \gamma_{\rm e}^2} the
critical synchrotron photon energy (\teq{\nu_c=\erg_c m_{\rm e} c^2/
(2\pi\hbar)= 4.199 \times 10^{6} B_\perp \gamma_{\rm e}^2} is the
critical frequency) for an electron of energy \teq{\gamma_{\rm e}},
with \teq{B_\perp} being the component of the magnetic field (in Gauss)
perpendicular to the line of sight.  For these calculations we assume
that the magnetic field is highly tangled and essentially isotropic, as
is indicated by the low polarized fractions observed at radio
frequencies in most remnants (Reynolds \& Gilmore 1993).  We use
Pacholczyk's tabulation of \teq{F(x)} to perform the integral
numerically for shock acceleration distributions \teq{N_{\rm
e}(\gamma_{\rm e})}, since commonly used approximations such as
assuming that each electron radiates all its synchrotron power at
\teq{\nu_c} can be poor for electron distributions with cutoffs (e.g.
see Reynolds 1998).

\section{RESULTS}
 \label{sec:results}

The results of our calculations fall into two main categories.  First,
we discuss particle distributions generated by the Monte Carlo code:
their evolution in the Sedov phase and issues of electron injection and
cosmic ray production up to the ``knee.'' Second, we present photon
emission spectra, focusing first on predictions of gamma-rays, then
considering the test case of the remnant IC 443, and finally extending
the discussion to broad-band (radio to gamma-ray) spectra.

\subsection{Particle Spectra During SNR Evolution}
 \label{sec:spectra}

We first produce models where the environmental parameters are taken to
be:  $n_{\rm p,1}=1$ \hbox{cm$^{-3}$}, ${\cal E}_{\hbox{\sixrm
SN}}=10^{51}$ erg, $M_{\rm ej}= M_{\odot}$, and $B_1=3\times 10^{-6}$
G.  For these parameters, $V_{\rm trans} \simeq 8.2\times 10^{3}$
\hbox{km~s$^{-1}$}, $R_{\rm trans} \simeq 1.9$ pc, and $t_{\rm trans}
\simeq 90$ yr, and in addition,  we have model parameters which we take
to be $g=1/r$ and $\eta=10$.  We compare results at three ages during
the SNR lifetime:  300 years ($V_{\rm sk} \simeq 4000$
\hbox{km~s$^{-1}$}), 1000 years ($V_{\rm sk} \simeq 2000$
\hbox{km~s$^{-1}$}), and $10^4$ years ($V_{\rm sk}\simeq 490$
\hbox{km~s$^{-1}$}).  To obtain our steady-state shock solution, we
must also know $d_{\hbox{\sevenrm FEB}}$.  As indicated in
equation~(\ref{eq:FEBratio}), $d_{\hbox{\sevenrm FEB}}$ depends on the
shock compression ratio, $r$, and $r$ is not known until the non-linear
solution is found.  Therefore, as we iterate toward a solution,
changing both the shock structure and the overall compression ratio,
$d_{\hbox{\sevenrm FEB}}$ will be iterated with $r$ using
equation~(\ref{eq:FEBratio}). For each of the three ages, we thus
obtain the self-consistent shock structure and complete ion spectra
which show the absolute injection and acceleration efficiency.  With
the additional parameter $E_{\rm crit}$ for electron injection and the
observed downstream electron temperature, $T_{\rm e,DS}$ (or our
parameter $f_{\rm e}$), we obtain the electron spectrum as well.  The
parameters for these models ($a$, $b$, and $c$) are listed in
Table~1.

\placefigure{fig:fourprofiles}
In Figure~\ref{fig:fourprofiles} we show the final smooth shock
structure for the three ages with the parameters just listed.  A fourth
shock (heavy dotted line) will be discussed in Section~4.2 below.  In
each case, we have iterated to a solution for the shock profile, as
well as the overall compression ratio, and the final smooth shock
conserves (to within a few percent) mass, momentum, and energy fluxes
at all positions from far upstream, through the nearly discontinuous
subshock (at $x \sim 0$), into the downstream region where we hold  all
parameters constant (i.e. we do not include adiabatic cooling).  For a
description of how the iteration process is performed, see Ellison \&
Reynolds (1991) or Ellison, Baring, \& Jones (1996).  The
self-consistent compression ratios we obtain (see labels in
Figure~\ref{fig:fourprofiles}), which decline with $t_{\hbox{\fiverm
SNR}}$, are well above the Rankine-Hugoniot value of four in all
cases.  The escape of particles at the FEB, as well as the lowering of
the ratio of specific heats from the contribution of relativistic
particles to the total pressure (see Ellison \& Reynolds 1991), causes
the increase in $r$.  This increase is indicated by the downstream
portions of the flow profiles where the test-particle speed (light
dashed line) is $0.25 \, V_{\rm sk}$ (i.e. $r=4$), while the downstream
flow speed for the non-linear shocks is always less than $0.25 \,
V_{\rm sk}$.  For each age, the shock is smoothed on the diffusion
length scale $\sim \kappa(E_{\rm max})/V_{\rm sk}$ of the highest
energy particles in the system.  This upstream precursor increases in
size as the remnant evolves.  Note that the distance scale is measured
in units of $\eta \, r_{\rm g1}$, where $r_{\rm g1} = m_{\rm p} V_{\rm
sk} c/(e \, B_1)$, so that distance {\it units} are proportional to
$V_{\rm sk}$ and are different for each profile (see Table~1 for values
of $d_{\hbox{\sevenrm FEB}}$ in pc).  Despite this extreme smoothing, a
distinct subshock persists in all cases with an abrupt transition to
the downstream state occurring over a length scale of about one thermal
ion gyroradius.  The subshock strength is the main determinant of the
downstream thermal ion and electron temperatures.

\placefigure{fig:SpectraThreePanels}
In Figure~\ref{fig:SpectraThreePanels} we plot omni-directional
particle spectra, $dJ/dE$ (i.e. particles per cm$^2$ per sec per
steradian per MeV/A), obtained with the smooth shocks shown in
Figure~\ref{fig:fourprofiles}.  This figure presents a time history of
evolution of particle distributions during a remnant's expansion.  In
all panels, the solid and dashed histograms represent the proton and
He$^{+2}$ spectra, respectively, with the helium injected far upstream
from the shock at cosmic abundances, i.e. $n_{\rm He,1} / n_{\rm p,1} =
0.1$.  The shock structure is determined including the helium
contribution self-consistently.  All spectra are calculated in the
shock reference frame at a position downstream from the shock, the
region of enhanced density where the gamma-ray emission is expected to
be greatest.  The spectra here are all normalized such that $n_{\rm
p,1} V_{\rm sk} = 1$ cm$^{-2}$ s$^{-1}$\ ($n_{\rm p,1}$ is the far
upstream proton number density).  The number density per unit energy
$N(E)$ is $(4\pi/v) dJ/dE$, where $v$ is the particle speed.  The
dotted lines in Figure~\ref{fig:SpectraThreePanels} are electron
spectra accelerated by the same shock as the protons and helium (note
that for electrons, the abscissa scale is energy, not energy per
nucleon).  Our approximation that the electrons are test particles will
be valid as long as the $e/p$ ratio at relativistic energies is much
less than unity.  All other assumptions concerning diffusion properties
are the same for electrons and ions.  Note that we have set $n_{\rm
e,1} V_{\rm sk} =  n_{\rm e,2} V_{\rm sk}/r = 1.2$ cm$^{-2}$
s$^{-1}$\ (i.e., charge neutrality with fully-ionized helium is
assumed), where $n_{\rm e,2}$ is the downstream number density of
electrons and is calculated by integrating $(4\pi/v) dJ/dE$ over {\it
all energies}.  For all three electron examples,  $E_{\rm crit}=100$
keV and $f_{\rm e} = 1$.

The spectra in Figure~\ref{fig:SpectraThreePanels} are ``complete'' in
that they are entire distributions from thermal energies to the highest
energies where the spectra turn over due to particles escaping at the
FEB.  Electron losses are not important in any of these examples, but
we will show examples later where they are.  The spectra possess an
enhancement of He$^{+2}$ that comes about because the smooth shock
naturally accelerates particles with large mass-to-charge ratios more
efficiently (e.g. Jones \& Ellison 1991), as they possess longer
diffusion lengths and therefore sample larger effective compression
ratios.  This enhancement, which is discussed at length in Ellison,
Drury, \& Meyer (1997), permits He$^{2+}$ to dominate the proton
contribution to the energy density, except for the highest energies per
nucleon.  In general, and in agreement with other non-linear models
(e.g. Berezhko, Yelshin, \& Ksenofontov 1996), we find very high
efficiencies, easily putting over 50\% of the total energy density in
particles above 1 GeV.  Equally important, the contribution of
He$^{+2}$ to gamma-ray production is further enhanced by the factor in
equation~(\ref{eq:alphacorr}) and can be comparable to that from
protons even though helium is injected far upstream with only 10\% of
the number density.

Another important feature of the spectra in
Figure~\ref{fig:SpectraThreePanels} is that they are not strictly
power-laws (even if plotted on a momentum scale), but show an upward
curvature, becoming harder at higher energies.  This effect is masked
somewhat for the ions because of the kinematic break at $\sim mc^2$,
but shows up more dramatically for the electrons between $\sim 10$ MeV
and 10 GeV.  The smooth shock, combined with our assumption that the
upstream diffusion length is an increasing function of energy, causes
high energy particles to be accelerated more efficiently than low
energy ones, producing the upwardly curved spectra (e.g. Eichler 1984;
Jones \& Ellison 1991), and very different spectral shapes for protons
and electrons below a few GeV.  In general, when compared to
test-particle results with $r=4$, the non-linear spectra are
considerably {\it steeper} at the lowest energies because of the weak
subshock with compression ratios less than 4.  At the highest energies,
the non-linear spectra are flatter than the test-particle ones because
the overall compression ratio is greater
\footnote[2]{
Note that the spectral indices at the highest energies, but below the
cutoff from the FEB, are larger (i.e. the spectra are steeper) than
$\sigma = (r + 2)/(r - 1)$, the value expected from unmodified shocks
with a compression ratio of $r$. This is a purely non-linear effect
from particle escape and our results are quite close to the analytic
estimate of Berezhko (1996), i.e. $\sigma \simeq 3.5 + [ (3.5 - r_{\rm
sub}/2) / (2 r - r_{\rm sub} -1) ]$.  Malkov (1997) obtained a similar
result.
}
than 4.  At intermediate energies, between $\sim 1$ MeV and $10$ GeV
(i.e. electron energies responsible for radio synchrotron emission),
the non-linear spectra, particularly electrons, are considerably
steeper than the test-particle predictions.

Finally, in Figure~\ref{fig:SpectraThreePanels} the cutoff from the FEB
occurs at energies proportional to the particle charge
(equation~\ref{eq:EmaxSize}) so the helium spectrum extends to a total
energy a factor of two higher than the electrons or protons (a factor
of two lower in energy per nucleon).  Note also that in cases where
electron cooling losses are important, cooling-generated structure can
appear in the electron distribution just below the cutoff energy.

While our Monte Carlo implementation of non-linear shock acceleration
uniquely includes the self-regulation of injection, giving the full ion
spectrum self-consistently, some of the other results of our
calculation will be properties of any non-linear shock model.  In
particular, the concave spectra will result in any modified shock if
the diffusion length increases with energy, since more energetic
particles will then see a larger effective compression ratio.
Compression ratios larger than 4 should also always result from
non-linear models that include particle escape, so that the asymptotic
high-energy slope is flatter than the test-particle value.

\subsection{Acceleration of Particles to $10^{15}$ eV}
 \label{sec:cosmicrays}

If SNRs are the main sources of galactic cosmic rays, they must be
capable of accelerating ions up to at least $\sim 10^{15}$ eV where the
so-called ``knee'' in the all-particle cosmic-ray spectrum is
observed.  Here we consider how acceleration to energies considerably
higher than in our previous examples will influence the gamma-ray
emission. This addresses the crucial question of whether shell-type
remnants can both supply the observed galactic cosmic ray population
and explain the emission from the handful of EGRET unidentified sources
that have SNR associations.  From equation~(\ref{eq:EmaxSize}) we see
that $E_{\rm max}^{\rm size}$ will increase for increased magnetic
field, increased ${\cal E}_{\hbox{\sixrm SN}}$, decreased ambient
density, and/or decreased $\eta$.  The sensitivity of the maximum
energy to $r$, $g$, or $t_{\hbox{\fiverm SNR}}$ is relatively small.
The dependence of \teq{E_{\rm max}} on density strongly suggests that
it may be difficult for a given remnant to simultaneously generate
cosmic rays out to the knee {\it and} radiate sufficiently to support
the detections by EGRET.

To obtain a high maximum energy, we choose $\eta = 1$ (strong
scattering), $g=1/r$ (the scattering mean free path is inversely
proportional to the plasma density), $n_{\rm p,1} = 10^{-3}$
\hbox{cm$^{-3}$}, $B_1 = 10^{-5}$ G, ${\cal E}_{\hbox{\sixrm SN}} =
10^{52}$ erg, and $M_{\rm ej}=10 M_{\odot}$.  These choices result in
$V_{\rm trans} \simeq 8240$ \hbox{km~s$^{-1}$}, $t_{\rm trans} \simeq
1950$ yr, and $R_{\rm trans} \simeq 41$ pc.  Referring to
Figure~\ref{fig:figEmax} \ (top curve), we determine our solution at
$t_{\hbox{\fiverm SNR}} \simeq 4\times 10^{4}$yr, near the peak in the
maximum acceleration energy, where the acceleration changes from being
time-limited to space-limited.  This optimization gives a maximum
energy of the cosmic rays of $E_{\rm max} \simeq 4\times 10^{15}$ eV.
At this age, $V_{\rm sk}\simeq 1340$ \hbox{km~s$^{-1}$}\ and $R_{\rm
sk} \simeq 137$ pc (Model {\it d} in Table~1), and our  non-linear
shock solution yields a compression ratio, $r\simeq 6.5$.
Figure~\ref{fig:figEmax}\ shows that electrons will experience severe
losses for these parameters, so that the contributions of
bremsstrahlung and inverse Compton scattering at the highest energies
will be suppressed.  The maximum electron energy at $t_{\hbox{\fiverm
SNR}}= 4\times 10^{4}$ yr is about $3\times 10^{13}$ eV.

\placefigure{fig:SpectraEmax}
In Figure~\ref{fig:SpectraEmax} we show the proton (solid line) and
helium (dashed line) spectra for this cosmic ray knee energy example.
We also depict the electron spectrum (dotted line) with $E_{\rm
crit}=0$  and $f_{\rm e}=0.05$ (i.e. the electrons are injected with
thermal distributions at $T_{\rm e,inj}= 1.0\times 10^{6}$ K).  These
$E_{\rm crit}$ and $f_{\rm e}$ values have been chosen to provide an
$(e/p)_{\rm 10GeV} \simeq 0.02$ consistent with cosmic ray observations
(e.g. M\"uller et al. 1995).  The cutoff in the electron spectrum from
combined synchrotron and inverse Compton losses is clearly seen as is a
slight pile-up of electrons just below the cutoff.
Figure~\ref{fig:SpectraEmax} reveals the important result that it is
difficult to obtain much higher cosmic ray energies than these using
normally-accepted ISM parameters.  Significant juggling of the various
parameters was necessary to increase \teq{E_{\rm max}} to above $\sim
10^{15}$ eV, including a requisite decrease in the density to an almost
untenably low value.  The accompanying increase in \teq{E_{\rm max}}
came at the expense of large decreases in photon emissivity: pion decay
and bremsstrahlung are proportional to $n_{\rm p,1}^2$ while inverse
Compton is proportional to $n_{\rm p,1}$.  Such trade-offs are inherent
in the problem of simultaneously producing super-100 TeV cosmic rays
and copious GeV-TeV gamma-ray emission in individual remnants.  Hence,
in accord with many previous expositions, we find it extremely
difficult to generate cosmic rays beyond \teq{10^{15}}eV with normal
Fermi acceleration in the homogeneous ISM.

The dotted line in Figure~\ref{fig:fourprofiles} shows the shock
structure for Model $d$ and it is quite different from the three other
examples.  The main reason for this, besides the much higher $E_{\rm
max}$ and consequently, longer precursor, is the extremely low Alfv\'en
Mach number that results from a high $B_1$ and a low ISM density. For
this case, ${\cal M}_{\hbox{\sixrm A}} \simeq 2.3$, the Alfv\'en wave
heating in the precursor is very strong, and $v_{\hbox{\sixrm
A}}/V_{\rm sk} \simeq 0.5$.  The combination of strong precursor
heating and a high $v_{\hbox{\sixrm A}}$ (the scattering centers move
through the upstream plasma at $v_{\hbox{\sixrm A}}$) results in a
lowering of the acceleration efficiency and the overall compression
ratio, which is only $\sim 6.5$ for this case. Furthermore, the
subshock compression ratio is quite large; $r_{\rm sub} \simeq 4.4$.
The combination of a large $r_{\rm sub}$ and relatively small $r$
results in less shock smoothing than our previous examples which, in
turn, results is little or no $A/Q$ enhancement of helium over protons,
as is evident in Figure~\ref{fig:SpectraEmax}.

\subsection{Examples of Photon Production}
 \label{sec:examples}

\placefigure{fig:PhotonsThreePanels}
In the top panel of Figure~\ref{fig:PhotonsThreePanels}, we depict
photon spectra produced by pion decay (dashed lines), bremsstrahlung
(dot-dashed line), and inverse Compton scattering off the CMB radiation
(dotted line) by particles with power-law momentum distributions,
$dN/dp \propto p^{-2}$.  We note that the p-He contribution to the pion
decay emission is of the same shape as the p-p one illustrated, but
with a simple multiplicative factor that combines the relative
abundance of He and the factor in equation~(\ref{eq:alphacorr}).  Also,
e-e bremsstrahlung contributes a virtually identical spectrum above 10
MeV to the e-p one shown.  The ``\teq{\nu F_{\nu}}'' format of the
figure is chosen to illustrate at what energy the peak power of the
gamma-rays emerges.  This depiction follows the work of Gaisser,
Protheroe, \& Stanev (1998), and to facilitate comparison with their
results, we use the same spectral shape and normalization they use for
their Figure~3, i.e.
\begin{equation}
   \left( \dover{4 \pi}{v} \right)\, \dover{dJ}{dE}\; =\;
     \dover{1}{V} \dover{dN}{d{\cal E}}\; =\;
     \dover{a}{V} \left( \dover{{\cal E}}{1 {\rm GeV}} \right)
     \left( \dover{p}{1 {\rm GeV/c}} \right)^{-3}
       \exp \left( -\dover{\cal E}{{\cal E}_{\rm cut}} \right)
     \quad {\rm GeV}^{-1} \ {\rm cm}^{-3}\ ,
\end{equation}
including an exponential cutoff with ${\cal E}_{\rm cut} = 80$ TeV.
Here, $\cal{E}$ is the total particle energy, $E$ is the kinetic
energy, $V$ is the volume of the emitting source, and the normalization
of Gaisser et al.  of $a/V = 1$ GeV$^{-3}$ \hbox{cm$^{-3}$} for both
electrons and protons is used (the electron to proton ratio is set to
one at fully relativistic energies).  For this example only, we neglect
helium (or heavier ion species), as in Figure~3 of Gaisser et al.
(1998).  Alternatively, this distribution can be expressed as:
\teq{n(\gamma )= {\cal N} \gamma^{-2}\beta^{-3} \exp
[-\gamma/\gamma_{\rm cut}]}, where $\gamma$ is the particle Lorentz
factor and the normalization constant is \teq{{\cal N}= 1 \ {\rm
GeV}/(mc^2)}, i.e. \teq{\simeq 1957} for electrons and \teq{\simeq
1.066} for protons.  A prominent feature of this particular example is
that the radiation is dominated by inverse-Compton emission, which is
intrinsically flatter than bremsstrahlung and pion decay radiation due
to the Compton scattering kinematics.  The relative importance of the
various processes depends strongly on the ambient density, electron
losses, and the $(e/p)$ ratio, as discussed below.

The power-law portions of the particle distributions can be used to
derive asymptotic limits as checks on our computations.  For IC
scattering, equation~(7.31) of Rybicki \& Lightman (1979) can be used
to derive an analytic approximation to the spectrum.  For e-p and e-e
bremsstrahlung, since both cross-sections asymptote to the expression
for \teq{\sigma_1} in equation~(A2) for ultrarelativistic electrons,
this form can be integrated over the power-law to obtain
\teq{dn_{\gamma}(\erg_{\gamma} )/dt\simeq 4 \, {\cal N}\alpha \, r_0^2
\, c\, [7/2+\log_e(2\erg_{\gamma} )]\,\erg_{\gamma}^{-2}}.  For pion
decay radiation, we note that well above threshold, the photon spectrum
traces that of the parent proton population (e.g.  see Baring \&
Stecker 1998), which results in an \teq{\erg_{\gamma}^{-2}} spectrum
for the case in the top panel of Figure~\ref{fig:PhotonsThreePanels}.
The normalization of this tracing is called a {\it spectrum-weighted
moment}, and is 0.16 for a \teq{\gamma^{-2}} proton distribution
(Gaisser 1990; see also Drury, Aharonian, \& V\"olk 1994; Gaisser,
Protheroe, \& Stanev 1998).  The asymptotic form for pion decay
radiation is then \teq{dn_{\gamma}(\erg_{\gamma} )/dt\simeq 0.16 \,
{\cal N}\sigma_{pp\to \pi^0X} \, c \, \erg_{\gamma}^{-2}}.  As
\teq{\gamma_{\rm cut}\to\infty}, our numerical results smoothly
approach these asymptotic forms for all three processes, thereby
providing confirmation that our integration routines were working
correctly.  Furthermore, we find good (though not perfect) agreement of
our results with the curves in Figure~3 of Gaisser, Protheroe, \&
Stanev (1998), with the slight differences being attributable to
assumptions made in the modeling of pion production and
bremsstrahlung.  Finally, note that Klein-Nishina corrections to
inverse Compton scattering off CMB photons become important for
electron energies exceeding around 20 TeV.

\subsection{Gamma Ray Spectra and IC 443 as a Test Case}
 \label{sec:IC443}

In the middle panel of Figure~\ref{fig:PhotonsThreePanels} we show the
individual components for the $V_{\rm sk} = 2000$
\hbox{km~s$^{-1}$}\ example of Figures~\ref{fig:fourprofiles} and
\ref{fig:SpectraThreePanels} (Model {\it b}:  $E_{\rm crit}= 100$ keV
and $f_{\rm e}=1$).  In the lower panel we exhibit the photon emission
for our extreme maximum energy example with $V_{\rm sk} = 1340$
\hbox{km~s$^{-1}$}\ (i.e. Model {\it d}:
Figure~\ref{fig:SpectraEmax}).  It is clear that the relative
importance of the various emission mechanisms can vary greatly
depending on the parameters, with the two most important being the
ambient density, $n_{\rm p,1}$, and the $(e/p)$ ratio at fully
relativistic energies, i.e. $(e/p)_{\rm 10GeV}$.  The cutoff energy,
$E_{\rm max}$, is also important for fitting the constraints imposed by
observations at TeV energies, and in particular the overall flux via
more subtle feedback effects of the non-linearity of the Fermi
acceleration process.  It also influences the radio synchrotron cutoff
which will be discussed below.  The main determinant of $E_{\rm max}$
in our model is $\eta$, which is basically a free parameter within the
broad range $1 \lesssim  \eta \lesssim 100$, though evidence from a
variety of origins suggests values of \teq{\eta \sim 1-10} apply to
cosmic plasmas.  In a particular source, it may be possible to restrict
$n_{\rm p,1}$ somewhat from X-ray and gamma-ray observations and
$(e/p)_{\rm 10GeV}$ by EGRET observations (discussed below).  An
important feature of Model $b$ is the importance of He$^{2+}$ pion
decay emission. The short dashed lines in
Figure~\ref{fig:PhotonsThreePanels} show the gamma-ray emission from
helium, while the long dashed lines show the contribution from protons.
At photon energies below $\sim 100$ GeV, the helium contribution is
approximately equal to the proton contribution for Model $b$. The same
is not the case for Model $d$ since little $A/Q$ enhancement arises, as
discussed above.

It is clear from the large number of parameters that only a limited
amount of information will be obtained from fitting a particular source
unless the parameters can be constrained either by better observations
or improved understanding of the plasma physics of shock acceleration.
Nevertheless, it is fruitful to investigate how the various parameters
influence the overall photon spectrum.  First, as $n_{\rm p,1}$ is {\it
decreased}, the importance of inverse Compton increases relative to
bremsstrahlung and pion decay since inverse Compton emission is
proportional to the electron density, while bremsstrahlung and pion
decay depend on the square of the ambient density.  The strength of a
pion decay bump at $\sim 100$ MeV gives important clues to $n_{\rm
p,1}$ and $(e/p)_{\rm 10GeV}$: a weak or non-existent bump implies a
low $n_{\rm p,1}$ and/or a large $(e/p)_{\rm 10GeV}$.  The quality of
the data in Esposito et al. (1996) for the EGRET unidentified sources
with shell-type SNR ``counterparts'' is insufficient to confirm or
exclude the existence of such a feature.  For this reason, the approach
of Gaisser, Protheroe, \& Stanev (1998) in using this data to constrain
the \teq{e/p} ratio is presently unrealistic.  Besides the existence
(or otherwise) of a pion decay bump, the overall slope of the photon
distribution gives information on the relative importance of inverse
Compton radiation compared to bremsstrahlung since the inverse Compton
component possesses a flatter slope.  The cutoff energy, $E_{\rm max}$,
if it can be determined by TeV observations, also gives useful
information: a low $E_{\rm max}$ implies some combination of large
$\eta$, high $n_{\rm p,1}$, and/or large $(e/p)_{\rm 10GeV}$.  In
principle, information on the background magnetic field strength can be
obtained if electron losses become important and the electron spectrum
cuts off at a lower energy than the proton spectrum, e.g. as in the
bottom panel of Figure~\ref{fig:PhotonsThreePanels}.  In this case, the
pion decay spectrum may extend beyond the inverse Compton and
bremsstrahlung spectra and a spectral feature may be present.
Furthermore, the consideration of X-ray synchrotron cutoffs (e.g.
Reynolds 1996; Allen et al. 1997) can constrain {\bf B}, and in
conjunction with TeV gamma-ray data, can restrict the permissible
regions of the density/magnetic field strength parameter space.

We now apply our model to one specific shell-type SNR with gamma-ray
detections reported in the Esposito et al. (1996) collection, namely IC
443.  While Gaisser, Protheroe, \& Stanev (1998) also use $\gamma$
Cygni as a test case, the likelihood that its counterpart EGRET source
is truly associated with shell-related emission is reduced by the small
EGRET error circle reported in Esposito et al. (1996), and all but
discounted by the refined localization performed by Brazier et al.
(1996).  Hence, we regard $\gamma$ Cygni as being a weak candidate for
producing detectable shell-associated gamma-ray emission.  Based on the
discussion of Lozinskaya (1992), Gaisser, Protheroe, \& Stanev (1998)
estimate for IC 443 that the ambient density is $n_{\rm p,1} \simeq
0.3$ \hbox{cm$^{-3}$}, the distance to IC 443 is about 1.5 kpc, the
radius is about 10 pc, and the remnant age is $t_{\hbox{\fiverm SNR}}
\simeq 5000$ yr.  With these parameters, we estimate (for standard SN
parameters and expansion into a homogeneous medium, which is clearly
not the case for IC 443) a current shock speed $V_{\rm sk} \simeq 940$
\hbox{km~s$^{-1}$}\ (Model {\it e} in Table~1), and an \teq{E_{\rm
max}} somewhat below 10 TeV.

\placefigure{fig:GammasNine}
In Figure~\ref{fig:GammasNine} we present a grid of nine models varying
$n_{\rm p,1}$ from 0.1 to 1 to 10 \hbox{cm$^{-3}$}, and $(e/p)_{\rm
10GeV}$ from 0.01 to 0.1 to 1.  In these plots, we also depict the
EGRET observations of the IC 443 region (2EG J0618+2234 data points:
Esposito et al. 1996) plus the upper limits from the Whipple telescope
(Buckley et al.~1997, see also Lessard et al.~1995) and the HEGRA
array.  Note that there exist upper limits above 10 TeV from the
scintillator array experiment in Tibet (Amenomori et al. 1997); these
are not displayed.  All plots show the number of photons per cm$^2$ per
sec incident at Earth (i.e., flux) assuming a standard normalization: a
source at 1 kpc with an emitting volume of $V_{\hbox{\fiverm SNR}} = 1$
pc$^3$.  For the distance (1.5 kpc) and radius (10 pc) estimates of
IC443 given above, the photon fluxes in Figure~\ref{fig:GammasNine}
should be multiplied by $\sim 10^3/(1.5)^2 \sim 400$ (assuming a
fractional shell thickness of 0.1, or a volume filling factor of about
0.25).  In all of these models, we use $B_1=3\mu$G and $\eta=10$,
noting that a higher $B_1$ or lower $\eta$ would yield a higher cutoff
energy.

In generating this range of emission spectra, we computed a single
non-linear shock solution using the parameters just described (i.e.
fixing $E_{\rm max}$) plus $E_{\rm crit} = 0$ and $f_{\rm e}=1$ for the
electron injection (i.e. Model {\it e} Table~1), and then simply scaled
the electron spectral normalization to give the $(e/p)_{\rm 10GeV}$
values quoted (this amounts to varying \teq{E_{\rm crit}} and
\teq{T_{\rm e,inj}}) and calculated the photon emission using the
densities shown.  This approximation will not be accurate for electrons
well below 1 GeV, but the {\it shape} (as opposed to normalization) of
the electron spectrum above GeV energies is insensitive to variations
in Sedov solution parameters and $E_{\rm crit}$ and $f_{\rm e}$ for a
given self-consistent shock solution; hence the gamma-ray components
exhibited in Figure~\ref{fig:GammasNine} are representative of results
of self-consistent shock simulation runs.  One caveat is that this is
not entirely true for the \teq{(e/p)_{\rm 10GeV} =1} cases, since then
it becomes necessary to include the effect of the electrons on the
shock dynamics.

It is clear from this set of models that the lack of a prominent
pion-decay bump centered at around 67 MeV in the EGRET data for IC 443
(true also for other sources listed in Esposito et al. 1996) can only
be matched with $n_{\rm p,1} \gtrsim 3$\hbox{cm$^{-3}$} and $(e/p)_{\rm
10GeV} \gtrsim 0.1$.  This $(e/p)_{\rm 10GeV}$ ratio is larger than is
believed to be the case for galactic cosmic rays: $(e/p)_{\rm
10GeV}\sim 0.02$ is inferred from the measured local cosmic ray
abundances in the 1-10 GeV range (e.g. M\"uller et al. 1995), and also
from modeling of the diffuse galactic gamma-ray background radiation
(Bertsch et al.  1993; Hunter et al. 1997).  Thus, the EGRET
observations provide significant constraints on the modeling of IC
443.  If $(e/p)_{\rm 10GeV} \gtrsim 0.1$, then values of $n_{\rm
p,1}\gtrsim 3$ \hbox{cm$^{-3}$} can provide an adequate fit to the {\it
shape} of the observed EGRET spectrum.  The slope of the EGRET data
argues against parameter regimes that yield a dominant (flat) inverse
Compton component, with bremsstrahlung possessing a spectral index
appropriate to the data; such a conclusion applies to most of the EGRET
unidentified sources in Esposito et al.  (1996), and was made by
Gaisser, Protheroe, \& Stanev (1998).  It's also clear from
Figure~\ref{fig:GammasNine} that our models predict fluxes slightly
above the Whipple upper limit.  However, if we had chosen a larger
value of $\eta$ (i.e., $\eta=50$--100), or a lower magnetic field, the
maximum energy would have been less and the Whipple point could have
been comfortably accommodated.  On the other hand, as we show below in
Figure~\ref{fig:GammasAll}, virtually any $n_{\rm p,1} \gtrsim 1$
\hbox{cm$^{-3}$}\ predicts radio emission well below observed fluxes.
While this might result if the $\gamma$-ray emission volume is
considerably less than that inferred for radio, or if there is
significant clumping of the magnetic field, it must be emphasized that
all of the above conclusions are based on the assumption that the EGRET
detection of IC 443 is of {\it shell-related emission} and this may not
be the case.

It is important to note that all of the densities referred to in this
paper (such as in
Figures~\ref{fig:PhotonsThreePanels}--\ref{fig:GammasMax}) with a
subscript ``1'' are true upstream ISM values.  The simulation produces
downstream densities for use in the emissivity calculations, and these
are the upstream ISM values multiplied by the total compression ratio.
As such, we establish correct normalizations so that the photon
emission spectra actually correspond to the stated ISM densities.  This
contrasts with the work of Gaisser, Protheroe, \& Stanev (1998), and
Sturner et al.  (1997), who used power-law distributions, with the
stated ISM density being used as a {\it coefficient} for infinite
power-laws; no connection between the power-law normalization and the
ISM density can be made in these two papers.  Drury, Aharonian, \&
V\"olk (1994) did introduce bounds to proton distributions, but chose
lower limits around 1--10 MeV, well in excess of the thermal values
expected from dissipational heating of shocks (e.g. see
Figure~\ref{fig:SpectraThreePanels}).  Thus, their normalizations  (say
at 1 GeV/nucleon) and corresponding photon fluxes are greater than ours
--- this can be seen by a comparison of Figure~4 of Drury et al. (1994)
and the middle row of Figure~\ref{fig:GammasNine} here.

\subsection{Broad-Band Photon Spectra}
 \label{sec:broad-band}

Our focus so far has been gamma-ray emission from SNR shells.  However,
important information and constraints can be gained from broad-band
studies of emission throughout the electromagnetic spectrum.  This has
been the approach of Mastichiadis \& de Jager (1996) and de Jager \&
Mastichiadis (1997), who have examined the remnants SN1006 and W44.
While non-thermal nucleons are only important for producing gamma-rays
from pion decay, electrons produce photons from radio to gamma-ray
energies.  In Figure~\ref{fig:GammasAll} we show all of the various
photon spectral components from Model {\it e} (Table~1) that formed the
basis for the array of examples in Figure~\ref{fig:GammasNine}.  The
individual components are again normalized to $d=1$ kpc and
$V_{\hbox{\fiverm SNR}} = 1$ pc$^3$, while the total photon emission
(heavy solid line) has been multiplied by 500 for a rough match to the
EGRET data.  We depict radio and X-ray observations in addition to the
previously illustrated EGRET, Whipple, and HEGRA data, but omit the
OSSE upper limits in the 50 keV--1 MeV band that are presented in
Sturner et al. (1997), since they do not significantly constrain our
continuum spectra.  It is important to emphasize that in this plot, we
are not attempting a detailed fit to the data, but rather aiming to
illustrate how the various components relate to one another.

\placefigure{fig:GammasAll}
There are several features to observe in Figure~\ref{fig:GammasAll}.
First of all, for the particular density of this model ($n_{\rm p,1} =
0.3$  \hbox{cm$^{-3}$}), inverse Compton (dotted line) emission
contributes to the spectral flattening in the EGRET range.  Such a
flattening, if seen in some source, is therefore not necessarily
indicative of the presence of cosmic ray nucleons, unlike a pion decay
bump.  Second, normalizing the overall continuum to approximately match
the flux level for the EGRET unidentified source 2EG J0618+2234 that is
associated with IC 443 conflicts slightly with the Whipple upper limit
but not the HEGRA array limit (note that the HEGRA imaging telescope
upper limits at 500 GeV reported by Hess (1997) are comparable to those
of Whipple).  This result depends almost totally on the maximum energy
obtained (i.e. equations~[\ref{eq:Etrans}], [\ref{eq:Emax}], and
[\ref{eq:EmaxSize}]) which, in turn, is a decreasing function of the
parameter $\eta$.  A large value of $\eta$ could result from
environment effects such as the SNR being contained in a partially
ionized region, thereby permitting the shock acceleration model to
comfortably accommodate the constraints imposed by atmospheric
\v{C}erenkov telescope measurements.  Third, the electrons that produce
the IC gamma-rays also generate the radio to optical synchrotron
emission (light solid line) and the synchrotron spectrum does not
extend into the X-rays, due to the maximum electron energy being in the
TeV range.  As a consequence, it is actually the bremsstrahlung from
non-relativistic electrons that dominates the X-ray signal.  This could
potentially provide an alternative explanation to synchrotron emission
for the non-thermal X-rays seen in IC 443 (Keohane et al.~1997) and Cas
A (Allen et al.~1997).  We obtain X-ray indices (at around 10 keV) in
the 2.3--2.7 range (e.g. see Figure~\ref{fig:GammasAll}), which would
nicely describe the index ASCA obtained for IC 443, but are generally
flatter than in the X-ray observations of Cas A and SN1006 (\teq{\sim
3}).  Furthermore, the absence of X-ray lines in SN1006 (Koyama et al.
1995), normally excited by electron impact, indicates a paucity of
electrons with energies of a few keV, whether from  thermal or
nonthermal distributions.

The flattening bremsstrahlung spectral shape at hard X-ray energies
(e.g. see Figure~\ref{fig:GammasMax} below) strongly contrasts with the
sharpness of X-ray synchrotron cutoffs (Reynolds 1996), providing a
potential observational discriminant; evidence for this steepening in
RXTE data for SN1006 (Allen et al.~1998, in preparation), supports the
synchrotron interpretation.  However, we caution against automatically
assuming that non-thermal X-rays from shells are synchrotron
radiation.  We can obtain quite steep X-ray bremsstrahlung spectra
because of the curvature in the electron distribution that results from
our non-linear treatment of the acceleration process; power-law
distributions generate much flatter X-ray spectra (e.g. see Sturner et
al.  1997), as suggested by the top panel of
Figure~\ref{fig:PhotonsThreePanels}.  The curvature in the electron
distribution disguises the break that naturally arises in the
Bethe-Heitler cross-section at \teq{\erg_{\gamma}\sim m_{\rm e} c^2}.
It is evident in Figure~\ref{fig:GammasAll} that the overall steepness
of the bremsstrahlung spectrum precludes any attempt to simultaneously
fit both EGRET and Ginga data.

Another obvious property of our particular model is that it falls well
below the synchrotron radio spectrum of Erickson \& Mahoney (1985), and
as we mentioned above, this may imply that the $\gamma$-ray emission
volume is considerably less than the radio.  As for the spectral {\it
shape} however, the model can reproduce the unusually flat (\teq{\sim
0.35}) synchrotron radio spectrum, as indicated by the upper dotted
line in Figure~\ref{fig:GammasAll}, which is the synchrotron emission
multiplied by $6\times 10^{4}$.  Such a flat radio spectral index is
also present in W44 (though in virtually none other of the $\sim 200$
Galactic shell remnants), and was used in the test-particle model of de
Jager \& Mastichiadis (1997) to argue that it is too flat to be
explained by a shock-accelerated electron population.  This is not
necessarily the case, given that non-linear solutions to the Fermi
acceleration problem can generate flat distributions (i.e. flatter than
\teq{E^{-2}}).  However, the conjecture of de Jager \& Mastichiadis
(1997) that a pulsar may inject electrons with the required
distribution via its relativistic wind, thereby circumventing the need
to invoke Fermi acceleration at the remnant's outer shock, may still be
correct.  For IC 443, Sturner et al. (1997) retained shock
acceleration, but included free-free absorption which produces a
flattening at the lower radio frequencies roughly matching the radio
data.  Presumably, adding free-free absorption could also provide
better compatibility between our model and the observed radio
emission.

\placefigure{fig:GammasMax}
The usefulness of considering broad-band emission comes from the fact
that a variation in any single model parameter impacts several
wavebands.  For example, the radio intensity depends on the square of
the magnetic field, $B_1$, but increasing $B_1$, also makes electron
losses more severe, lowering the energies where the bremsstrahlung and
IC emission cut off.  Variations in the ISM density impact all
wavebands. As density declines, the maximum particle energy increases,
and the gamma-ray continuum extends to higher energies, but the overall
flux at sub-TeV energies decreases accordingly.  With lower densities
the inverse Compton component becomes more prominent in the gamma-ray
band, flattening the spectral index.  This prominence was emphasized by
Mastichiadis and de Jager (1996) and Pohl (1996) in their predictions
that SN1006 would be a TeV gamma-ray source.  As suggested by
Mastichiadis and de Jager (1996), TeV upper limits or positive
detections can constrain the parameter $\eta =\lambda r_{\rm g} $ to
values signifying departure from Bohm diffusion (i.e. $\eta \gg 1$).
If steep X-ray emission is interpreted as coming from a synchrotron
cutoff, this determines \teq{E_{\rm max}^2 B} and also a combination of
\teq{B} and the electron density.  Through Equation~(\ref{eq:Etrans}),
\teq{\eta} therefore couples to \teq{B} and the gamma-ray inverse
Compton flux must anti-correlate with both \teq{B} and \teq{\lambda
/r_g}.  Hence lower bounds to \teq{\eta} are derivable from TeV
observational constraints.  These features are illustrated by our
cosmic ray knee example shown in Figure~\ref{fig:GammasMax}, whose
particle distributions are exhibited in Figure~\ref{fig:SpectraEmax}.

In addition to all this, of course, is the fact that the SNR
environment is likely to be far from homogeneous.  Rayleigh-Taylor
instabilities behind expanding SNR shocks (e.g. Jun \& Norman 1996) may
produce localized non-cospatial clumping of the magnetic field and/or
density, as exemplified by the complexity of spatial maps of remnants
such as Cas A, and different processes (e.g. radio synchrotron and
pion-decay) may have different emission volumes.  Furthermore, while we
assumed that the remnant shock is everywhere plane-parallel, a given
remnant shock is expected to be oblique over a sizable fraction of its
surface where the downstream (interior to the shock) magnetic field and
consequently the synchrotron emissivity are enhanced accordingly.
Clearly, the surface brightness of the radio flux is highly sensitive
to field or density clumping.  Given the complexity of the situation
and the interplay of the various parameters, we believe more will be
learned about a particular source by combining a general fit to
observations over the widest possible frequency band, with detailed
fits to narrow band observations.

\placefigure{fig:GamEvolve}
To conclude this subsection, we display in the top panel of
Figure~\ref{fig:GamEvolve} our remnant evolutionary sequence (i.e. the
spectra shown in Figure~\ref{fig:SpectraThreePanels}) in an $E^2 dN/dE$
format.  This illustrates how the highest energy ions dominate the
energy density of the system and emphasizes the differences in the
electron and proton spectra.  The spectra in the top panel are all
normalized to $n_{\rm p,1} V_{\rm sk} = 1$ cm$^{-2}$ s$^{-1}$, and
since $V_{\rm sk}$ is decreasing, the population densities decline with
time.  If we assume, however, that the emission volume is $\propto
R_{\rm sk}^3$, a remarkable property emerges.  In the bottom panel of
Figure~\ref{fig:GamEvolve} we show the total photon spectra with the
emission volume set to $V_{\hbox{\fiverm SNR}}=R_{\rm sk}^3$
(corresponding roughly to a shell between $0.9R_{\rm sk}$ and $R_{\rm
sk}$) and the total emission flux at earth is approximately constant
over the time span from 300 yr to $10^4$ yr.  This result is very
similar to the behavior reported by Drury, Aharonian, \& V\"olk (1994),
whose time-dependent two-fluid model generated a more-or-less constant
luminosity in the Sedov phase, and Berezhko \& V\"olk (1997) (where the
integrated $\gamma$-ray flux varies by less that a factor of 3 during
$1 < t_{\hbox{\fiverm SNR}}/t_{\rm trans} < 100$) and probably results
from the evolutionary properties of the Sedov solution.

\section{DISCUSSION}
 \label{sec:discussion}

\subsection{Previous Gamma-Ray Models}
 \label{sec:previous}

Reviews of previous models of gamma-ray emission from SNRs can be found
in Baring (1997) and de Jager and Baring (1997).  Briefly, Drury,
Aharonian, \& V\"olk (1994) calculated gamma-ray emission from protons
using the time-dependent, two-fluid analysis (thermal ions plus cosmic
rays) of Drury, Markiewicz, \& V\"olk (1989).  They assumed a power-law
proton spectrum with an arbitrary maximum energy cutoff; no
self-consistent determination of temporal or spatial limits to the
maximum energy of acceleration was made.  In some sense, our results
can be considered complementary to those of Drury et al., since their
model includes global spherical shock dynamics, but does not
self-consistently yield an energetic particle spectrum, while our model
provides a fairly self-consistent calculation of the total shock
acceleration spectrum, but does not treat time-dependent dynamics in
detail.  We find that during much of Sedov evolution, maximal diffusion
length scales are considerably less than a remnant's shock radius,
concurring with the findings of Drury, Aharonian, \& V\"olk (1994).

Gaisser, Protheroe, \& Stanev (1998) computed emission from
bremsstrahlung, inverse Compton scattering, and pion-decay from proton
interactions and only slight differences exist between our treatment
and theirs of the physics of bremsstrahlung and pion production
processes, other than that we include helium.  Gaisser et al. did not
consider non-linear shock dynamics or time-dependence and assumed
test-particle power-law distributions of protons and electrons with
arbitrary $e$/$p$ ratios.  In order to suppress the flat inverse
Compton component and thereby accommodate the EGRET observations of
$\gamma$ Cygni and IC443, Gaisser et al. assumed a high matter density
to enhance the ratio of bremsstrahlung and \teq{\pi^0}-decay flux to IC
flux.  We have shown (Figure~\ref{fig:GammasNine}) that the same effect
can be achieved without a high density if the primary $e/p$ ratio is
reduced.

A time-dependent model of gamma-ray emission from SNRs using the Sedov
solution for the expansion was presented by Sturner et al. (1997). They
numerically solved equations for electron and proton distributions
subject to cooling by inverse Compton scattering, bremsstrahlung,
\teq{\pi^0} decay, and synchrotron radiation and included all the
radiation processes of Gaisser, Protheroe, \& Stanev (1998) plus
synchrotron emission to supply a radio flux.  Expansion dynamics and
non-linear acceleration effects were not treated, and power law spectra
were assumed.  One feature of their model is the general dominance of
inverse Compton emission.  This arises because they often have the same
energy density in non-thermal electrons and protons, thereby assuming
high \teq{e/p} ratios; this appears hard to reconcile with galactic
cosmic ray observations.  Sturner et al.'s work marks a significant
advance over previous work by introducing cutoffs in the distributions
of the accelerated particles (actually first done by Reynolds 1995,
1996; Mastichiadis \& de Jager 1996; de Jager \& Mastichiadis 1997),
which are defined by the limits on the achievable energies in Fermi
acceleration discussed in Section~2.3.  Hence, given suitable model
parameters, Sturner et al.  can accommodate the constraints imposed by
Whipple's upper limits to $\gamma$ Cygni and IC 443.

To date, the most complete model coupling the time-dependent dynamics
of the SNR to cosmic ray acceleration comes from Berezhko \& V\"olk
(1997) (based on the model of Berezhko, Yelshin, \& Ksenofontov 1996).
They numerically solve the gas dynamic equations including the cosmic
ray pressure and Alfv\'en wave dissipation, following the evolution of
a spherical remnant in a homogeneous medium.  Only pion decay is
considered and the gamma ray spectra, spatially integrated over the
remnant, exhibit some curvature.  There are a number of similarities
between this model and ours; we both obtain maximum efficiencies near
and above 50\% and we both obtain overall compression ratios well above
standard Rankine-Hugoniot values.  However, Berezhko \& V\"olk argue
that systems will naturally be driven to the Bohm limit (i.e. $\eta
\sim 1$), giving them higher upper limits to the maximum cosmic ray
energy than we estimate.  Another important difference between our work
and Berezhko \& V\"olk's, comes from the treatment of particle
injection:  while this is automatic in our Monte Carlo technique,
affording an elegant connection between the thermal and non-thermal
populations, it is specified by a free parameter in Berezhko \& V\"olk
(1997).  Berezhko \& Ellison (1998, in preparation) demonstrate that,
for most parameter regimes of interest, the shock dynamics are
relatively insensitive to the efficiency of injection, and furthermore
that there is good agreement between the two approaches when the Monte
Carlo output specifies injection for the model of Berezhko et al.
Output particle spectra produced by the two models are then essentially
identical for a remnant's free-expansion phase, though minor
differences do arise during the Sedov phase because our Monte Carlo
model does not include the influence of particles accelerated prior to
the Sedov phase.  A significant advance in our work here is the
inclusion of electrons.

\subsection{The Observational Status Quo}
 \label{sec:observ}

Several questions of interest are raised by the current observational
situation, the first being how real are the proposed associations
between EGRET unidentified sources and young shell-type remnants like
IC 443, W44, $\gamma$ Cygni, W28, the Monoceros Loop, and CTA 1
(Sturner \& Dermer 1995; Esposito et al. 1996; Yadigaroglu \& Romani
1997)?  Second, if the associations are true, is the gamma-ray emission
connected with particle acceleration at the shell?  Furthermore, is the
signal above 100 MeV produced by cosmic ray ions or electrons?

The sources detected so far above a few hundred GeV are the nebulae
surrounding several pulsars, namely the Crab, Vela, PSR 1706-44 and PSR
1509-58, and the high latitude supernova remnant SN1006.  With the
exception of SN1006, these are likely to be associated with plerionic
emission.  The situation is, however, more confused in the GeV and
sub-GeV ranges.  The candidate associations identified by Esposito et
al. (1996) all suffer from large uncertainties in the EGRET source
positions, localizations that were derived using the point-source
assumption.  Statistically, the chance probability of spatial
coincidence with the candidate remnants is small:  0.1\% for IC 443,
1.4\% for $\gamma$ Cygni, 6\% for W28, and 7.4\% for W44 (Yadigaroglu
\& Romani 1997).  Yet, these remnants are often found in active
star-forming sites, amid numerous massive stars and HII regions, and
the chance probability of associating the EGRET source with a close-by
OB star or a radio pulsar is equally small or even smaller (Yadigaroglu
\& Romani 1997).  For the unidentified EGRET source 2EG J1801-2312, for
example, the chance probability of an association with the OB star Sgr
1c or the pulsar PSR B1758-23 is 1.2\% instead of 6\% for the remnant
W28.  The relative dimensions of the remnant and of the gamma-ray error
circle may also raise difficulties for the identification of EGRET
sources with known remnants.  The 95\% confidence positions of 2EG
J0008+7307 and 2EG J2020+4026, as measured at GeV and super-GeV
energies (Brazier et al. 1996, 1998), are constrained to within 11 and
8 arcmin, respectively, so that the sources appear unmistakeably {\it
inside} their respective remnants, namely CTA 1 and $\gamma$ Cygni,
which are much larger.  Furthermore, note that systematic errors in the
gamma-ray position due to the highly structured background along the
lines of sight to the candidate remnants are very likely.  These can be
reduced by taking advantage of the narrower angular resolution of EGRET
above 1 GeV, given the significant flux of the candidates sources above
this energy.  Reimer et al. (1997) recently adopted this approach to
determine more accurate positions that differ by 1 or \teq{2\sigma}
from the positions listed in the 2nd EGRET catalogue (Thompson et al.
1995).

Such improved locations led Brazier et al.~(1996) to conjecture that
2EG J2020+4026 is perhaps associated with a distinct ROSAT source with
no optical counterpart that lies within the EGRET error box: they
suggest that this source may be a radio-weak pulsar or a plerion.
There is also the recent proposal (Brazier et al. 1998) of a
pulsar/plerion counterpart to the CTA 1 remnant's EGRET source 2EG
J0008+7307.  In addition, de Jager \& Mastichiadis (1997) contend that
the EGRET source 2EG J1857+0118 associated with W44 may be of plerionic
nature due to the presence of a radio pulsar and its wind nebula within
the 95\% EGRET confidence circle.  A pattern seems to be emerging,
namely that pulsars or plerionic activity may generate the gamma rays
in half of the remnants tentatively associated with EGRET sources.  The
statistical conceivability that pulsars could account for {\it most} of
the unidentified EGRET sources near the Galactic plane (Kaaret \&
Cottam 1996; Yadigaroglu \& Romani 1997; Mukherjee, Grenier, \&
Thompson 1997) currently precludes any assertion stronger than just
weak suggestions of GeV emission from two shell remnants, namely IC 443
and W 28.  For this reason, we have used only data for 2EG J0618+2234
in association with IC 443 in some of our spectral plots, principally
as a general guide for the reader in considerations of the emission
mechanism.

The spectral properties of the EGRET detections cannot presently
determine whether the emission is of ionic or electronic origin.  All
of the candidate remnants except the source toward CTA 1 present
differential photon spectra in the EGRET band quite consistent with
\teq{E^{-2}} (Merck et al. 1996), the source toward CTA 1 being much
harder (\teq{E^{-1.58\pm 0.18}}, Brazier et al. 1998).  Based on the
spectra presented in this paper, it seems probable that these slopes
exclude inverse Compton emission as the dominant mechanism operating,
and furthermore that pion decay emission is not overwhelmingly
prevalent.  Since all of the candidates are known to have massive
clouds in their vicinity and, in some cases, to interact with molecular
clouds, perhaps the very proximity of such gaseous regions can cast
light upon the nature of the energetic particles.  Besides H$_2$ line
emission, OH maser emission, collisionally excited by H$_2$ molecules
heated by a non-dissipative shock, acts as a good tracer of shock/cloud
interactions.  In this respect, it is interesting to note that the
three closest remnants with OH masers, namely IC 443, W28, and W44,
belong to the candidate gamma-ray list (Claussen et al. 1997).  The
Monoceros loop and $\gamma$ Cygni are also classical examples of
remnants colliding with clouds (Pollock 1985; Huang \& Thaddeus 1986).
While Drury, Aharonian, \& V\"olk (1994) concluded that, in the EGRET
energy range, \teq{\pi^0} decay gamma rays in a 1 cm$^{-3}$ medium
would be drowned in the diffuse Galactic emission, the proximity of
large target masses may give support to a cosmic-ray origin of the
emission and its visibility.  Sturner \& Dermer (1995) showed that
\teq{\pi^0} decay emission meets the bulk energy requirements for IC
443.  But enhanced bremsstrahlung gamma-ray emission in the compressed
gas has also been advocated by Pollock (1985) to account for the COS-B
sources seen toward $\gamma$ Cygni and W28.  Hence, the mere presence
of clouds cannot resolve the ambiguity between electronic and hadronic
emission.

The complexity of the morphology and broad-band emission properties of
shell-type remnants, as exemplified by IC 443, currently precludes a
comprehensive understanding of the relationship of shell-associated
shock acceleration to the various emission components.  However there
is one distinctive property of radio emission that can be discussed
concisely, namely the observation that the radio spectral index
flattens from 0.7--0.6 to around 0.3 with increasing brightness across
IC 443 (Green 1986).  This correlation is consistent with our model:
with increasing density, there is less pressure in the highest-energy
particles and the subshock compression ratio increases toward the
canonical value of 4.  Therefore, the radio-emitting (GeV band)
electrons, which sample the length scales not dramatically larger than
those on which the subshock is established, present a flatter spectrum
concomitant with a rise in synchrotron luminosity.  However, it should
be noted that while modified shocks can in principle provide radio
synchrotron spectra flatter than 0.5, it is difficult (but not
impossible) for our model to explain such a large amplitude in spectral
index variations in one source.

Turning now to the TeV band, the possibility of either shell-related or
plerionic origin of gamma-rays from the EGRET sources quickly propelled
the TeV gamma-ray astronomy community into an observational program.
The absence of any positive detections from the ensuing monitoring of
sources in the Esposito et al. collection, or from other prominent
remnants such as Tycho, spawned a number of potentially constraining
upper limits first from the Whipple team (Lessard et al.~1995, updated
in Buckley et al.~1997), and then from HEGRA (Prosch et al.~1995; Hess
1997 for IC 443; and Prosch et al.~1996 for $\gamma$ Cygni).  As
discussed earlier in the results section, SNR shells can generate GeV
gamma-rays at flux levels that would be detectable by EGRET without
conflicting with TeV upper limits if the SNR resides in a region of the
ISM of moderate to high density (\teq{n_{\rm p,1}\gtrsim 1}cm$^{-3}$).
Hence, dense remnant environs, as might be expected for most of the low
galactic latitude sources in the Whipple and HEGRA surveys, might
result in luminous emission in the sub-GeV range coupled with a
simultaneous absence of TeV gamma-rays.

Not surprisingly, the first reported detection of probable
shell-related TeV emission from a remnant came from an entirely
different type of SNR, SN1006, a southern hemisphere source at high
galactic latitude that was accessible to the CANGAROO experiment
(Tanimori et al.~1997, described in detail in Tanimori et al.~1998).
This impressively symmetric barrel-shaped remnant (see Moffett, Goss,
\& Reynolds 1993, for example, for a radio image) is probably embedded
in a low density, unclumpy medium, presumptions based upon its
geometrical symmetry and its high latitude.  This source had recently
provided the first evidence of the presence of super-100 TeV electrons
in SNR shells via the observation (Koyama et al. 1995) of non-thermal
X-ray emission by ASCA.  Its low environmental density (possibly
\teq{\lesssim 0.1}cm$^{-3}$) clearly enhances the plausibility of Fermi
acceleration to such high energies (as described in Section~2.3 here).
This fact, combined with the ASCA discovery, prompted Mastichiadis \&
de Jager (1996) and Pohl (1996) to predict a strong inverse Compton
signal in the TeV band for SN1006, motivated by the comparative
efficiency of this process.  It seems probable that the TeV signal from
SN1006 is inverse Compton emission, and therefore confirms the
production of cosmic ray electrons by a SNR blast wave, but offers
little evidence for cosmic ray ions.  The picture for SN1006 is not
entirely simple:  Tanimori et al. (1998) note that their data is
strongly asymmetric, with a positive detection of a flux of $\sim
4.6\times 10^{-12}$ cm$^{-2}$ s$^{-1}$ at energies greater than around
1.7 TeV from the NE rim, and an upper limit of less than half this
value for emission from the SW rim of the shell.  CANGAROO is the first
atmospheric \v{C}erenkov telescope to possess such angular resolution
capabilities, a property that is crucial to the inference of the shell
association of the emission, and which will form a major goal for
future experimental design.  We note that the recent suggestions of
non-thermal super-10 TeV electrons from X-ray observations of Cas A
(Allen et al. 1997), IC 443 (Keohane et al. 1997) and W44 (Harrus et
al. 1997) may identify them as prime candidates for future searches
with atmospheric \v{C}erenkov telescopes.

\subsection{Super-TeV Cosmic-Ray Production in Gamma-Ray SNRs?}
 \label{sec:CRs}

Two questions are of paramount importance to cosmic ray physicists.
First, can individual supernova remnants simultaneously generate GeV
gamma rays detectable by EGRET {\it and} produce cosmic rays to
energies above $\sim 10^{14}$ eV?  Second, if not, could some remnants
be gamma-ray bright while others supply the cosmic ray population?  We
believe the answer to the first question may be no.  Generally, for
SNRs in a homogeneous environment, the circumstances that favor intense
gamma-ray production in the EGRET and sub-TeV bands (namely high ISM
density) limits acceleration of particles to energies well below the
knee.  Given the total lack of detections in the TeV band of remnants
with associated EGRET unidentified sources, combined with the only
positive TeV detection coming from a source (i.e. SN1006) with no
associated EGRET emission, there seems little doubt that there is an
anti-correlation between sub-10 GeV gamma-ray luminosity and super-10
TeV cosmic ray production in individual sources.

For the supernova remnant population as a whole, the situation is much
less clear, and the second question above remains open.  If those
remnants associated with EGRET sources are actually emitting at such
intensities, which may be unlikely given the discussion just above,
then they may well represent a bright minority of cosmic-ray producing
remnants.  Their peculiarity may be coupled to their unusually dense
environments (as deduced from optical and microwave band observations),
thereby enhancing sub-TeV gamma-ray emission.  Thus, since the
pion-decay bump at $\sim 70$ MeV cannot be determined with current
EGRET data, there is still no direct for the acceleration of cosmic ray
{\it ions} to energies of 100 TeV or above.  However, whether or not
any of the EGRET detections amount to observations of shell-related
emission, the contention that the vast majority of shell-type remnants
can produce cosmic rays out to the knee still remains tenable.  It is
also possible that the remnants with EGRET identifications may actually
be quite representative of the SNR population as a whole, producing
energetic cosmic rays, but emitting at flux levels below EGRET's
sensitivity; in such a case, alternative origins for the EGRET sources
must be sought.

\subsection{Issues and Prospects for TeV Gamma-Ray Astronomy
            of Shell Remnants}
 \label{sec:TeVastron}

Several issues surrounding the comparison of our models with data
require clarification.  First, we remind the reader that the curves
depicted in Figures~\ref{fig:GammasNine}--\ref{fig:GamEvolve} are
arbitrarily normalized to an emitting region of 1 pc$^3$.  It is
unlikely that the gamma-ray emission in any real source would be
confined to such a small volume.  The actual angular size of a source
affects the detection sensitivity:  the sensitivities of telescopes
like Whipple and HEGRA are typically 3-4 times better for sources
smaller than a few tenths of a degree, which they see as point-like,
than for extended ones ($\gtrsim 0.5^\circ$) due to better
hadron-rejection capabilities and associated improved sensitivities.
Therefore, strictly, the model results for the various components in
these figures should be compared with upper limits pertaining to
point-sources.  However, we note that such subtleties make little
impact on most comparisons, since small emission volumes underpredict
both point and extended source bounds in the TeV band.  Second,
observational upper limits are generally {\it integral limits} above a
threshold energy, obtained by assuming a power-law spectrum (typically
\teq{E^{-2}}) for the underlying source.  Hence they cannot strictly be
compared with flux spectra, but we chose to do so remembering that such
upper limits, and their associated flux bounds, may vary with spectral
assumptions by factors of at most 2--3.  It should be borne in mind
that if the source spectra are steeply declining with energy in the
sub-TeV and TeV band, as is suggested by a number of our models, then
the sensitivity of atmospheric \v{C}erenkov telescopes to such sources
drops and the upper limits rise accordingly.

The CANGAROO observations of SN1006 have dramatically bolstered the
prospects for future positive detections in the 0.3--1 TeV range, but
these will depend on telescope sensitivity, threshold energy, and
angular resolution.  Improved sensitivities and lower thresholds in
future experiments such as CAT (Rivoal 1997), HESS (Hofmann et al.
1997), VERITAS (Weekes et al. 1997), and MAGIC (Lorenz 1997) are
obviously desirable requirements.  For example, it is anticipated
(Aharonian et al. 1997) that the HESS experiment should achieve a
conservative \teq{5\sigma} upper limit on IC443 of \teq{3\times
10^{-12}}cm$^2$ s$^{-1}$ above 100 GeV, assuming an emitting region of
$0.4^\circ$ in radius, almost an order of magnitude below the
\teq{\gtrsim 250}GeV limits of Whipple (Buckley et al. 1997) depicted
in Figures~\ref{fig:GammasNine}--\ref{fig:GamEvolve}.  Given the
general property of rapidly declining spectra near the maximum energies
of emission in most of our models, lower thresholds are of paramount
importance.  At the same time, our models and those of Drury,
Aharonian, \& V\"olk (1994), yield luminosities that scale with the
source volume so that the {\it total flux} from remnants is virtually
constant in time during the Sedov phase (as in
Figure~\ref{fig:GamEvolve}).  Limb-brightening concentrates this flux
somewhat towards the extremities of the remnant, less for IC emission
than for the other three processes (due to their different
density-dependences: this could provide an observational discriminant
among the emission mechanisms as gamma-ray imaging capabilities
improve).  However, as long as the remnant evolution remains roughly
self-similar, the surface brightnesses of both rims and remnant
interiors will decline similarly with age (basically as \teq{R_{\rm
sk}^{-2}\sim t_{\hbox{\fiverm SNR}}^{-4/5}}), so that detectability
should be enhanced in younger remnants.

Of equal or greater importance to the resolving of some of the open
questions surrounding gamma-ray emission from supernova remnants will
be improved telescope angular resolution.  This was emphasized by the
ability of CANGAROO to unequivocally connect its reported TeV flux from
SN1006 to the NW rim of the remnant, a feat that was not generally
possible for the EGRET data on other remnants.  The most detailed study
of the angular resolution for future telescopes is found in Aharonian
et al. (1997), with estimates varying from a \teq{0.1^\circ} radius for
68\% acceptance for gamma rays at 100 GeV down to \teq{0.05^\circ} at 1
TeV. This implies that spatial features of the order of 1 to 2 pc that
are 1 kpc distant could, in principle, be resolved given reasonable
data accumulation times.  The CAT, HESS, and VERITAS telescopes will
both have angular resolutions close to this.  One big difference
between the existing single dish telescopes (e.g. CAT, WHIPPLE,
CANGAROO, etc.) and future stereo arrays is with the improved hadronic
background rejection (as demonstrated by HEGRA, the only existing
stereo experiment).  An improvement by a factor of 10 to 20 is afforded
by stereoscopic techniques, which directly translates into an increase
in sensitivity by a factor of $\sim 4$.  Given additional bonuses such
as greater mirror area and/or finer imaging resolution, sensitivity
increases of factors of 5 to 6 may be attainable.  The unquestionable
scientific impact of resolving gamma-ray SNR emission should amply
motivate developmental programs towards achieving this technical goal.

\section{CONCLUSIONS}
 \label{sec:conclusion}

In this paper we have applied a well-documented and tested steady-state
Monte Carlo simulation of non-linear Fermi acceleration at
plane-parallel shocks to the problem of gamma-ray and broad-band
emission from shell-type supernova remnants.  By coupling the
simulation to a standard Sedov model, which estimates the spherical
shock dynamics as a function of remnant age in a homogeneous
environment, we have obtained a reasonably self-consistent description
of a spherical SNR as it evolves and accelerates particles.  A simple
parametric model of electron injection and acceleration allows us to
determine, for the first time, complete shock accelerated ion (proton
and helium) and electron distributions, and to use them to predict
broad-band emission from synchrotron, pion decay, bremsstrahlung, and
inverse Compton processes from radio through the X-ray continuum into
the super-TeV gamma-ray band.  The uncertainty in ISM parameters
provides a broad range of model predictions for individual sources.

Our principal results are as follows.  (1) We comprehensively treat
non-linear feedback effects between accelerated particles and the shock
structure that result in particle spectra (both ion and electron) which
deviate significantly from power-laws.  Such deviations are not only
crucial to overall efficiency considerations, but also impact photon
intensities and spectral shapes at all energies, producing GeV/TeV
intensity ratios that are quite different from test particle
predictions.  (2) We address the electron injection problem in the
context of SNRs with a simple and coherent prescription for connecting
observational inferences of the \teq{e/p} ratio to quantities that
relate to plasma wave and dissipational properties of the shock.  The
treatment of electrons is particularly important in the light of the
recent observations of SN1006, since it now appears that emission
(X-ray and TeV gamma-ray in addition to radio) from cosmic ray
electrons may dominate those from ions in some (or perhaps most)
shell-type remnants.  (3) We connect, in a reasonably self-consistent
way, the spectral shape and intensity of gamma-ray emission to the ISM
density without arbitrary coefficients of infinite or semi-infinite
power-laws, and for ions, without ad-hoc injection parameters.  (4) We
determine the relative acceleration efficiencies of different ion
species and find that, generally, the pion-decay contribution from
helium (at cosmic abundances) is {\it comparable to that of protons}, a
byproduct of the enhancement of heavy ions in non-linear Fermi
acceleration.  (5) Our results exhibit a general {\it anti-correlation}
between the maximum energy of gamma-ray emission, and the source
luminosity in the super-MeV band; this should be a property of any
self-consistent model of particle acceleration and associated radiation
at SNR shocks.  (6) Finally, we also describe the parameters required
to give maximum particle energies of a few TeV for ISM densities near
$\sim 1$ cm$^{-3}$, so as to provide spectral consistency with current
upper limits from the Whipple and HEGRA atmospheric \v{C}erenkov
telescopes for remnants with unidentified EGRET source associations.
These parameters are fully consistent with Fermi shock acceleration and
do not produce unacceptably steep particle spectra.  At the same time,
we can generate models with low densities (i.e. $n_1 \lesssim 0.01$
\hbox{cm$^{-3}$}) that accelerate cosmic rays to above 100 TeV, but
with fluxes below the EGRET and Whipple sensitivities in the 100
MeV--10 TeV band.

While the results we present here are still preliminary in many ways,
we believe they form a basis for future studies of non-linear shock
acceleration and gamma-ray emission from shells of supernova remnants.
Unresolved issues include the radial extent of emission, the modeling
of flat spectrum radio sources, spatial variations in radio, X-ray and
gamma-ray spectral indices, the physical processes responsible for the
non-thermal X-ray and gamma-ray flux, the role of field obliquity
around the shell, the \teq{e/p} ratio and a more complete description
of electron injection, and cosmic ray abundances and production up to
the knee.  Such studies would anticipate the development of the next
generation of space- and ground-based gamma-ray telescopes with greater
flux sensitivity and spatial resolution.

\acknowledgments
We thank Frank Jones for many insightful discussions on shock
acceleration and its application to SNRs, and Joe Esposito, Jonathon
Keohane, and Glen Allen for informative discussions on the EGRET
unidentified sources, and on X-ray and radio observations of remnants.
We also thank the anonymous referee for comments helpful to the
improvement of the manuscript.  DCE and MGB thank the Service
d'Astrophysique, CE-Saclay, for hospitality during part of the period
when this work was performed.  This work was supported by the NASA
Space Physics Theory Program.

\clearpage

\appendix
\section{APPENDIX}
 \label{sec:appendix}

\centerline{\eightrm APPROXIMATIONS TO ELECTRON-ELECTRON BREMSSTRAHLUNG\rm}
\vskip 5pt

In this Appendix, we present approximations to the differential
cross-section for electron-electron bremsstrahlung that expedite our
computations, circumventing use of the unwieldy and lengthy exact
results for this process given in Haug (1975).  Consider first the
relativistic case.  We adapt the approximation derived by Baier, Fadin,
\& Khoze (1967) which has the form
\begin{equation}
   \sigma_{\rm e-e} (E_{\rm e} ,\,\erg_{\gamma} )\; =\;
   (\sigma_1 + \sigma_2) \ A(\displaystyle{\erg_{\gamma}}, \gamma_{\rm e}) \ ,
 \label{eq:bfkbrems}
\end{equation}
where $\gamma_{\rm e} =(E_{\rm e} + m_{\rm e} c^2)/m_{\rm e} c^2$ is
the electron Lorentz factor,
\begin{equation}
   \sigma_1\; =\;\dover{4r_0^2\alpha}{\displaystyle{\erg_{\gamma}}}
   \left[ 1+ \Bigl( \dover{1}{3} -
      \dover{\displaystyle{\erg_{\gamma}}}{\gamma_{\rm e}} \Bigr)\,
   \Bigl(1 -\dover{\displaystyle{\erg_{\gamma}}}{\gamma_{\rm e}}\Bigr) \right]
   \left\{ \log_e \left[ 2\gamma_{\rm e} \dover{(\gamma_{\rm e} - 
      \displaystyle{\erg_{\gamma}} )}{\displaystyle{\erg_{\gamma}} } \right]
      -\dover{1}{2} \right\}\ ,
 \label{eq:bfkterms}
\end{equation}
and
\begin{equation}
   \sigma_2\; =\; \dover{r_0^2\alpha}{3\displaystyle{\erg_{\gamma}}}\;
   \cases{
\Biggl[ 16 (1 - \erg_{\gamma} + \erg_{\gamma}^2)
\log_e{ \left ( \dover{\gamma_{\rm e}}{\displaystyle{\erg_{\gamma}}} \right ) }
    - \dover{1}{\displaystyle\erg_{\gamma}^2} +     
    \dover{3}{\displaystyle{\erg_{\gamma}}} - 4 + 4 \erg_{\gamma}
    - 8 \erg_{\gamma}^2 & \cr
\quad - 2 (1 - 2 \displaystyle{\erg_{\gamma}} ) \log_e{(1 - 2 \displaystyle{\erg_{\gamma}})}\;
\biggl( \dover{1}{4 \displaystyle\erg_{\gamma}^3} - \dover{1}{2 \displaystyle\erg_{\gamma}^2} +
\dover{3}{\displaystyle{\erg_{\gamma}}} - 2 + 4 \displaystyle{\erg_{\gamma}} \biggr) \Biggr ] \ , &
$ \quad \erg_{\gamma} \leq \dover{1}{2}\ ,$\cr
\dover{2}{\displaystyle{\erg_{\gamma}}}
\left [
\left ( 4 - \dover{1}{\displaystyle{\erg_{\gamma}}} + \dover{1}{4\displaystyle\erg_{\gamma}^2} \right )
\,\log_e{(2 \gamma_{\rm e})}
 - 2 + \dover{2}{\displaystyle{\erg_{\gamma}}} - \dover{5}{8\displaystyle\erg_{\gamma}^2}\;\right ]\ , &
$ \quad \erg_{\gamma} > \dover{1}{2}\ .$\cr }
 \label{eq:bfktermstwo}
\end{equation}
Here $r_0 = e^2/(m_{\rm e} c^2)$ is the classical electron radius,
$\alpha\simeq 1/137$ is the fine structure constant, and the
ultrarelativistic result of Baier, Fadin, \& Khoze (1967) sets
\teq{A(\erg_{\gamma}, \gamma_{\rm e}) =1}.

As it stands, this expression is accurate only for ultrarelativistic
energies: this can be deduced from Figure~10 of Haug (1975), though it
appears that Haug's numerical computations of Baier, Fadin, \& Khoze's
formula are slightly in error.  Therefore, we add a mildly relativistic
correction factor to equation~(\ref{eq:bfkbrems}):
\begin{equation}
   A(\erg_{\gamma},\,\gamma_{\rm e}) \; =\;
   1 - \dover{8}{3}\, \dover{(\gamma_{\rm e} - 1)^{1/5}}{\gamma_{\rm e} + 1}
   \left( \dover{\erg_{\gamma}}{\gamma_{\rm e}} \right)^{1/3}.
 \label{eq:Afunc}
\end{equation}
With this factor included, our expression for $\sigma_{\rm e-e}$ is
well within 10\% of the exact result of Haug (1975) for electron
energies above 5 MeV.

Equation~(\ref{eq:bfkbrems}) is suitable for the consideration of
bremsstrahlung contributions to gamma-rays from SNRs.  However, it
becomes inappropriate for X-ray and lower energies as the electrons
become non-relativistic.  In such regimes, we adopt a modification of
the standard non-relativistic asymptotic forms obtained by Fedyushin
(1952) and Garibyan (1953).  In the rest frame of the ISM electrons,
their expression for the angle-integrated differential (in photon
energy) cross-section is
\begin{equation}
   \sigma_{\hbox{\fiverm NR}}\; =\;
   \dover{4r_0^2\,\alpha}{15\displaystyle{\erg_{\gamma}}}\;
   F\Bigl(\dover{4\erg_{\gamma}}{\gamma_{\rm e}^2-1}\Bigr)\ ,
   \qquad 0\, <\,\erg_{\gamma}\, <\,\dover{1}{4}\,
   (\gamma_{\rm e}^2-1)\ ,
 \label{eq:nrbrems}
\end{equation}
where, for \teq{0<x<1},
\begin{eqnarray}
   F(x)\; &=&\; B(\gamma_{\rm e} )\, \biggl[ 17 -\dover{3x^2}{(2-x)^2}
     -C(\gamma_{\rm e},\, x)\biggr]\,\sqrt{1-x} \nonumber\\[-5.5pt]
 && \label{eq:fgfunc}\\[-5.5pt]
   && \quad +\biggl[ 12(2-x)-\dover{7x^2}{2-x}-\dover{3x^4}{(2-x)^3} \biggr]
     \log_e\dover{1+\sqrt{1-x}}{\sqrt{x}}\ ,\nonumber
\end{eqnarray}
specifically with \teq{B(\gamma_{\rm e} )\equiv 1} and
\teq{C(\gamma_{\rm e},\, x)\equiv 1} for extreme non-relativistic
energies.  Haug (1975) notes that such a form is accurate to a few
percent for cosmic ray electron energies below around 10 keV.
Accordingly, we add the mildly-relativistic correction factors
\begin{equation}
   B(\gamma_{\rm e} )\; =\; 1+\dover{1}{2}(\gamma_{\rm e}^2-1) \ \ ;\quad
   C(\gamma_{\rm e},\, x)\; =\; \dover{10x\,\gamma_{\rm e}\beta_{\rm e}    
      (2+\gamma_{\rm e}\beta_{\rm e} )}{1+x^2(\gamma_{\rm e}^2-1)} \ ,
 \label{eq:BCfunc}
\end{equation}
which render the cross-section in equation~(\ref{eq:nrbrems}) accurate
(compared with Haug's numerical evaluations of the full cross-section)
to within 10\% for \teq{E_{\rm e} < 500}keV, and is also of comparable
accuracy for all but the highest photon energies (which yield
insignificant contributions to the total bremsstrahlung spectrum) for
\teq{E_{\rm e}} up to a few MeV.  Hence,
equations~(\ref{eq:nrbrems})--(\ref{eq:BCfunc}), together with
equations~(\ref{eq:bfkbrems})--(\ref{eq:Afunc}), provide a description
of \teq{\sigma_{\rm e-e}} that is suitable for the purposes of this
paper, for all \teq{E_{\rm e}}; we switch between the two asymptotic
regimes at \teq{E_{\rm e} =2}MeV.

%\vskip24pt

\newpage

\centerline{}
   \centerline{\psfig{figure=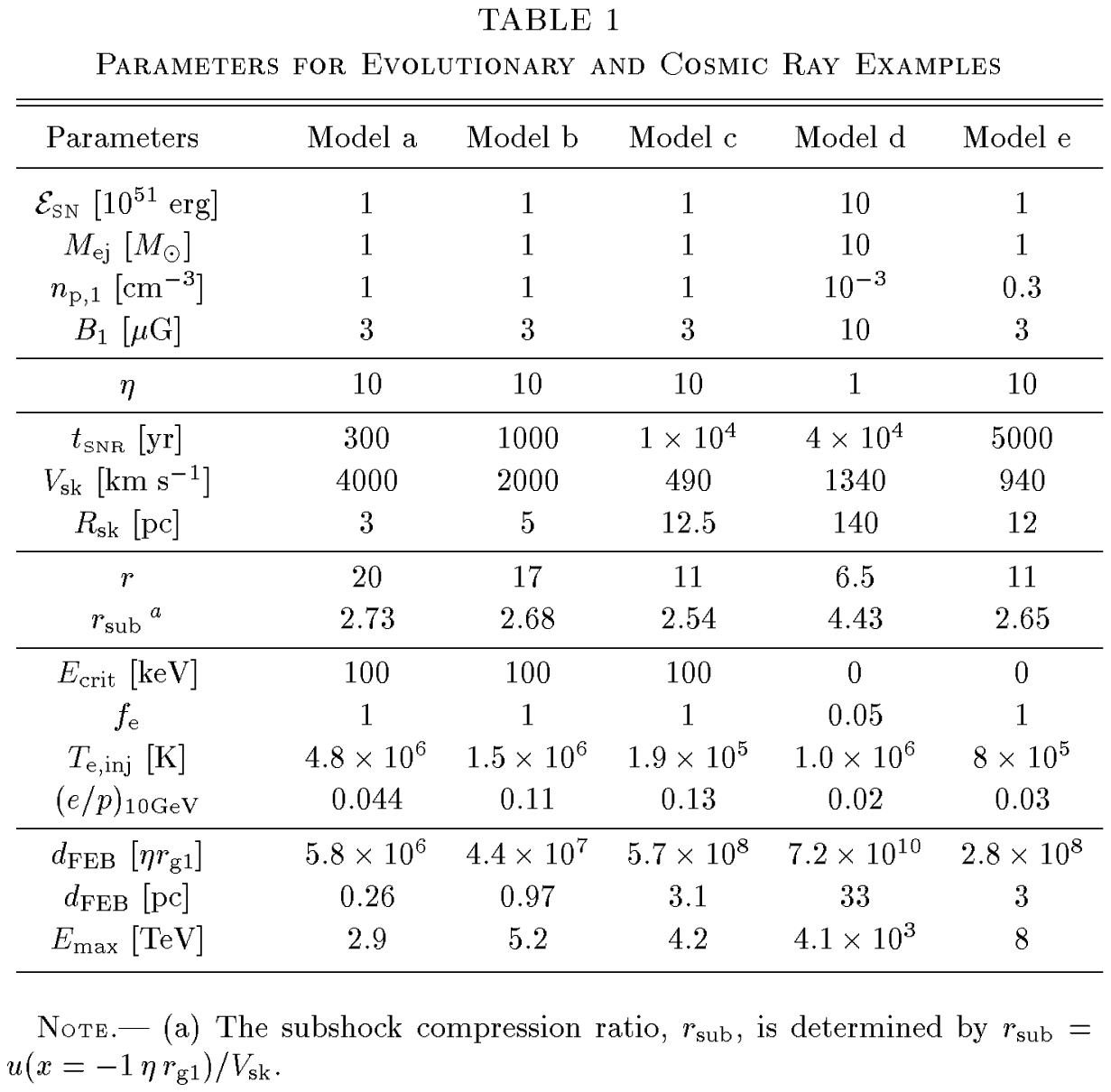,width=5.5in}}
\clearpage

% Figure Captions
% 
%\begin{center}
%   \bf FIGURE CAPTIONS\rm  \bigskip
%\end{center}
%
%\vphantom{p}
%\vskip 5.6truein
%
\figureout{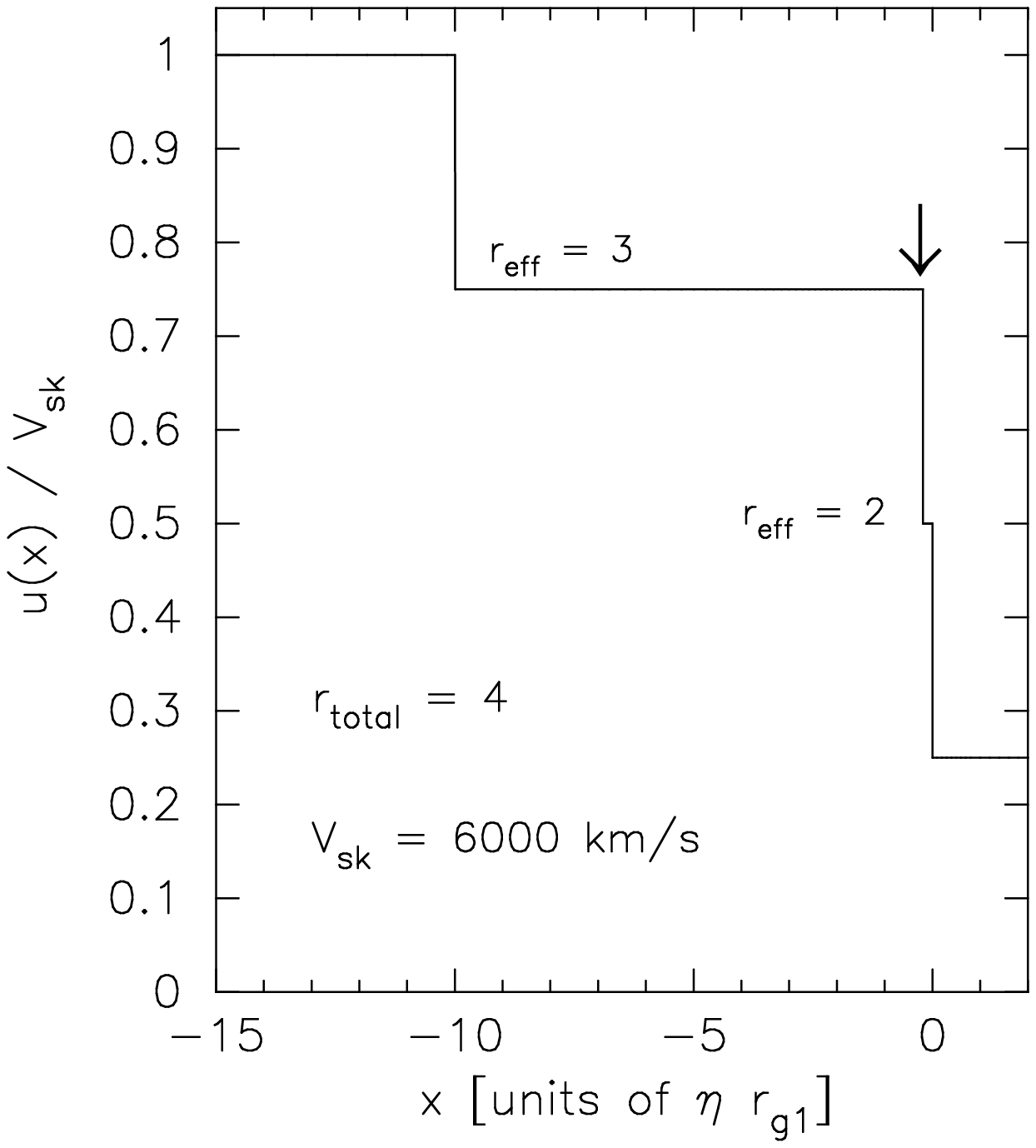}{
A schematic depiction of a non-linear shock profile, which shows the
flow speed versus distance normal to the shock in its rest frame. This
artificial shock profile has an  overall compression ratio of $r_{\rm
total}=4$ and two subshocks with $r_{\rm eff} = 3$ at $x= -10 \, \eta
r_{\rm g1}$ and $r_{\rm eff} = 2$ at $x= -0.2 \, \eta r_{\rm g1}$.  The
shock speed is  $V_{\rm sk} = 6000$  \hbox{km~s$^{-1}$}, which just
acts as a scale to the system.   The arrow indicates the minimum
upstream diffusion length for the $p_{\rm crit}=1.5\times 10^{-3} \,
m_{\rm p} c$ electron example shown with the dotted line in
Figure~\ref{fig:TestSpectra}. All electrons from injection energies
upward diffuse farther upstream that $-0.2 \, \eta r_{\rm g1}$ for this
example.
 \label{fig:TestProfile} }       % fig. 1

\figureout{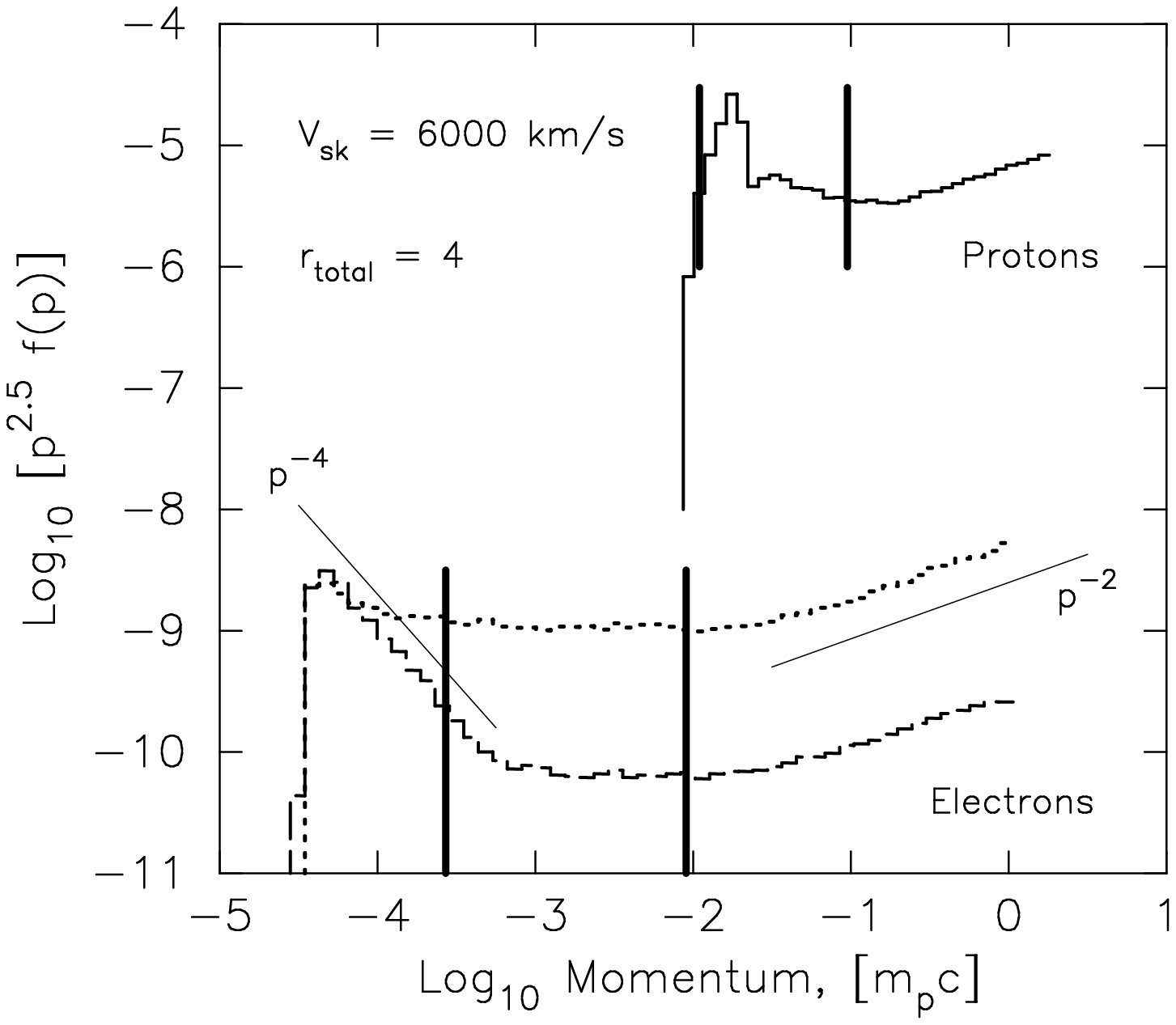}{
The number density in scalar momentum space, $f(|{\bf p}|)$, versus
$|{\bf p}|$.  We have plotted $|{\bf p}|^{2.5} \, f(|{\bf p}|)$ to
flatten the spectra.  The momentum is in units of $m_{\rm p} c$. The
upper solid curve is the proton spectrum, while the two lower curves
are electron spectra.  The dashed electron curve results from $p_{\rm
crit} =0$, while the dotted curve results from $p_{\rm crit}=1.5\times
10^{-3} \, m_{\rm p} c$, for which \teq{e^-} injection is more
efficient.  In all cases, particles are injected at the shock with a
$\delta$-function distribution at 1 keV.  The heavy vertical lines
indicate the momenta corresponding to upstream diffusion lengths,
$-L_{\rm D} = -0.2 \, \eta r_{\rm g1}$ and $-10 \, \eta r_{\rm g1}$.
The slopes of the power-law portions reflect those obtained from
equation~(\ref{eq:FofP}) with the values of $r_{\rm eff}$ in
Figure~\ref{fig:TestProfile}.
 \label{fig:TestSpectra} }       % fig. 2

\figureout{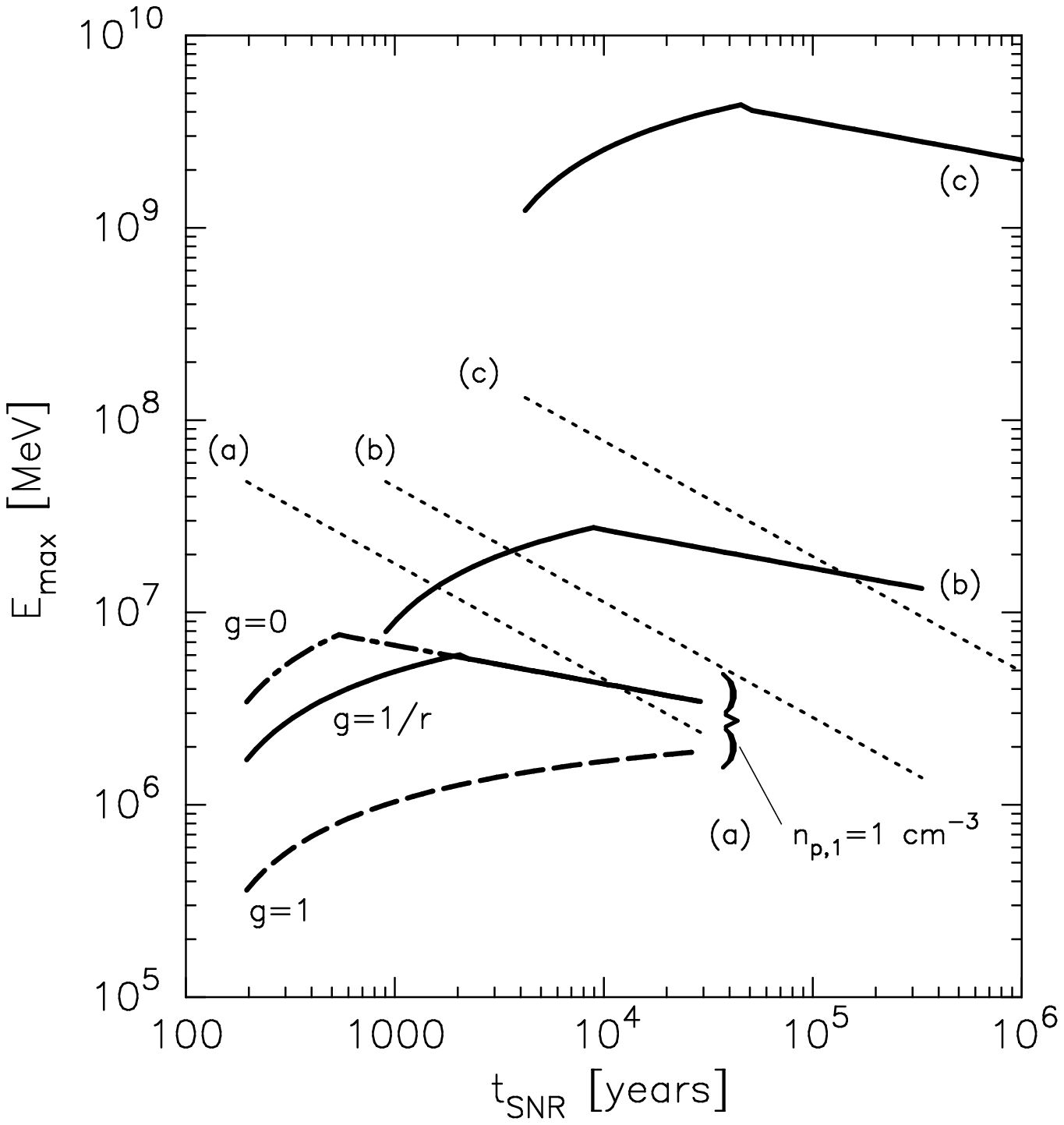}{
Estimates of maximum particle energy during the Sedov phase versus SNR
age, $t_{\hbox{\fiverm SNR}}$.  The lower three heavy curves (a) are
calculated with $n_{\rm p,1}=1$ \hbox{cm$^{-3}$} using $g=0$
(dot-dashed curve, corresponding to no time spent downstream),
$g=1/r\simeq 0.12$ (solid curve), and $g=1$ (dashed curve).  The heavy
curve labeled (b) uses $n_{\rm p,1} = 0.01$ \hbox{cm$^{-3}$} with
$g=1/r$. The upper most heavy curve has parameters  chosen to obtain a
high maximum energy.  The light dotted lines show the maximum energy
versus SNR age electrons will obtain under the influence of synchrotron
and inverse Compton losses; see equation~(\ref{eq:Cutoff}).  The light
dotted line at the lower left applies to the lower three heavy curves
(a), the middle light dotted line applies to the middle heavy solid
line (b), and the rightmost light dotted line applies to the top solid
line (c).
 \label{fig:figEmax} }       % fig. 3

\figureoutsmall{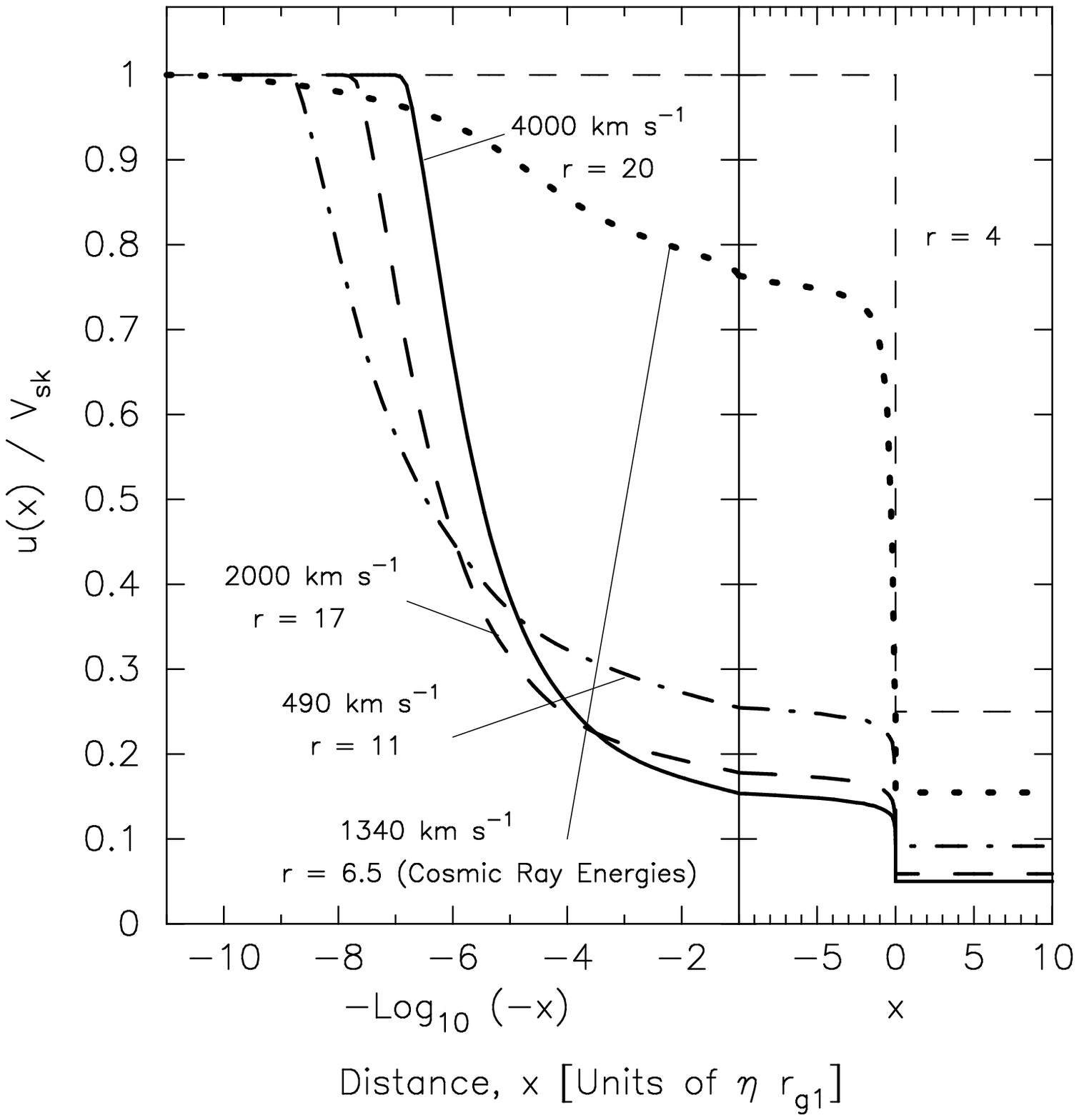}{
The bulk flow speed versus distance (i.e. shock velocity profile),
obtained in the Monte Carlo simulation.  Three of the four profiles
correspond to Sedov evolution of a remnant's shock with given ISM
parameters, namely $n_{\rm p,1}=1$ \hbox{cm$^{-3}$},
$B_1=3\mu\hbox{G}$, and with $\eta=10$ and $g=1/r$.  These comprise the
heavy solid curve (Model {\it a} in Table~1, with $r=20$), the heavy
dashed curve (Model {\it b} in Table~1; $r=17)$, and the heavy
dot-dashed curve (Model {\it c} in Table~1; $r=11$).  The shock
weakens slightly with time, and for comparison, we depict a standard
linear (test-particle) strong shock profile with a compression ratio
$r=4$ as the light dashed step-function.  As a separate example, the
heavy dotted curve shows the structure for a shock capable of
accelerating particles to the cosmic ray ``knee'' at $\sim 10^{15}$ eV
(Model {\it d} in Table~1), with different ISM parameters.  In first
three examples, a distinct subshock exists with a compression ratio
$r_{\rm sub} \sim 2.5$.  The cosmic ray energy shock (dotted line),
however, has a much stronger subshock ($r_{\rm sub} \sim 4.4$) due to
the strong Alfv\'en wave heating in the precursor.  Notice that the
distance is plotted with a logarithmic scale for $x<-10 \, \eta r_{\rm
g1}$ and a linear scale for $x> -10 \, \eta r_{\rm g1}$.
 \label{fig:fourprofiles} }       % fig. 4

%\figureout{apj99bergg_snr_f5.ps}{
\vphantom{p}
\vskip 0.5truein
\centerline{\psfig{figure=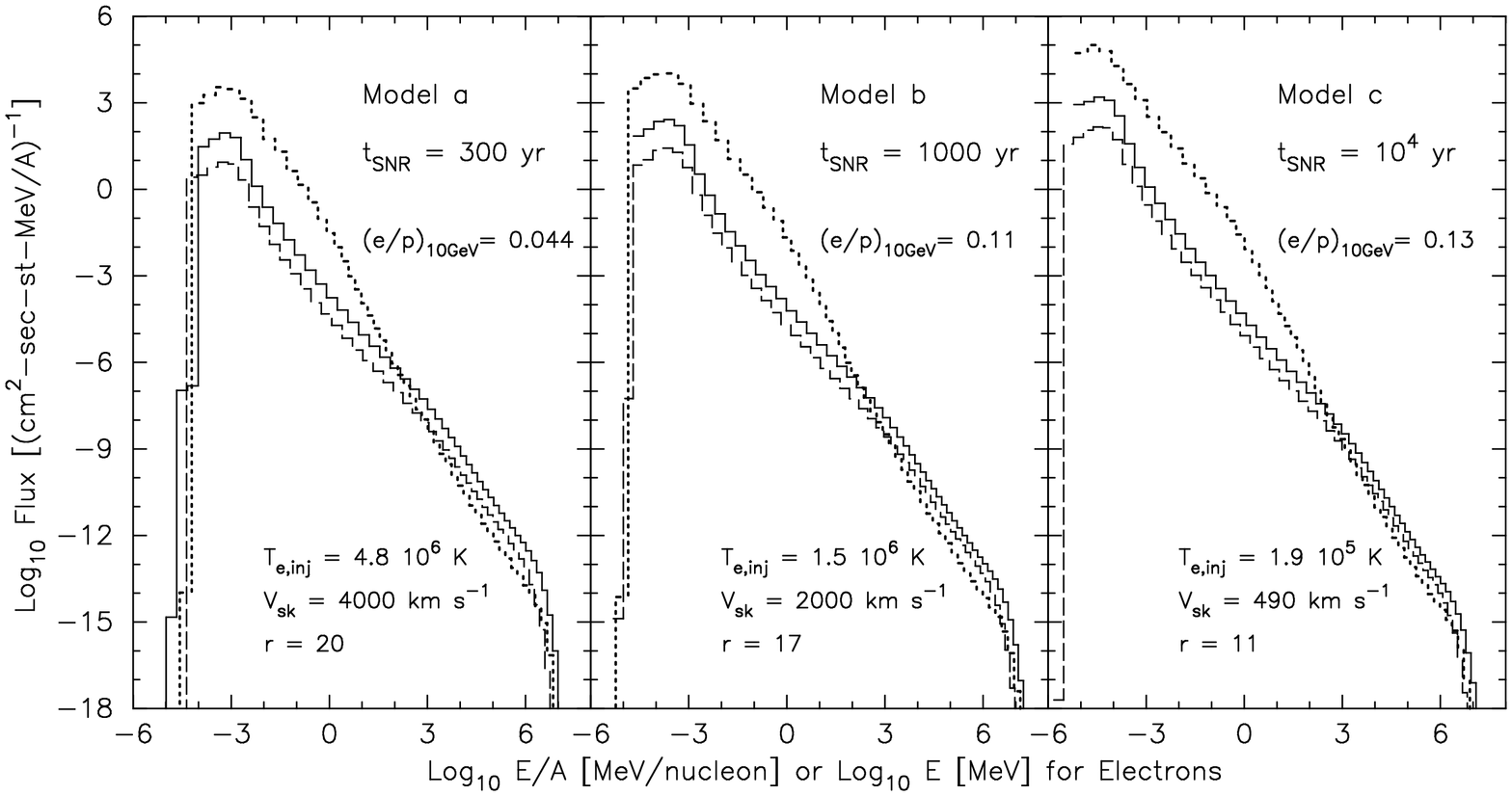,width=7.5in,angle=270}}
\figcaption{
Particle omni-directional fluxes, $dJ/dE$
[particles/(cm$^2$-s-ster-MeV/A)], versus energy per nucleon for ions
and versus energy for electrons ($A \equiv 1$ for electrons), obtained
from our example of an expanding remnant in the Sedov phase (see
Table~1 for model parameters).  All spectra are calculated downstream
from the shock in the shock rest frame and are obtained as explained in
the text with a steady-state approximation.  In each panel, the solid
and dashed lines show the hydrogen and He$^{+2}$ spectra, respectively,
and the dotted line shows the electron spectrum.  Both ionic species
contribute to the shock smoothing and the far upstream number density
of helium is 1/10 that of hydrogen.  The curves are normalized such
that $V_{\rm sk} n_{\rm p,1} = 1$ cm$^{-2}$ s$^{-1}$.  The electron
spectra are obtained with $E_{\rm crit}=100$ keV and $f_{\rm e} = 1$.
As the remnant evolves, the shock slows and weakens, and the injected
electron temperature \teq{T_{\rm e,inj}} diminishes in accordance with
the decline in the dissipative heating of ions (for fixed $f_{\rm e}$)
in the shock layer.
 \label{fig:SpectraThreePanels} }       % fig. 5
\clearpage

\figureout{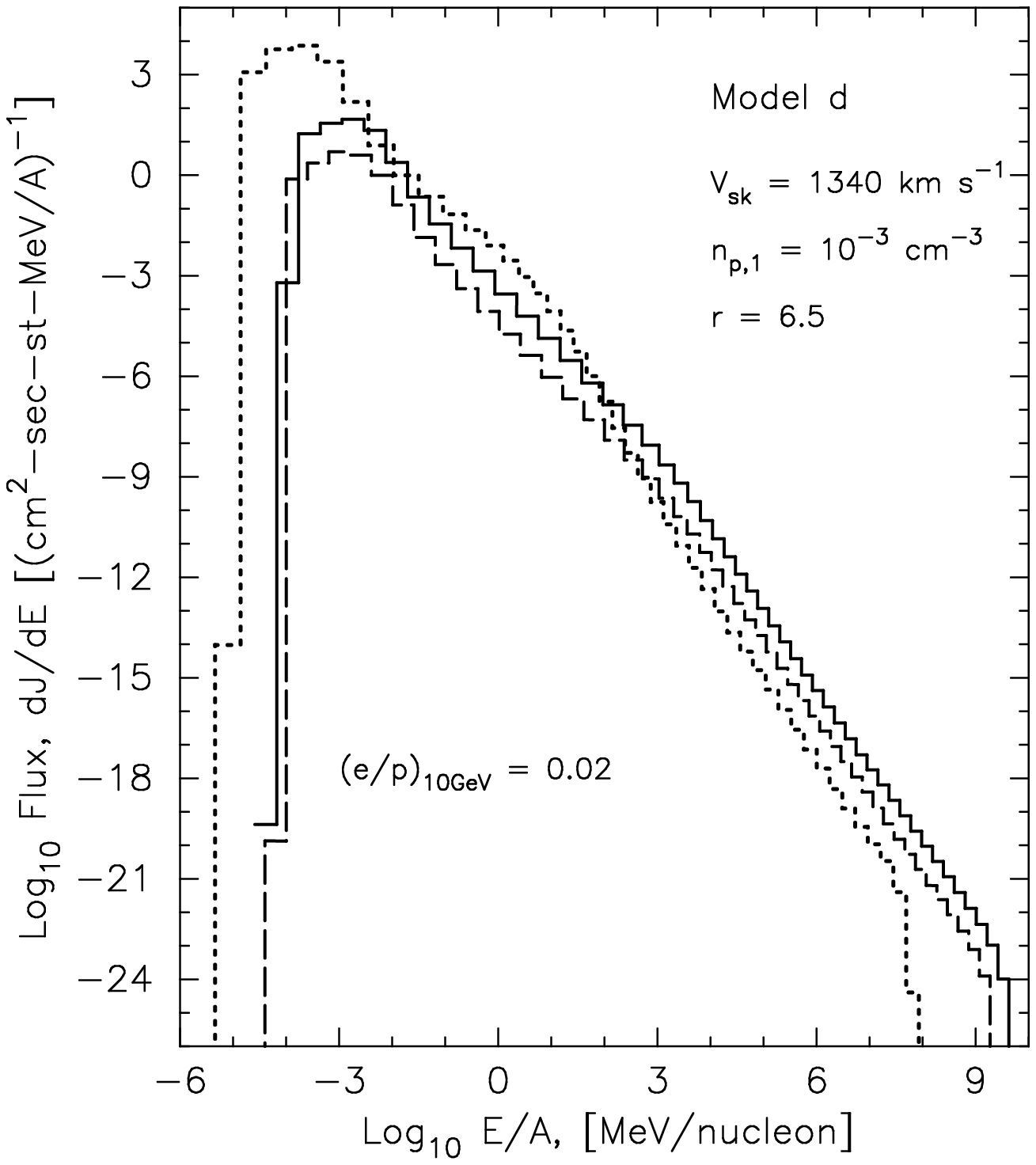}{
Particle spectra, $dJ/dE$ [particles/(cm$^2$-s-ster-MeV/A)], versus
energy per nucleon for ions (or energy for electrons with $A \equiv
1$).  The solid line is the proton spectrum, the dashed line is the
He$^{2+}$ spectrum, the dotted line is the electron spectrum, and all
spectra are calculated downstream from the shock in the shock rest
frame.  The parameters (i.e. Model {\it d}) have been chosen to produce
particles with energies above  $10^{15}$ eV  to account for cosmic rays
up to the knee.  The electron spectrum cuts off at lower energies than
the proton or helium because of significant synchrotron and inverse
Compton losses (see Figure~\ref{fig:figEmax}).
 \label{fig:SpectraEmax} }       % fig. 6

\figureoutvsmall{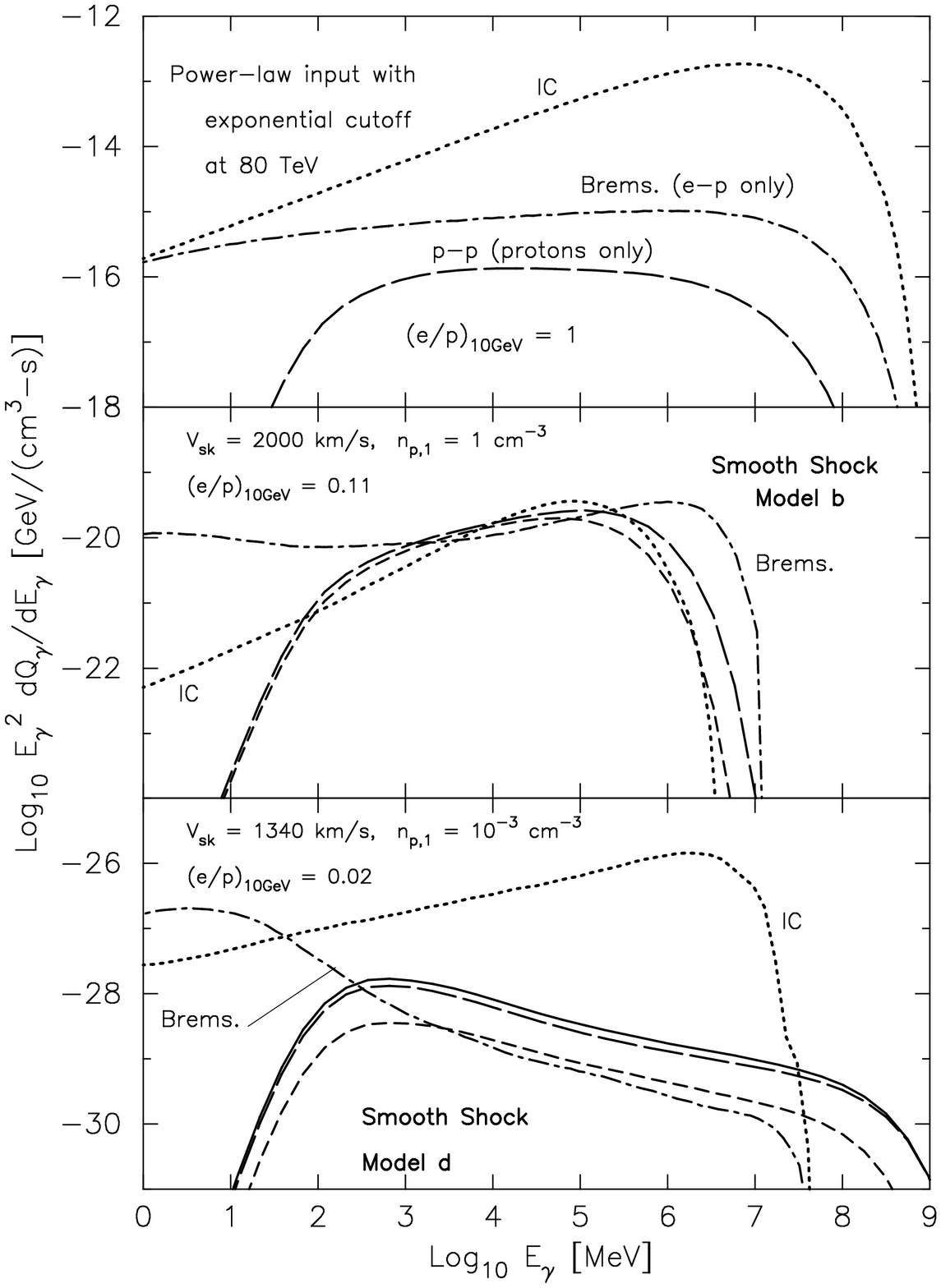}{
Examples of photon emission spectra, plotted as \teq{E_{\gamma}^2\,
dn_{\gamma}(E_{\gamma}) /dt} to emphasize the peak power of emission.
The top panel is produced using power-law electron and ion spectra with
identical normalization to that used in Figure~3 of Gaisser, Protheroe,
\& Stanev (1998).  These ``template'' curves reproduce their results
quite well, with small differences in the pion decay emission (p-p) and
e-p bremsstrahlung due to different assumptions in modeling these
components.  The bottom two panels depict sample photon spectra
produced by our self-consistent shock-accelerated electron and ion
(proton and He$^{2+}$) distributions.  The middle panel is Model {\it
b} of our evolving remnant trio, and the bottom panel is Model {\it d},
which produces cosmic rays up to the knee.  Comparison of these two
models indicates that density and other changes can strongly influence
the relative importance of inverse Compton scattering versus
bremsstrahlung and pion decay radiation.  In all panels, dotted lines
are inverse Compton (IC), dot-dashed lines are bremsstrahlung,
long-dashed lines are pion-decay from protons, short-dashed lines are
pion-decay from helium, and the solid line in the bottom panel is the
total pion-decay emission.
 \label{fig:PhotonsThreePanels} }       % fig. 7

\figureoutsmall{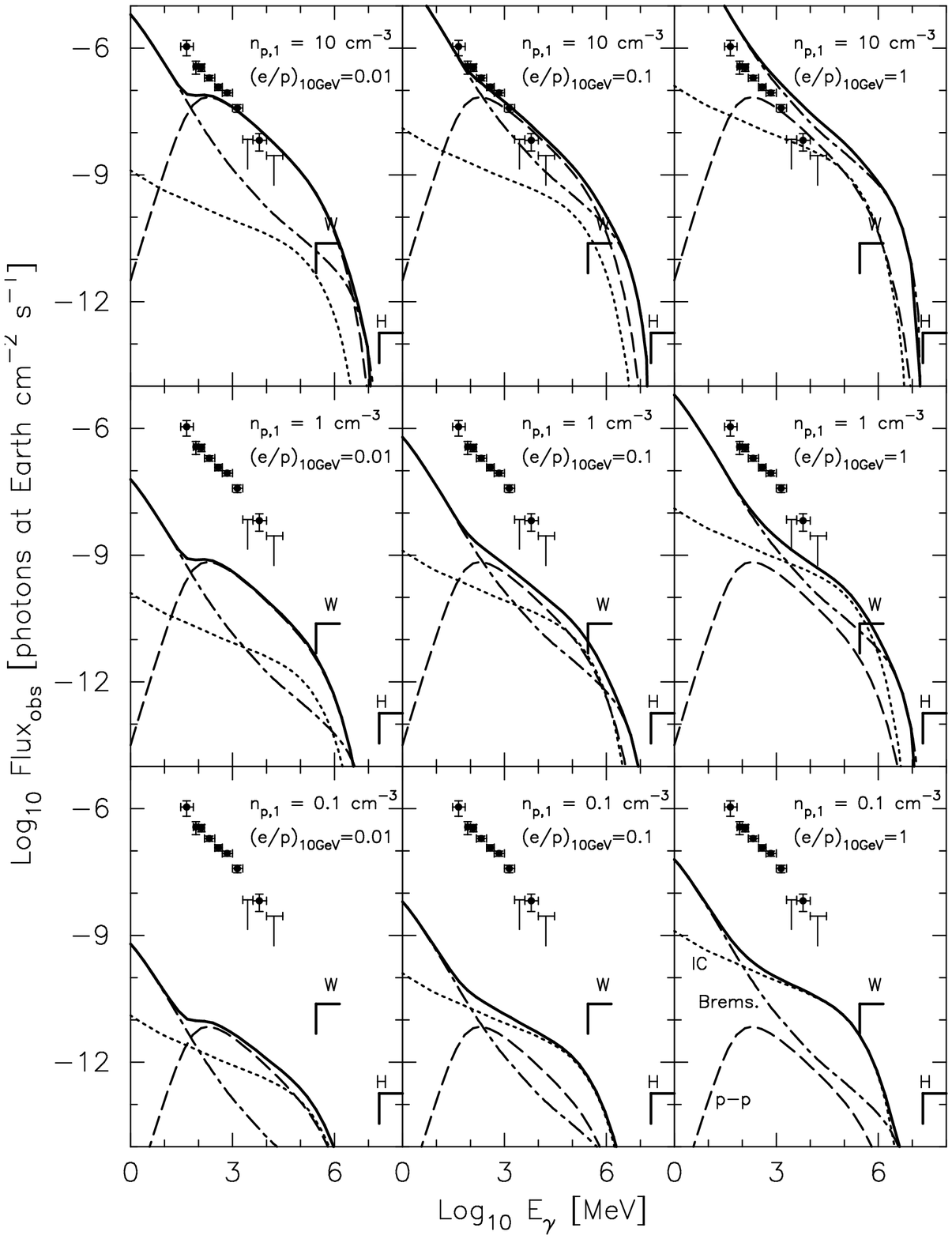}{
An array of emission spectra generated from Model {\it e} (Table~1),
illustrating trends in the parameters \teq{n_{\rm p,1}} and
\teq{(e/p)_{\rm 10GeV}}; these are compared with observations of the
shell remnant IC 443.  The data points are from EGRET observations of
2EG J0618+2234 (Esposito et al. 1996), and the upper limits are from
the Whipple imaging telescope (Buckley et al. 1997) and the HEGRA {\it
array} (Prosch et al.  1995) as marked.  In all panels, dotted lines
are inverse Compton, dot-dashed lines are bremsstrahlung, dashed lines
are the total pion-decay emission from protons and helium (denoted
hereafter by p-p), and solid lines are the sums of the three
components.  The model spectra are normalized to a source at 1 kpc with
emission volume = 1 pc$^3$.
 \label{fig:GammasNine} }       % fig. 8

\figureout{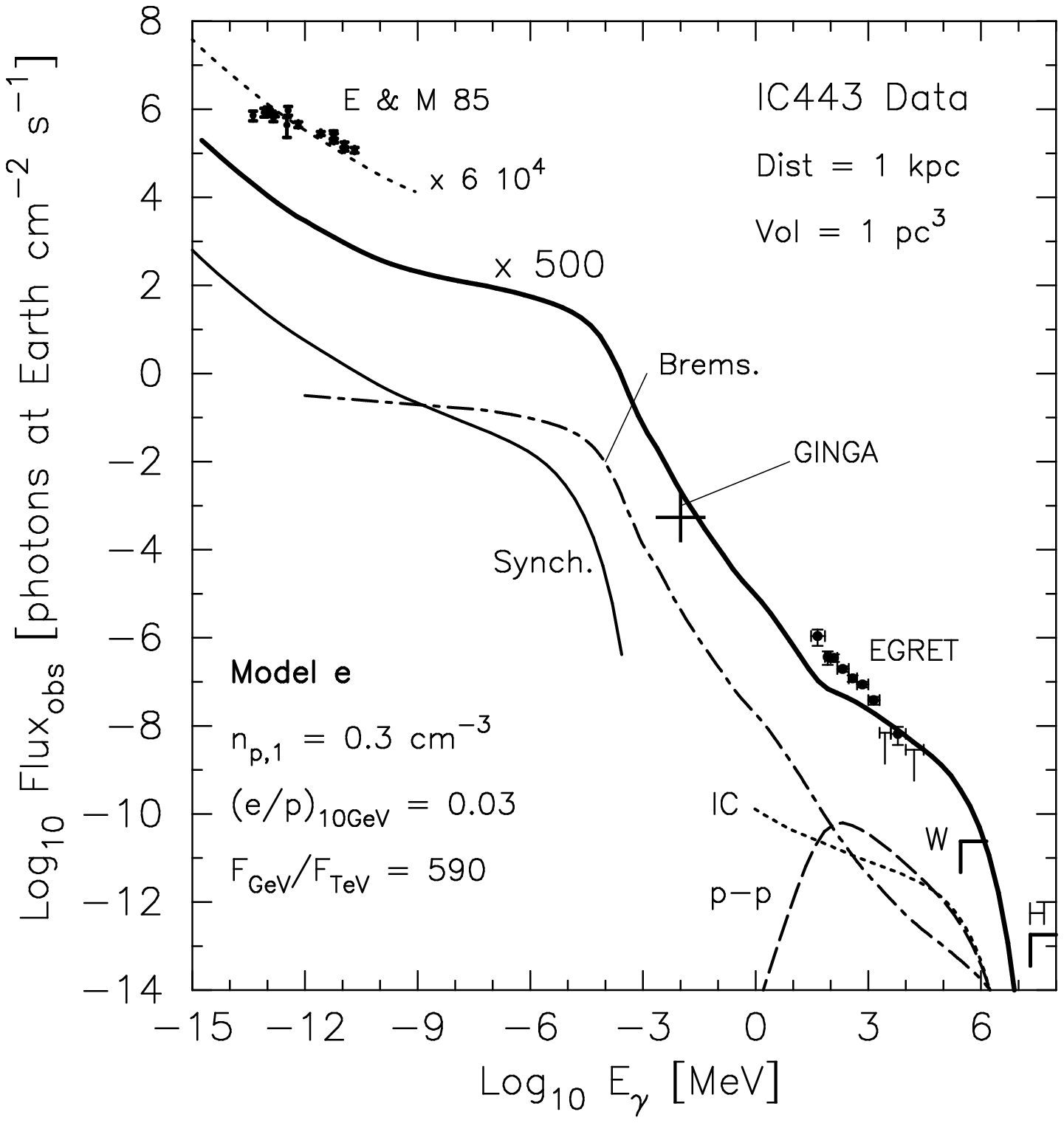}{
Photon spectra for the various emission processes (pion decay from p-p,
p-He, {\it and} He-He collisions [denoted by p-p], bremsstrahlung,
inverse Compton, and synchrotron radiation, as labelled, with the same
line styles as in Figure~\ref{fig:GammasNine}) for our $n_{\rm p,1} =
0.3$ \hbox{cm$^{-3}$} Model {\it e} used to generate the examples in
Figure~\ref{fig:GammasNine}.  The component spectra are all normalized
to a source at 1 kpc with emission volume = 1 pc$^3$, but the total
spectrum (heavy solid line) is multiplied by 500 to roughly match the
EGRET flux.  Whipple (W) and HEGRA (H) upper limits are referenced in
the text and in Figure~\ref{fig:GammasNine}.  The Ginga data point is
from Wang et al. (1992) and the radio data (labelled E \& M 85) are
from Erickson \& Mahoney (1985).  The GeV/TeV flux ratio \teq{F_{\rm
GeV}/F_{\rm TeV}= 590} obtained in this model is slightly lower than
that expected for an \teq{E_{\gamma}^{-1}} {\it flux} power-law due to
the prominence of the IC contribution.
 \label{fig:GammasAll} }       % fig. 9

\figureoutsmall{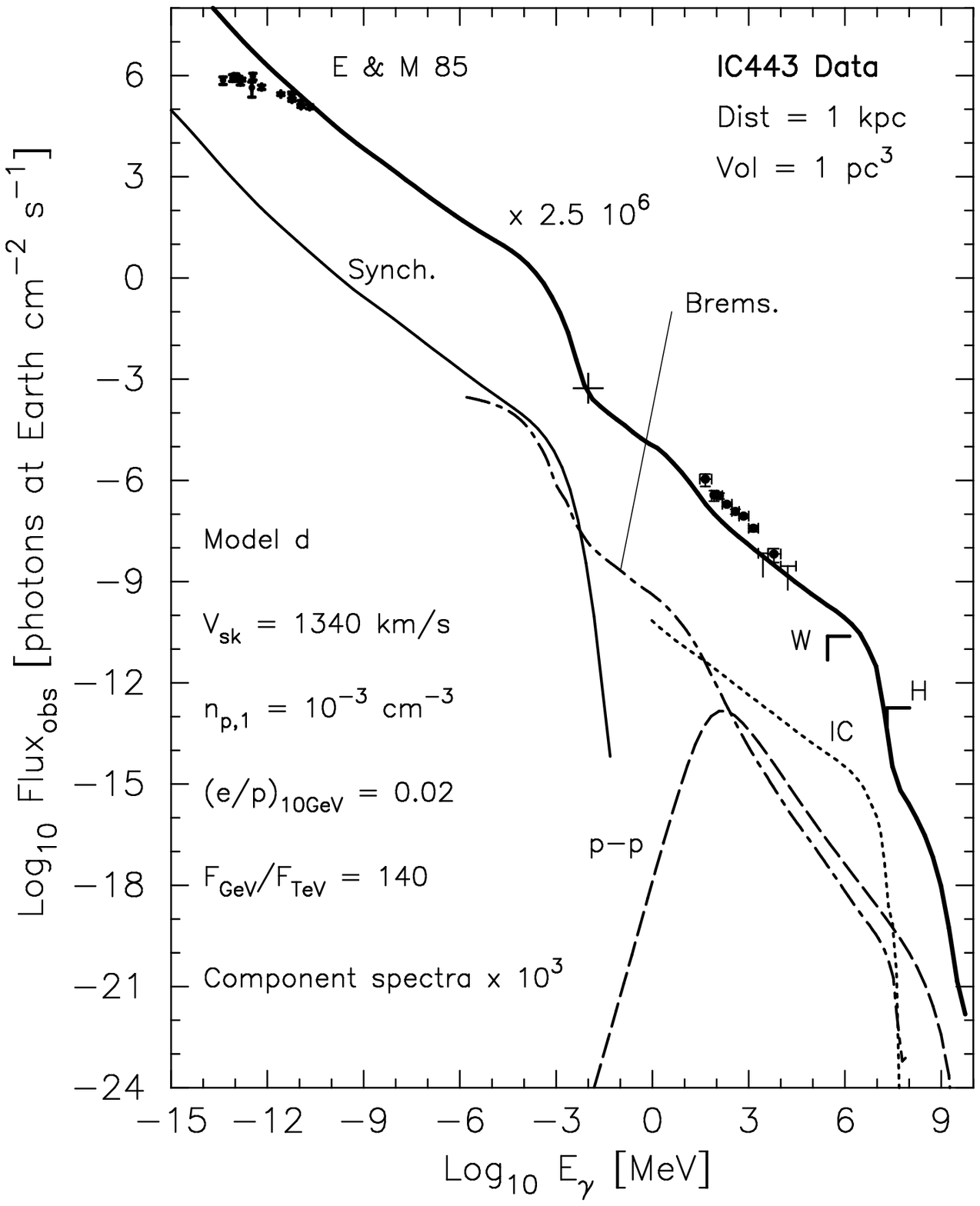}{
Photon spectra for our model producing cosmic rays up to the ``knee''
(i.e. Figure~\ref{fig:SpectraEmax}).  The line styles for the various
component spectra are as in Figure~\ref{fig:GammasAll}, as are the
data.  The component curves are all normalized to 1000 times a source
at 1 kpc with emission volume = 1 pc$^3$, but the total spectrum (heavy
solid line) is multiplied by \teq{2.5\times 10^6} (bottom panel) to
give fluxes more-or-less comparable to the EGRET levels for IC 443.
Note that since the density is very low in this example, in order to
give a high \teq{E_{\rm max}}, the inverse Compton component is very
prominent, yielding a low \teq{F_{\rm GeV}/F_{\rm TeV}} flux ratio.  As
in Figure~\ref{fig:GammasAll}, the p-p pion decay spectrum includes
contributions from p-He and He-He collisions.
 \label{fig:GammasMax} }       % fig. 10

\figureoutvsmall{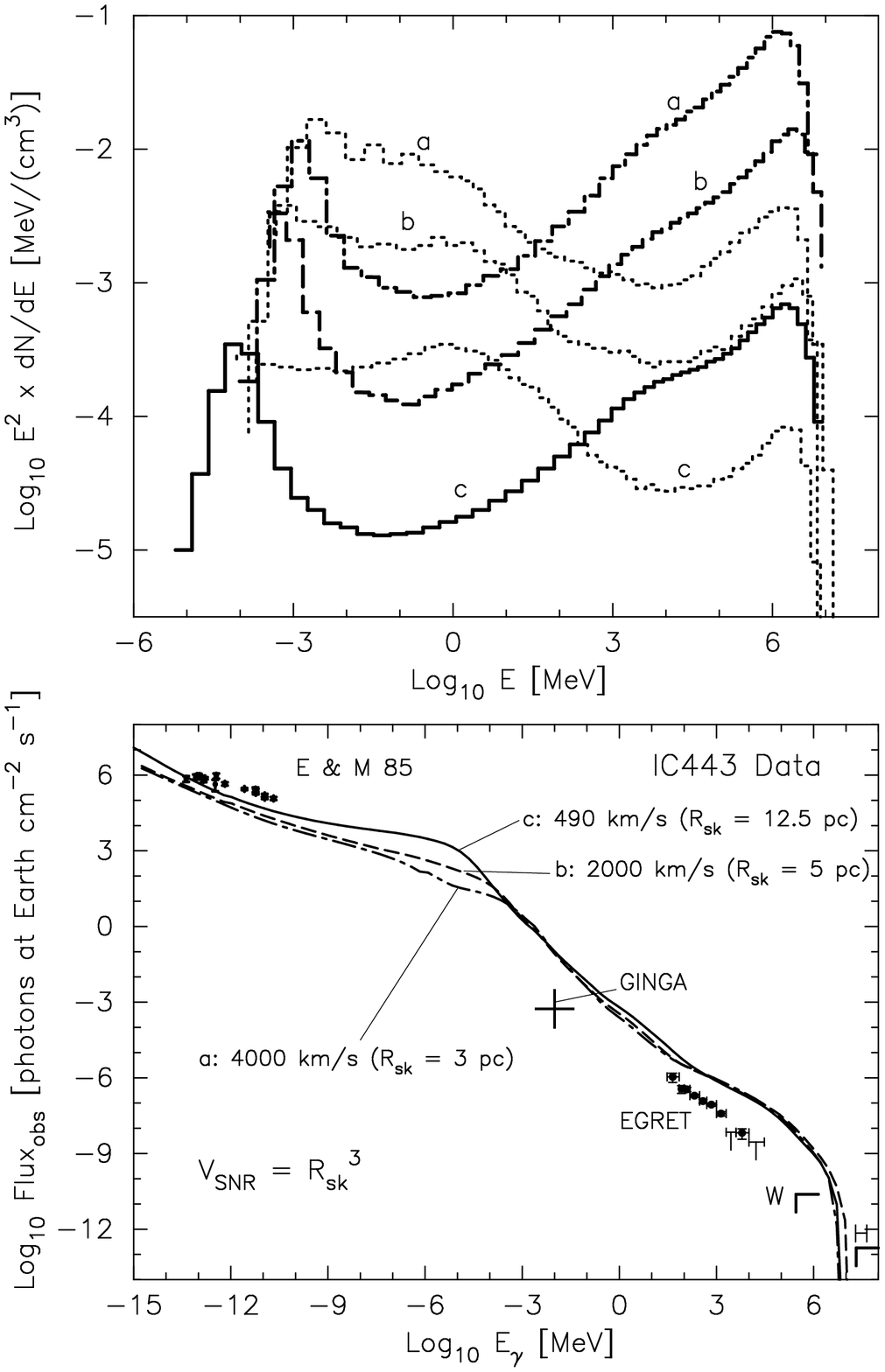}{
The evolutionary sequence corresponding to three of the profiles in
Figure~\ref{fig:fourprofiles}, and the particle distributions in
Figure~\ref{fig:SpectraThreePanels} (i.e. Models {\it a}, {\it b}, and
{\it c}).  The top panel shows the same proton (solid, dashed, and
dash-dot histograms) and electron (dotted histograms) spectra shown in
Figure~\ref{fig:SpectraThreePanels}, but multiplied by $E^2$ to
illustrate that the maximum energy density is in the highest energy
protons.  The bottom panel shows the total photon emission (i.e. the
sum of the bremsstrahlung, inverse Compton, pion decay, and synchrotron
emission) for these models with a source volume, $V_{\hbox{\fiverm
SNR}} = R_{\rm sk}^3$ (for each time) and a source distance \teq{d =
1}kpc.  This illustrates a property, probably the consequence of Sedov
evolution of a SNR, that the X-ray to hard gamma-ray photon spectra are
virtually independent of time between \teq{300} and \teq{10,000} years
of age.
 \label{fig:GamEvolve} }       % fig. 11

\end{document}